\documentclass[prx,twocolumn,superscriptaddress,longbibliography]{revtex4-2}
\usepackage{bm}
\usepackage{amsmath, amsfonts}
\usepackage{amssymb}
\usepackage{times}
\usepackage{graphicx}
\usepackage[table]{xcolor}
\usepackage{booktabs}
\usepackage{mathtools}
\usepackage{float}
\usepackage{xcolor}
\usepackage[colorlinks=true,allcolors=blue]{hyperref}
\usepackage[capitalise,nameinlink]{cleveref}
\usepackage{tikz}
\usepackage{soul}
\crefname{section}{Sec.}{Figs.}
\graphicspath{{Figures/}}
\crefname{equation}{Eq.}{Eqs.}
\Crefname{equation}{Equation}{Equations}
\crefname{figure}{Fig.}{Figs.}
\Crefname{figure}{Figure}{Figures}
\crefname{section}{Sec.}{Secs.}
\Crefname{section}{Section}{Sections}
\crefname{appendix}{Appendix}{Apps.}
\Crefname{appendix}{Appendix}{Apps.}
\crefname{paragraph}{Sec.}{Secs.}
\crefname{table}{Table}{Tables}

\begin{document}

\title{Moving beyond the transmon: Noise-protected superconducting quantum circuits}
\author{Andr\'as Gyenis}
\address{Center for Quantum Devices, Niels Bohr Institute, University of Copenhagen, 2100 Copenhagen, Denmark}
\address{Department of Electrical, Computer \& Energy Engineering, University of Colorado Boulder, CO 80309, USA}
\author{Agustin Di Paolo}
\address{Institut quantique \& D\'epartement de Physique, Universit\'e de Sherbrooke, Sherbrooke J1K2R1, Quebec, Canada}
\address{Research Laboratory of Electronics, Massachusetts Institute of Technology, Cambridge, MA 02139, USA}
\author{Jens Koch}
\address{Department of Physics and Astronomy, Northwestern University, Evanston, Illinois 60208, USA}
\author{Alexandre Blais}
\address{Institut quantique \& D\'epartement de Physique, Universit\'e de Sherbrooke, Sherbrooke J1K2R1, Quebec, Canada}
\address{Canadian Institute for Advanced Research, Toronto, M5G1M1 Ontario, Canada}
\author{Andrew A. Houck}
\address{Department of Electrical Engineering, Princeton University, Princeton, New Jersey 08544, USA}
\author{David I. Schuster}
\address{The James Franck Institute and Department of Physics, University of Chicago,  Chicago, Illinois 60637, USA}

\begin{abstract}
Artificial atoms realized by superconducting circuits offer unique opportunities to store and process quantum information with high fidelity. Among them, implementations of circuits that harness intrinsic noise protection have been rapidly developed in recent years. These noise-protected devices constitute a new class of qubits in which the computational states are largely decoupled from local noise channels. The main challenges in engineering such systems are simultaneously guarding against both bit- and phase-flip errors, and also ensuring high-fidelity qubit control. Although partial noise protection is possible in superconducting circuits relying on a single quantum degree of freedom, the promise of complete protection can only be fulfilled by implementing multimode or hybrid circuits. This Perspective reviews the theoretical principles at the heart of these new qubits, describes recent experiments, and highlights the potential of robust encoding of quantum information in superconducting qubits.
\end{abstract}

\maketitle

\section{Introduction}

Superconducting circuits are one of the most promising candidates for large-scale quantum computers, as they can reliably implement strongly interacting artificial atoms that have reduced susceptibility to environmental noise~\cite{martinis2020,kjaergaard2020,blais2020,krantz2019,wendin2017,clarke2008,schoelkopf2008,devoret2013}. The substantial effort invested in understanding and designing numerous types of superconducting quantum circuits has given rise to small-scale quantum processors that are able to execute proof-of-principle implementations of quantum algorithms~\cite{otterbach2017,kandala2017,neill2018,havlicek2019,arute2019,kjaergaard2020b}.  And yet, despite more than two decades of progress, most superconducting systems today consist simply of arrays of harmonic oscillators - microwave cavities - and slightly anharmonic oscillators - \textit{transmon} qubits~\cite{koch2007}. We have only begun to scratch the surface of the potential versatility of this platform.

Looking ahead, we envision a truly heterogeneous superconducting architecture, where the components that support the varying roles of logic, gates, memory, ancilla, reset, and readout are not necessarily the same.  In a heterogeneous architecture, component circuits are optimized for specific functions based on coherence, gate speed, anharmonicity, tunability, interaction strength, noise polarization and more.  Given the differing requirements for these functions, it is likely that different types of circuits will play roles in the separate parts of such heterogeneous systems.

\begin{figure}
    \centering
    \includegraphics[width = \columnwidth]{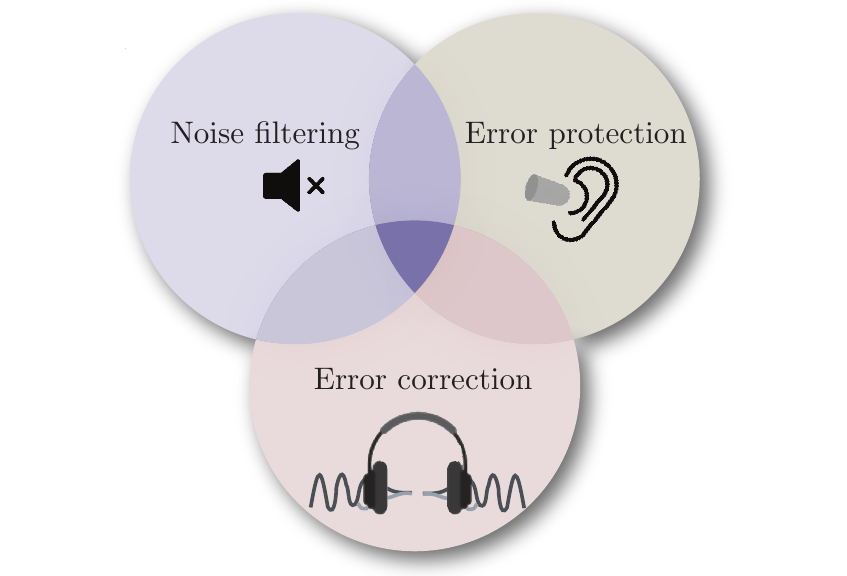}
    \caption{\textbf{Three common ways to enhance the lifetimes of superconducting qubits.} \textit{Noise filtering}: by utilizing various filtering techniques, shielding and quiet host materials, qubits can be isolated in noisy environments. \textit{Error protection}: by realizing systems where errors are energetically unfavorable, qubits can be protected from decoherence at the hardware level. \textit{Error correction}: by redundantly encoding the quantum information in multiple local objects, logical qubits can be implemented such that errors are actively detected and corrected. Quantum processors can benefit from all three approaches.}
    \label{fig:intro}
\end{figure}

How can we ensure that the various blocks of these systems are resilient against environmental noise? There are, quite generally, three strategies to reduce the rate at which errors occur (\cref{fig:intro}): (i) reducing the amplitude of the noise itself (\textit{noise filtering})~\cite{houck2008,oliver2013,quintana2014,braumuller2020,place2020,rosenberg2020}, (ii) boosting the qubits’ immunity against environmental noise (\textit{hardware-level error protection})~\cite{kitaev2003,ioffe2002a,ioffe2002b,doucot2002,doucot2005,gladchenko2008,doucot2012,bell2014}, and (iii) detecting and actively correcting errors (\textit{quantum error correction})~\cite{shor1995,fowler2012,reed2012,chow2014,saira2014,kelly2015,ofek2016,gambetta2017,gottesman2001,campagne2020}. These strategies are not mutually exclusive and can benefit from each other when employed in tandem. For example, using more reliable qubits and understanding the noise in these systems is paramount in reducing the physical qubit overhead in quantum error correction. Here, our discussion focuses on these protected qubits that have the potential as building blocks for heterogeneous quantum systems.

The evolution of the transmon itself shows the power of hardware-level noise protection. A critical feature that made this circuit exceptionally successful is its resilience against the most detrimental noise affecting solid-state qubits: charge noise. Its ancestor, the Cooper pair box~\cite{shnirman1997,bouchiat1998,nakamura1999}, which is composed of a Josephson junction shunted by a small capacitance, was used in the very first experiment that demonstrated quantum coherence in a solid-state qubit but was severely limited by charge-noise decoherence. The first stride towards enhanced coherence times was achieved by the quantronium circuit~\cite{vion2002}. This qubit showed a dramatic improvement of the coherence times when operated at a special working point--at the \textit{sweet spot}--with first-order insensitivity to charge fluctuations. By introducing a large shunt capacitor, the transmon qubit extended the quantronium's protection from first to all orders~\cite{koch2007}, providing excellent coherence but remaining susceptible to relaxation--a mechanism that ultimately limits the coherence of this device. For fifteen years, this qubit has remained ubiquitous, but with gradual improvement earned through the painstaking labor of noise filtering, microwave engineering and advancements in materials.

As the development of the transmon highlights, hardware-level protection is a cornerstone of the field of superconducting qubits. But why stop at protection against only dephasing? Can we realize and operate fully noise-protected devices where both bit-flip and phase-flip errors are suppressed? In this Perspective article, we discuss the potential of achieving such intrinsic noise protection in superconducting qubits, starting from the simplest circuits and expanding our discussion towards more complex devices. These next-generation superconducting qubits represent a new paradigm shift with emphasis on noise immunity rather than noise filtering. While such paradigm shifts often lead to trade-offs--as it was the case for the transmon with substantially reduced anharmonicity and weaker dispersive shifts--, we expect that future innovations will establish protected qubits as a critical component for heterogeneous superconducting quantum systems.

\section{Basics of noise protection \label{noise_protection}}

To understand the underlying principles of hardware-level protection, we first review the types of errors that affect qubits. If~$|0\rangle$ and~$|1\rangle$ are the two computational basis states of a qubit, we can encode the quantum information into the superposition of these states, such that~$|\psi\rangle = \cos\left(\theta/2\right)|0\rangle + \sin\left(\theta/2\right) e^{i\phi}|1\rangle$. Here,~$\tan\left(\theta/2\right)$ describes the relative amplitude and~$\phi$ the relative phase between the basis states. Thus, we need to preserve both the relative
\textit{phase} and \textit{amplitude} between the basis states to protect the superposition.  In the transmon, for example, phase-flip errors are suppressed by design: the frequency of this qubit is insensitive to fluctuations in its control parameters, resulting in little phase-flip errors.
However, there is no built-in protection against bit-flip errors as incoherent transitions between the qubit states are not suppressed. On the other hand, the situation is reversed for some variants of the flux qubit which demonstrate robustness against bit-flip errors but little or no protection against phase-flip errors~\cite{federov2011,earnest2018,lin2018}.

There are two different timescales that characterize the qubit's resilience against
bit- and phase-flips: the depolarization time~$T_1$ and the pure-dephasing time~$T_\varphi$. As the total coherence time of a qubit,~$T_2=(1/2T_1 + 1/T_\varphi)^{-1}$, is related to both mechanisms, it is necessary to simultaneously protect against both types of errors. We can quantify the timescale of these error processes with (1) the qubit's sensitivity to the noise and (2) the strength of the noise. To do so, we denote  $\lambda$ as an external parameter entering the circuit Hamiltonian that can tune the energy of the qubit. In superconducting circuits, for example, the most common external knobs are the gate voltage biasing an island of the circuit~$V_\mathrm{gate}$ or the magnetic flux in a loop of the device~$\Phi_\mathrm{ext}$. The qubit can be biased at a value $\bar{\lambda}$ of this control parameter, around which there are fluctuations $\delta\lambda=\lambda-\bar{\lambda}$. If the amplitude of the noise is small enough, it is possible to safely assume that the coupling between a qubit operator $\hat{\mathcal{O}}$ and the noise is linear, $\hat{H}_\mathrm{int}\propto\delta\lambda\times\hat{\mathcal{O}}$, where $\hat{\mathcal{O}}$ could be, for example, the charge or phase operator. In this situation, the corresponding dephasing and depolarization rates are~\cite{ithier2005}
\begin{subequations}
\begin{align}
    1/T_\varphi^\lambda & \propto \left|\frac{\partial E_{01}}{\partial\lambda}\right|^2 S_{\lambda}(\omega\to 0) \label{eq:coherence_phi},\\
    1/T_1^\lambda & \propto \left|\langle0|\hat{\mathcal{O}}|1\rangle\right|^2 S_{\lambda}(\omega=E_{01}/\hbar), \label{eq:coherence_1}
\end{align}
\end{subequations}
where~$S_{\lambda}(\omega)$ is the power spectral density of the noise as a function of frequency $\omega/2\pi$ and~$E_{01}$ is the qubit transition energy.

While noise filtering aims at reducing the amplitude of the noise by shielding qubits from its environment [i.e.~reduce the power spectral densities in the right-hand side of~\cref{eq:coherence_phi} and~\cref{eq:coherence_1}], the key idea behind hardware-level noise protection is to suppress the susceptibility of the qubit to the noise [i.e.~reduce the first terms on the right-hand side of~\cref{eq:coherence_phi} and~\cref{eq:coherence_1}].

\begin{figure}
    \centering
    \includegraphics[width = \columnwidth]{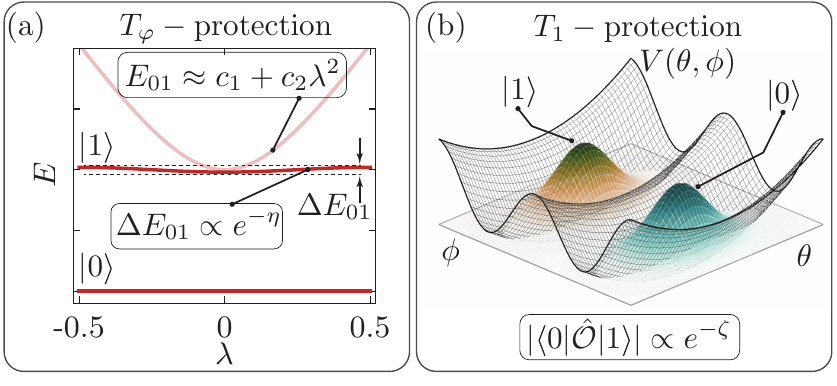}
    \caption{\textbf{Principles of noise protection: suppressed energy dispersion and wavefunctions with disjoint support.} (a) Energy levels of a qubit as a function of an external parameter~$\lambda$, such as external flux or offset charge. Dark red lines illustrate the case where the dispersion of the qubit energy level with respect to $\lambda$ is exponentially small, i.e.~the difference between the maximum and minimum transition energies $\Delta E_{01}\propto e^{-\eta}$, where~$\eta$ is related to the circuit parameters of the qubit and can be made large by design. On the same plot, the light red lines show a
    dispersion with a first-order sweet spot and quadratic dispersion, $E_{01}\approx c_1 + c_2 \lambda^2$, where~$c_1, c_2$ are constants. To ensure that sweet-spot protection is sufficient for reaching high-coherence times, the quadratic term needs to be small compared to the scale of the noise amplitude~$\delta\lambda$, such that~$c_2\delta\lambda^2/E_{01}(0)\ll 1$. (b) Schematics of the potential landscape of a qubit (black grid) with two quantum degrees of freedom~$\phi$ and~$\theta$. Protection against bit-flip errors can be achieved by localizing the qubit wavefunctions in distinct regions of the potential. Such states with disjoint support have exponentially small transition matrix elements,~$\left|\langle0|\hat{\mathcal{O}}|1\rangle\right|\propto e^{-\zeta}$, where due to the design and parameter choices of the circuit~$\zeta$ is a large number.}
    \label{fig:noise_protection}
\end{figure}

We illustrate the concept of error protection by first considering phase-flip errors
[\cref{fig:noise_protection}(a)]. As~\cref{eq:coherence_phi} shows, dephasing arises as a consequence of the dispersion of the qubit energy~$E_{01}(\lambda)$ with respect to the external parameter~$\lambda$. Invoking the series expansion~$E_{01}(\lambda) = E_{01}(\bar\lambda) + \sum_{k=1}^\infty \partial_\lambda^k E_{01}(\bar{\lambda})  \delta\lambda^k/k!$, we see that, generally, the first-order term dominates as long as the amplitude of the fluctuation is weak. The first strategy~\cite{vion2002,ithier2005} to increase the dephasing time is to operate the qubit at a sweet spot where~$\partial_\lambda E_{01}(\bar{\lambda})=0$, such that the leading-order contribution to the qubit dispersion is of second order [light red lines in~\cref{fig:noise_protection}(a)]. Further protection against dephasing requires that the higher-order contributions become negligible, which can be guaranteed by exponentially vanishing dispersion of the qubit transition energy [dark red lines in~\cref{fig:noise_protection}(a)].

On the other hand, bit-flip protection requires the transition matrix element~$\left|\langle0|\hat{\mathcal{O}}|1\rangle\right|$ in~\cref{eq:coherence_1} to vanish. Central to this requirement is the concept of disjoint support of the qubit eigenfunctions.
Whenever two wavefunctions localize in separate ``spatial" regions [\cref{fig:noise_protection}(b)], typical matrix elements between them are dominated by the exponentially small tails and hence significantly reduced in magnitude. This generally leads to long relaxation times [\cref{eq:coherence_1}]. Useful terminology to describe this situation is the notion of (nearly) disjoint support~\cite{dempster2014}. Recall that the support of a function $f\colon X\to \mathbb{C}$ is the subset $S\subset X$ on which the function has non-zero values. Outside the support, $f$ vanishes, $f(X\setminus S)=0$. In the case of two wavefunctions $\psi_0(\phi)$ and $\psi_1(\phi)$ localized in separate regions, their respective supports are nearly disjoint, $S_0\cap S_1\approx \emptyset$. The (near-)disjointness can be quantified via the integral $\int d\phi\, |\psi_0(\phi)|^2|\psi_1(\phi)|^2\ll1$, and results in suppressed matrix elements $|\langle0|\hat{\mathcal{O}}|1\rangle|\ll1$ as long as $\hat{\mathcal{O}}$ is a local operator like position or momentum operator (as opposed to non-local operators such as displacement operators). Since the common noise channels in superconducting qubits are associated with such local operators, engineering logical qubit wavefunctions with disjoint support is a highly effective strategy to reach bit-flip protection.

Alternatively, symmetries can prohibit qubit transitions when the qubit states are encoded into different eigenspaces of the symmetry operator of the Hamiltonian~\cite{kitaev2003,ioffe2002a}. In this case, the transition matrix elements vanish as long as the noise operators preserve the symmetries of the system.

\begin{figure}[!b]
    \centering
    \includegraphics[width = \columnwidth]{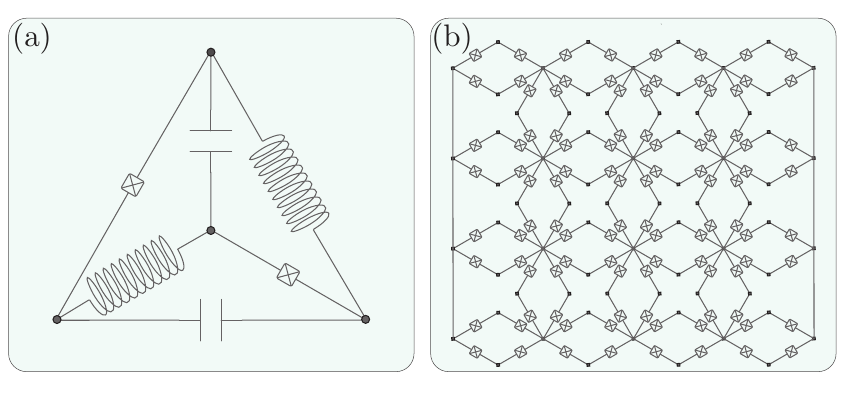}
    \caption{\textbf{Strategies proposed to realize noise-protected circuits.} (a) A compact qubit composed of a few large elements can host protected states if the circuit components fulfill stringent energy requirements~\cite{brooks2013}. (b) Periodic arrays of Josephson junctions and capacitors offer robust quasi-degenerate ground states when the number of unit cells is large~\cite{kitaev2006}.}
    \label{fig:encoding_strategies}
\end{figure}

Circuit-level protection against these two types of error mechanisms shares similarities to the redundant encoding utilized in quantum error correction. Instead of using multiple qubits or multiple energy levels for redundancy, here, large circuit elements or circuits with many degrees of freedom lead to protection against local noise. Accordingly, we consider two possible strategies for realizing noise-protected superconducting circuits: \textit{compact protected qubits}~\cite{brooks2013,smith2020,kalashnikov2020} and \textit{protected multi-junction arrays}~\cite{kitaev2003,ioffe2002a,ioffe2002b,doucot2002,doucot2005,gladchenko2008,doucot2012,bell2014}. Compact protected qubits utilize only a few circuit elements which are challenging to realize (for example, large linear inductors so that the inductive energies define the smallest energy scale in the circuit)  [\cref{fig:encoding_strategies}(a)]. In contrast, arrays involve a larger number of elements but in convenient parameter regimes [\cref{fig:encoding_strategies}(b)]. In this Perspective article, our discussion focuses on the first approach, while the article by Dou{\c{c}}ot and Ioffe has extensively covered the latter case~\cite{doucot2012}.
Finally, we also note that qubits relying on continuous microwave tones can exploit the combination of hardware-level error protection and autonomous error correction~\cite{ofek2016,mirrahimi2014,puri2017,campagne2020,grimm2020,mundada2020,huang2020}.

\section{Limitations of single-mode superconducting qubits \label{design}}

The simplest superconducting qubits are single-mode devices that contain one or more superconducting elements connected in parallel between two nodes. We dedicate this section to qualitatively study their coherence properties and limitations. We will see that--although their layout is simple and fixed--changing the parameters of their circuit elements can drastically change their coherence properties. These circuits offer protection against only one error channel (either bit-flip or phase-flip processes), and cannot accommodate the requirements for simultaneous protection against both~\cite{dipaolo2020}.  Larger multimode circuits or hybrid semiconducting-superconducting devices can mitigate the limitations of these simple qubits (\cref{full_protection} and ~\cref{hybrid}).

\subsection{Islands and loops as the charge and flux degrees of freedom}

\begin{figure}[!t]
    \centering
    \includegraphics[width = \columnwidth]{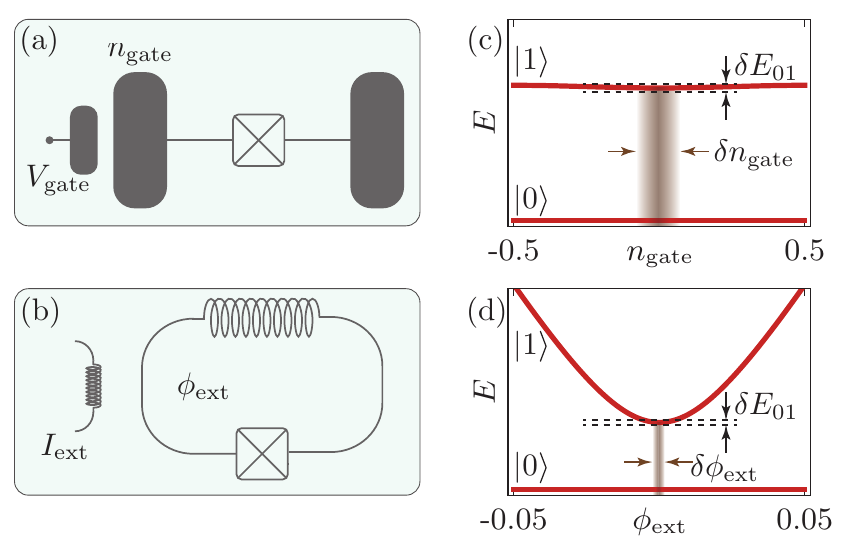}
    \caption{\textbf{The elementary superconducting qubits: the charge-mode and the flux-mode qubit.} (a) A circuit containing a pair of isolated islands connected by a Josephson junction represents a charge mode with sensitivity to offset charge~$n_\mathrm{gate}$. (b) The presence of a loop in a circuit leads to a flux mode, which is susceptible to external flux~$\phi_\mathrm{ext}$. (c)-(d) Typical energy dispersion of charge- and flux-mode qubits. In solid state systems, the amplitude of charge noise (illustrated with brown shaded regions) is about two orders of magnitude larger than that of flux noise, when compared to the periodicity of the qubit Hamiltonian,~$\delta n_\mathrm{gate}/\delta\phi_\mathrm{ext}\approx 10^2$~\cite{clerk2010,manucharyan2012,kou2017,slichter2012,yan2016,quintana2017,astafiev2004,shnirman2005,christensen2019,you2021}. Because of this, reaching high coherence times for charge modes requires exponentially suppressed charge dispersion, while for flux modes even sweet-spot protection can offer high coherence. At the same time, experimentally realizing suppressed flux dispersion is significantly more challenging than achieving charge-noise insensitivity.}
    \label{fig:charge_flux}
\end{figure}

Depending on the geometry of the underlying circuit, superconducting qubits utilize two types of quantum degrees of freedom. On the one hand, two galvanically disconnected pieces of superconductors define a pair of \textit{islands}, a capacitor [\cref{fig:charge_flux}(a)]. On the other hand, a closed superconducting wire identifies a \textit{loop}, an inductor [\cref{fig:charge_flux}(b)]. An island can store an integer number of Cooper pairs, while a loop can host an integer number of flux quanta. These elementary charge and flux states of the islands and the loop can be used to store quantum information. However, an additional nonlinear circuit component is necessary to isolate two states in the excitation spectrum
of these circuits to serve as the computational states of a qubit. Here, we focus on qubits that use Josephson junctions as the nonlinear elements, which are formed by growing a thin insulating layer between two superconductors, allowing for the tunneling of Cooper pairs or flux quanta through the layer.

These two types of circuit geometries define the two elementary superconducting qubits. A pair of superconducting islands with capacitance~$C$ connected by a Josephson junction constitutes a \textit{charge mode} [\cref{fig:charge_flux}(a)]. The Hamiltonian describing this mode is
\begin{equation}\label{eq:charge-mode}
\hat{H}_\mathrm{charge} = 4E_C(\hat{n}-n_\mathrm{gate})^2 -E_J\cos\hat{\theta},
\end{equation}
where~$E_C=e^2/2C$ is the charging energy,~$e$ is the electron charge,~$E_J$ is the Josephson energy associated with Cooper pair tunneling across the junction,~$n_\mathrm{gate}$ is the induced offset charge on the islands, while~$\hat{n}$ and~$\hat{\theta}$ are the canonical charge and phase operators. The charge mode couples to the environment through the offset charge~$n_\mathrm{gate}$. This quantity can be used to dc-bias the qubit and manipulate its logical state, but its value is also uncontrollably influenced by the surrounding electrostatic potential. To understand the different parameters regimes of this Hamiltonian, it is useful to interpret~\cref{eq:charge-mode} such that it describes a fictitious particle with position~$\theta$ and mass~$C$ moving in a cosine potential.

\begin{figure*}
    \centering
    \includegraphics[width = \textwidth]{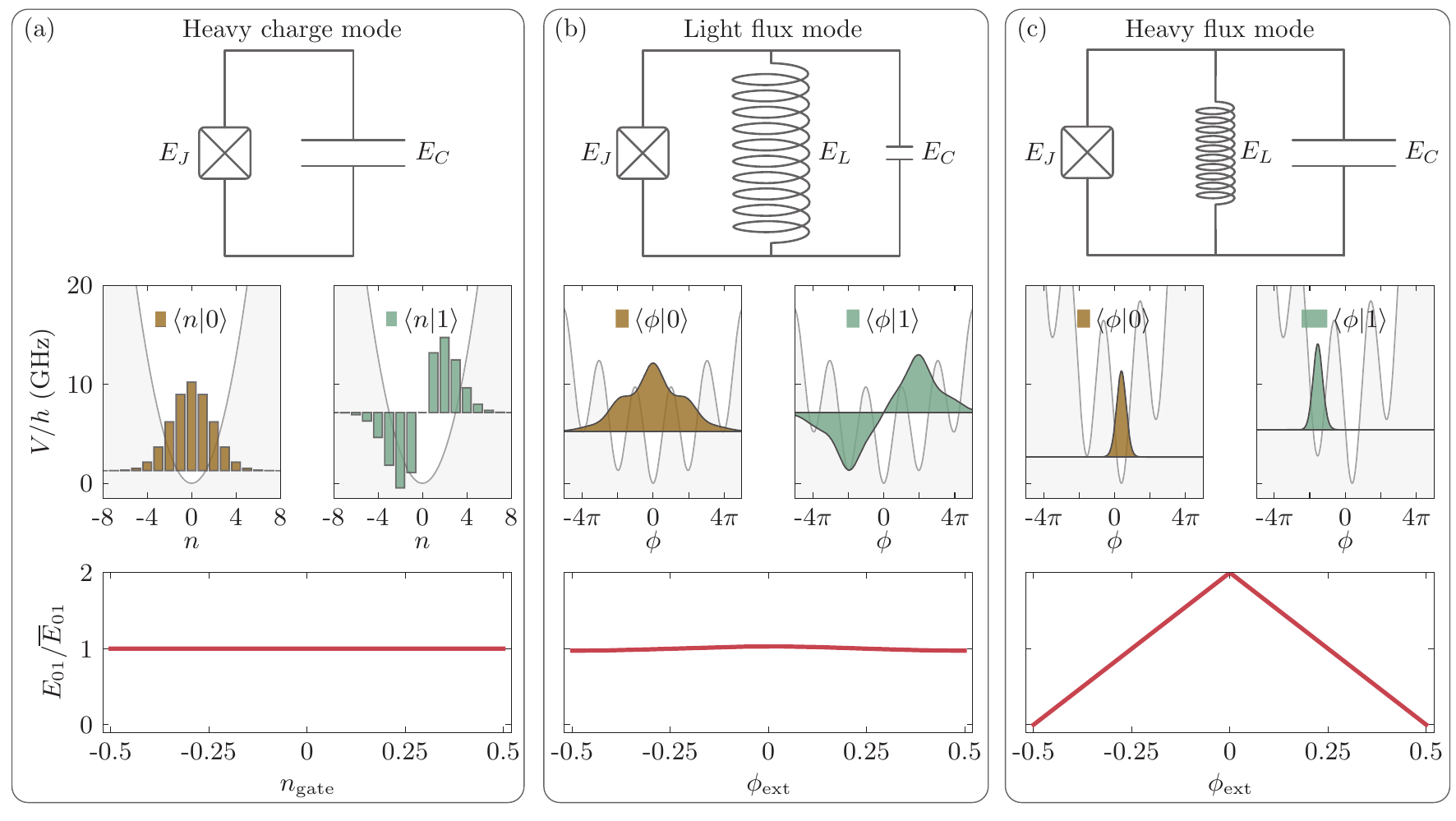}
    \caption{\textbf{Partial protection of charge and flux modes.} (a) The heavy charge mode (transmon qubit) consists of a single Josephson junction shunted by a large capacitor. The qubit is operated in the~$E_C\ll E_J$ regime, and the wavefunctions are distributed over many charge states (middle panels; shaded regions shows the charging energy), leading to reduced charge dispersion (bottom panel;~$\bar{E}_{01}$ corresponds to the mean transition energy). (b) The light flux mode (blochnium qubit) and (c) the heavy flux mode (heavy fluxonium qubit) have the same circuit layout but are operated in different regimes. In the light flux mode, the Josephson junction is shunted by a large inductance and small capacitance ($E_L\ll E_J<E_C$), and the wavefunctions are spread over multiple potential valleys (middle panel; shaded regions shows the potential energy). On the other hand, the heavy flux mode has a large shunting capacitor ($E_C\approx E_L\ll E_J$), and the qubit eigenfunctions are localized in separate wells at external flux values away from half flux quantum. This also results in much stronger flux dispersion for the heavy fluxonium (bottom panels). Parameters: (a)~$E_C/h$ = 0.2 GHz,~$E_J/h=$ 20 GHz, (b)~$E_C/h$ = 7.07 GHz,~$E_J/h=$ 4.7 GHz,~$E_L/h$ = 0.067 GHz~\cite{pechenezhskiy2020} and (c)~$E_C/h$ = 0.46 GHz,~$E_J/h=$ 8.11 GHz,~$E_L/h$ = 0.24 GHz~\cite{earnest2018}.}
    \label{fig:partial_protected_qubits}
\end{figure*}

Similarly, a loop made of a superconducting wire with inductance~$L$ and interrupted by a Josephson junction implements a \textit{flux mode} [\cref{fig:charge_flux}(b)]. The Hamiltonian of the circuit reads
\begin{equation}\label{eq:flux-mode}
\hat{H}_\mathrm{flux} = 4E_C\hat{n}^2 - E_J\cos\hat{\phi} +E_L(\hat{\phi}-2\pi\phi_\mathrm{ext})^2/2,
\end{equation}
where~$\hat{\phi}$ is the phase operator,~$E_L=\Phi_0^2/4\pi^2L$ is the inductive energy,~$\Phi_0=h/2e$ is the flux quantum,~$h$ is the Planck's constant,~$\phi_{\mathrm{ext}}$ is the reduced external flux defined as~$\phi_{\mathrm{ext}}=\Phi_\mathrm{ext}/\Phi_0$, and~$\Phi_\mathrm{ext}$ is the magnetic flux piercing the loop. The flux mode is coupled to the environment through the external flux $\phi_{\mathrm{ext}}$, which plays a similar role as the offset gate charge in~\cref{eq:charge-mode}. With the addition of the inductor, the potential energy of the effective phase particle is now a corrugated harmonic well, illustrated in~\cref{fig:partial_protected_qubits}(b) and (c) for two parameter regimes.

\subsection{High coherence of charge and flux modes}

How can we ensure that a charge or a flux mode is protected against environmental noise? Building on the discussion in~\cref{noise_protection}, we first consider the principles of noise protection against phase-flip errors and describe the conditions for realizing offset-charge insensitive charge modes and external-flux insensitive flux modes. To simplify the discussion, here we neglect bit-flip errors such as losses due to imperfect capacitors, inductors and Josephson junctions and assume that only the environmental fluctuations of the external bias parameters limit the qubit coherence.

By inspecting~\cref{eq:charge-mode} and~\cref{eq:flux-mode}, we can describe the coupling between charge or flux modes and the environment by the Hamiltonians~$\hat{H}_\mathrm{int}=-8E_C \hat{n}\times \delta n_\mathrm{gate}$ or~$\hat{H}_\mathrm{int}=-E_L\hat{\phi}\times \delta\phi_\mathrm{ext}$, respectively. Here~$\delta n_\mathrm{gate}$ and~$\delta\phi_\mathrm{ext}$ represent fluctuations of the control parameters. Focusing first on charge modes, a strategy to boost the coherence time relies on biasing the qubit at~$\bar n_\mathrm{gate}$ where the expectation value of the qubit's dipole operator is equal for the two logical states,~$\langle1|\hat{n}|1\rangle \approx \langle0|\hat{n}|0\rangle$. In this case, the change of the qubit transition frequency under the perturbation~$\hat{H}_\mathrm{int}$,~$\left|\langle1|\hat{H}_\mathrm{int}|1\rangle - \langle0|\hat{H}_\mathrm{int}|0\rangle\right|$, vanishes or, equivalently, the first derivative of the energy dispersion is zero.  Intuitively, in this situation, dephasing is minimized because the two qubit states cannot be distinguished (by an observer or the environment) from a measurement of the number of charges in the islands. For a charge mode, such sweet spots occur at biases where adjacent charge states are degenerate. Cooper pair tunneling hybridizes these two states leading to an avoided crossing with the usual quadratic dependence of the energy with the bias parameter.

By increasing the Josephson energy with respect to the charging energy--thus enhancing the rate at which Cooper pairs can tunnel--, we can achieve a more robust protection against dephasing~\cite{koch2007}. When~$E_J/E_C \gg 1$, hybridization occurs over several charge states resulting in qubit's logical states with support over many charges [\cref{fig:partial_protected_qubits}(a)]. A consequence of this delocalization over charge states is that the qubit energy levels are now exponentially flat at all bias points. For charge modes, this ``sweet-spot-everywhere'' condition is the transmon regime~\cite{koch2007,schreier2008}. In practice, with typical values of~$E_J/E_C\sim$ 50, transmon qubits feature dephasing times as large as~$T_{2}=100-200~\mu$s with energy relaxation times of~$T_1=200-500~\mu$s~\cite{place2020, wang2021}.

Coming back to the fictitious phase particle of mass~$C$, we refer to the transmon regime as the \textit{heavy} regime because it requires a large effective mass. Since flux modes couple to the environment through the phase operator (as opposed to the charge operator),  high-coherence requires delocalized eigenfunctions in phase space instead of in charge space. Thus, we arrive at the conclusion that protection against dephasing in a flux mode is realized by working in the opposite, \textit{light} regime, where~$E_J/E_C \ll 1$ and~$E_J/E_L\gg 1$~\cite{koch2009}. In this situation, the phase particle is delocalized over multiple wells of the corrugated harmonic potential leading to insensitivity to flux noise [\cref{fig:partial_protected_qubits}(b)]. As a result, while a heavy charge mode needs a large shunting capacitor, a light flux mode requires a large linear inductive element. Crucially, the large inductance should not be accompanied in this case by a large capacitance, something which can be challenging to realize in practice because of the stray capacitance of any superconducting circuit component. Large inductors with small parasitic capacitances are known as superinductors~\cite{manucharyan2009,peruzzo2020}. For instance, the experimentally realized light flux mode, the blochnium qubit relies on the Josephson-junction array being released from its substrate to minimize the capacitance to ground~\cite{pechenezhskiy2020}. To maximize the qubit coherence time, the superinductor must be engineered such that its self-resonant modes are far detuned from the qubit operation frequency~\cite{masluk2012,hazard2019}. Furthermore, in the case of Josephson-junction-array-based superinductances, special care must be taken to avoid introducing charge dispersion to the qubit transition freqeuncy due to quantum phase slips in the superinductance~\cite{manucharyan2012,di2021}.

\subsection{Fundamental asymmetry between charge and flux noise}

When identifying a desirable parameter regime of a superconducting qubit, we also need to take into account the amplitude associated with the charge and flux fluctuations [\cref{fig:charge_flux}(c)-(d)]. Experimentally, the amplitude of flux fluctuations (in units of~$\Phi_0$) is typically about two orders of magnitude smaller than that of charge fluctuations (in units of~$2e$)~\cite{clerk2010,manucharyan2012,kou2017,slichter2012,yan2016,quintana2017,astafiev2004,shnirman2005,christensen2019,you2021}. This is the consequence of the low-impedance of the electromagnetic environment surrounding the qubits~\cite{manucharyan2012superinductance}. As a result, the scale of the relevant noise fluctuations is smaller in flux modes than in charge modes, relative to the periodicity of their Hamiltonians [\cref{fig:charge_flux}(c)-(d)]. Thus, we expect the flux-mode qubits to have longer coherence times than charge-mode qubits for equal energy dispersion. Thanks to this observation, it is possible to operate flux-mode qubits away from the optimal light regime while maintaining large coherence times.  As an example, the \textit{fluxonium} qubit is defined in the regime~$E_J/E_C\sim 5-10$ and~$E_L/E_C\sim 0.5-1$~\cite{manucharyan2009}, which is not sufficient to reach exponential flux-noise insensitivity. Still, the coherence time at a magnetic sweet spot has been reported to be longer than in the transmon with best coherence times in the millisecond regime~\cite{nguyen2019,zhang2020,Somoroff2021}.

\subsection{Bit-flip limit in noise-insensitive single-mode devices}

In the noise-insensitive regime of both charge- and flux-mode qubits, the delocalization of the qubit states over charge or flux states results in a large transition matrix element. While this means that these qubits can be strongly coupled to electromagnetic modes to realize fast gates and readout, it also enhances the rate of bit-flip errors [\cref{eq:coherence_1}]. In these qubits, bit-flip errors can, in fact, become the main contribution to the qubit decoherence with~$T_2 \sim 2T_1$. For example, the higher coherence times in the fluxonium compared to the transmon is related to the longer relaxation times in the fluxonium circuit, which is mostly the result of its smaller charge transition matrix element~\cite{nguyen2019,Somoroff2021}.

An approach to further improve the situation is to artificially reduce the size of the qubit transition matrix element, essentially trading phase-flip protection for bit-flip protection. For example, in the heavy flux-mode qubit known as the \textit{heavy fluxonium}~\cite{federov2011,earnest2018,lin2018,hazard2019,zhang2020}, the larger~$E_J/E_C$ ratio significantly weakens the tunneling between neighboring potential valleys of the corrugated harmonic potential. Away from the flux sweet spot, this leads to long-lived states that are localized in single wells with lifetimes in the several milliseconds regime [\cref{fig:partial_protected_qubits}(c)]. As expected, this increase in~$T_1$ is obtained at the expense of reducing the pure dephasing times~$T_\varphi$: the heavy fluxonium shows a few microseconds coherence time away from the sweet spot due to the linear flux dispersion.

Moreover, the approach to realizing logical gates on heavy fluxonium highlight an important aspect of protected circuits: the disjoint nature of the qubit states complicates the control of the device. Indeed, while in the transmon, single qubit gates simply rely on driving the qubit at its transition energy, the suppressed tunneling matrix elements in the heavy fluxonium makes gates based on direct transition between the two logical state impractically long or outright impossible. An alternative approach is to break the protection during the gate, for example, by using Raman-type gates that rely on virtually populating higher lying levels that couple to both qubit states~\cite{earnest2018,hazard2019}.

\begin{figure}[t]
    \centering
    \includegraphics[width = \columnwidth]{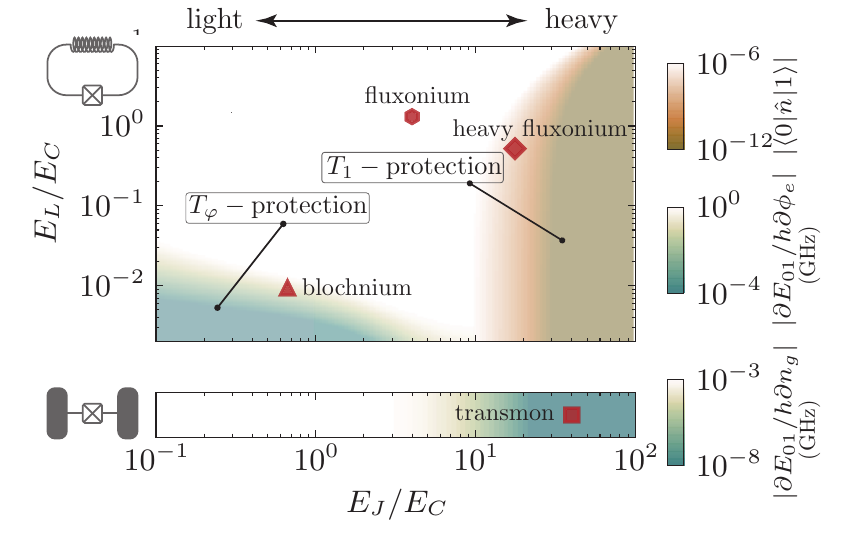}
    \caption{\textbf{Noise protection in single-mode devices.} Top (bottom) panel: coherence-time landscape for a flux- (charge-)mode qubit. Cyan colormap shows the slope of the energy dispersion at~$\phi_\mathrm{ext}=0.25$ (top panel) and at~$n_\mathrm{gate}=0.25$ (bottom panel), while the brown region in the top panel indicates the transition matrix element at~$\phi_\mathrm{ext}=0.45$ ($E_C/h=1$ GHz). In the case of the flux mode, the qubit is exponentially insensitive to flux fluctuations at small~$E_L/E_C$ and small~$E_J/E_C$ parameters (light flux mode), realizing phase-flip protection. For large~$E_J/E_C$ (heavy flux mode), instead, the qubit is protected against relaxation away from the sweet spot (bit-flip protection). In the case of the charge mode, the situation is reversed as the heavy regime supports offset-charge-noise insensitivity. There is no protection in single charge modes against bit-flip errors. The markers indicate experimentally realized devices: blochnium (triangle)~\cite{pechenezhskiy2020}, fluxonium (hexagon)~\cite{nguyen2019}, heavy fluxonium (diamond)~\cite{earnest2018} and transmon (square)~\cite{koch2007}.}
    \label{fig:partial_protected_phasediagram}
\end{figure}

The competition between bit-flip and phase-flip processes in single-mode devices can be illustrated by mapping the transition matrix element and energy dispersion as a function of the circuit parameters (\cref{fig:partial_protected_phasediagram}). While charge-mode qubits are protected against charge-noise dephasing in the heavy regime where~$E_J/E_C$ is large, flux modes are insensitive to flux noise in the light regime, where both~$E_J/E_C$ and~$E_L/E_C$ are small. Additionally, flux modes can also be protected against bit-flip processes but only in the heavy regime. This figure summarizes a key point of this discussion: simultaneous protection against energy relaxation and pure dephasing cannot be realized in single-mode qubits.

\section{Towards full noise protection with multimode circuits}\label{full_protection}

\begin{figure*}
    \centering
    \includegraphics[width = \textwidth]{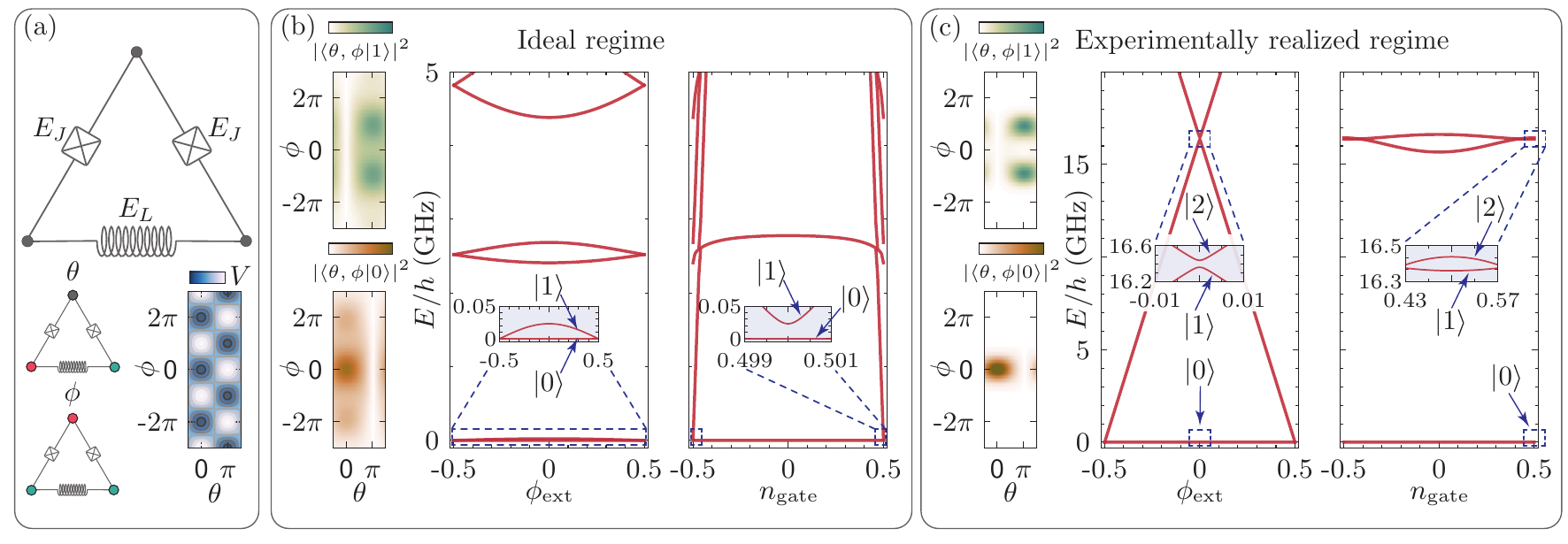}
    \caption{\textbf{The ideal and the experimentally realized bifluxon qubit.} (a) The bifluxon circuit contains two Josephson junctions and one superinductor in a loop and has two normal modes. (Bottom left: the colors of the nodes indicate the sign of the normal-mode amplitudes. Red is positive, green is negative and grey is zero.) Bottom right: the potential landscape of the qubit.  (b) The qubit eigenfunctions, as displayed in the two-dimensional phase space, spread along the~$\phi$ direction (left panel). In the hard parameter regime, the energy dispersion of the eigenstates as a function of external flux is exponentially suppressed (middle panel), while offset-charge dispersion is suppressed to first order at the sweet spot at~$n_\mathrm{gate}=0.5$ (right panel). Insets show the dispersion around the optimal operational point, where the qubit states are nearly degenerate. (c) Qubit eigenfunctions and energy dispersion of the experimentally implemented device. Flux insensitivity is achieved to first order when the qubit is operated at the sweet spot~$\phi_\mathrm{ext}=0$ and~$n_\mathrm{gate}=0.5$. Insets show the dispersion of the excited state around this point. Note that close to the excited state, there is an additional level. The disjoint support of the circuit eigenfunctions leads to large~$T_1$ times. Parameters: (b)~$E_C^\theta/h=E_C^\phi/h=E_J/h=10$ GHz,~$E_L/h=50$ MHz, (c)~$E_C^\theta/h=7.7$ GHz,~$E_C^\phi/h=2.5$ GHz,~$E_J/h=27.2$ GHz,~$E_L/h=1.88$ GHz~\cite{kalashnikov2020}.}
    \label{fig:bifluxon}
\end{figure*}

An approach to overcome the intrinsic limitations of single-mode qubits is to combine them into multimode circuits. For instance, by connecting two noise-insensitive modes, the resulting circuit can support simultaneous protection against bit-flip and phase-flip processes (see~\cref{tab:coherence_times}). The prototypical example of such a multimode circuit is the 0 {\textendash}~$\pi$ qubit, which combines a heavy charge mode and a light flux mode~\cite{kitaev2003,brooks2013,dempster2014,groszkowski2018,dipaolo2019,gyenis2019}. Other examples of protected qubits that leverage merged charge and flux modes are the bifluxon~\cite{bell2016,kalashnikov2020} and the ``$\cos2\theta$'' qubits~\cite{smith2020,doucot2002,protopopov2004,protopopov2006}.

It is instructive to investigate how charge and flux modes form composite circuits in the case of the 0 {\textendash}~$\pi$ and bifluxon qubits. Although these two devices have different circuit layouts and operate in distinct parameter regimes, the Hamiltonians describing their qubit degrees of freedom are identical. Both circuits rely on a charge-like mode~$\theta$ and a flux-like mode~$\phi$, which are strongly and nonlinearly coupled to each other. The charging energies of the two modes are~$E_C^\theta$ and~$E_C^\phi$, the inductive energy is~$E_L$, while the coupling term between the modes,~$\cos\hat{\theta}\cos\hat{\phi}$, is proportional to the Josephson energy~$E_J$. The Hamiltonian of these qubits takes the form
\begin{equation}
\begin{split}
    \hat{H}=& 4E_C^\theta (\hat{n}_\theta-n_\mathrm{gate})^2+4E_C^\phi \hat{n}_\phi^2 \\
    & + E_L(\hat{\phi}-\pi\phi_\mathrm{ext})^2 +2E_J\cos{\hat{\phi}}\cos{\hat{\theta}},
\end{split}
\end{equation}
which, in some sense, is the combination of the transmon and fluxonium Hamiltonians.  As we will see below, an important difference between the 0 {\textendash}~$\pi$ and bifluxon qubits is how they physically implement these coupled modes through the circuit layout.

When discussing the experimental realization of these ideas it is useful to contrast between two types of parameter regimes. First, there is the theoretically ideal but difficult to realize ``hard'' regime. Second, there is the ``soft'' regime, which is experimentally easier to implement but at the price of making compromises on the qubit protection. In practice, much of the difficulty in realizing the hard regime lies in making light flux modes as those require building large inductances and having very little stray capacitances in the circuit. As a result, while in the hard regime flux modes can be exponentially flux-insensitive, in the soft regime protection against flux noise relies on working at sweet spots.

In the hard parameter regime, the Hamiltonian of the circuits can often be approximated by an effective Hamiltonian with an underlying symmetry. For example, in the 0 {\textendash}~$\pi$ or~$\cos2\theta$ qubits, the effective Hamiltonian preserves the parity of Cooper-pair-numbers on the islands of the circuit. Similarly, in the bifluxon qubit, the symmetry corresponds to the parity of flux quanta contained in the loop of the circuit. In these cases, the qubit states become nearly degenerate and have odd or even parities. Such symmetry protection satisfies the requirements for high-coherence described in~\cref{noise_protection,design}. First, degenerate levels have vanishing energy dispersion that leads to protection against pure dephasing; second, transitions between states with opposite parity is prohibited as long as the noise respects the symmetry of the system (protection against energy relaxation).

In the soft parameter regime, however, the underlying parity symmetry is not present anymore and the protection originates from qubit wavefunctions with disjoint support and weak energy dispersion, as discussed in~\cref{noise_protection,design}. These devices still offer protection but instead of the simultaneous exponential
reduction in bit-flip and phase-flip errors, the logical states might be protected only to first order against these errors.

In this section, we discuss these ideas focusing on two experimentally realized multimode circuits, the bifluxon and the 0 {\textendash}~$\pi$ qubit. We review both the hard and soft regimes with emphasis on the underlying symmetries and the challenges associated with demonstrating full-noise protection.

\begin{figure*}
    \centering
    \includegraphics[width = \textwidth]{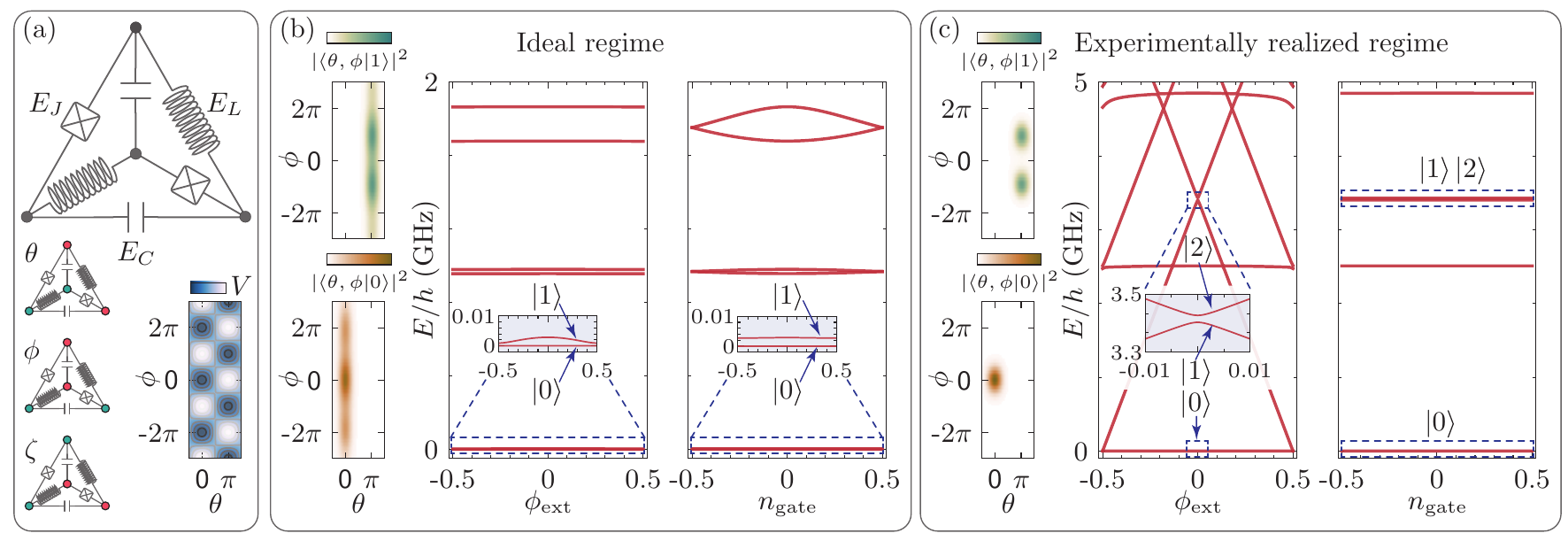}
    \caption{\textbf{The ideal 0 {\textendash}~$\pi$ qubit and its experimentally realized version.} (a) The circuit diagram of the 0 {\textendash}~$\pi$ qubit and its three relevant circuit modes.  Bottom right: the potential landscape of the qubit. (b) In the ideal regime, the wavefunctions of the qubit states have disjoint support, and the energy dispersion offers exponential protection against both charge and flux noise. Insets shows the energy dispersion of the nearly degenerate qubit states. (c) In the experimentally realized device, the degeneracy is split but the device is exponentially protected against charge noise and first-order protected against flux noise. The qubit has still disjoint wavefunctions, leading to exponential protection against energy relaxation. Inset shows the dispersion of the excited state around the flux sweet spot. Note that close to the excited state, there is an additional level. Parameters: (b)~$E_C^\theta/h=30$ MHz,~$E_C^\phi/h=20$ GHz,~$E_J/h=10$ GHz,~$E_L/h=50$ MHz, (c)~$E_C^\theta/h=92$ MHz,~$E_C^\phi/h=1.14$ GHz,~$E_J/h=6.0$ GHz,~$E_L/h=380$ MHz~\cite{gyenis2019}.}
    \label{fig:zeropi}
\end{figure*}

\subsection{The bifluxon qubit}

As illustrated in~\cref{fig:bifluxon}(a), the bifluxon qubit is a three-node device composed of a superinductor and two identical Josephson junctions forming a loop~\cite{kalashnikov2020,bell2016}. The bifluxon circuit has two degrees of freedom, a charge mode~$\theta$ and a flux mode~$\phi$. Correspondingly, the qubit has two control knobs: the offset charge~$n_\mathrm{gate}$ biasing the superconducting island, and the external flux~$\phi_\mathrm{ext}$ threading the loop of the device. Both the charge and the flux modes are light as~$E_L\ll E_J \approx E_C^\theta \approx E_C^\phi$, where~$E_L$ is the inductive energy of the superinductor,~$E_J$ is the Josephson energy,~$E_C^\theta$ and~$E_C^\phi$ are the charging energies of the~$\theta$ and~$\phi$ modes, respectively.

We can understand noise protection in this qubit by considering its logical states from the point of view of the flux mode~$\phi$ at~$n_\mathrm{gate}=0.5$. As a consequence of a destructive Aharonov-Casher-type interference~\cite{kalashnikov2020,bell2016,Friedman2002}, the logical states have opposite fluxon parities: the wavefunctions associated with the~$|0\rangle$ and~$|1\rangle$ states peak at the phase values~$\phi_{2m}=2m\times \pi$ and~$\phi_{2m+1}=(2m+1)\times \pi$, respectively, where~$m$ is a small integer [\cref{fig:bifluxon}(b)]. This can be also interpreted as the loop storing an odd or an even number of flux quanta depending on the qubit state. Since logical states with distinct fluxon parity cannot be connected by the phase operator, transitions are prevented. Furthermore, the large zero-point fluctuations of the~$\phi$ mode lead to exponential insensitivity against flux noise [\cref{fig:bifluxon}(b)].

The mechanism behind noise protection in this circuit can be further elucidated by focusing on the role of the charge mode~$\theta$. As the the charge mode in the bifluxon circuit operates in the Cooper-pair-box regime, the charge component of the low-lying eigenfunctions can be well approximated by a linear combination of only two charge states~$\{|n\rangle,|n+1\rangle\}$~\cite{kalashnikov2020}. At the sweet spot~$n_\mathrm{gate}=0.5$, the Josephson coupling between these charge states leads to eigenstates with bonding,~$(|n\rangle + |n+1\rangle)/\sqrt{2}$, and antibonding,~$(|n\rangle - |n+1\rangle)/\sqrt{2}$, symmetry in charge space. Effectively, the antibonding state introduces a flux bias of~$\phi_\mathrm{ext}=0.5$, favoring odd-fluxon-number states. The situation is the opposite for the bonding state, which favors even-fluxon-number states.

In the hard parameter regime at~$n_\mathrm{gate}=0.5$, the bifluxon qubit is protected against relaxation and displays first-order insensitivity against charge noise and exponential protection against flux noise. However, because it relies on a large superinductor, reaching the ideal parameter regime of the qubit is experimentally challenging. In the soft parameter regime~\cite{kalashnikov2020}, a heavier flux mode leads to a reduced first-order protection at the flux sweet spot while preserving bit-flip protection because the qubit wavefunctions have disjoint support [\cref{fig:bifluxon}(c)]. In this regime, an order of magnitude enhancement of~$T_1>100~\mu$s  was demonstrated with respect to the unprotected~$T_1$ value, although qubit coherence was limited to~$T_\varphi=700$ ns. The rather low dephasing time is due to the charge qubit nature of the device.

Similarly to the heavy fluxonium, protection against bit-flip errors unavoidably complicates qubit manipulation because direct transitions between the logical states are forbidden. Thus, in the first realization of the bifluxon, control is achieved by breaking the protection while the single qubit gate is applied: the gate bias is quickly detuned from the~$n_\mathrm{gate}=0.5$ value to~$n_\mathrm{gate}=0$, and tuned back to the protected regime after the end of the gate.

\subsection{The 0 {\textendash}~$\pi$ qubit}

The 0 {\textendash}~$\pi$ qubit is characterized by protected logical states with exponentially suppressed relaxation and dephasing rates against both charge and flux noise. The device was proposed by Brooks, Preskill and Kitaev~\cite{brooks2013} building on an earlier design known as the current mirror qubit~\cite{kitaev2006,weiss2019}.

To realize full noise protection, the 0 {\textendash}~$\pi$ circuit combines a heavy charge mode~$\theta$ and a light flux mode~$\phi$ based on a four-node circuit layout [\cref{fig:zeropi}(a)]. The wavefunctions are localized along the~$\theta$ coordinate and delocalized along the~$\phi$ direction [\cref{fig:zeropi}(b)]. A first requirement to reach this regime is to ensure that the ratio of charging energies of the two modes,~$E_C^\phi/E_C^\theta$, is large. Moreover, for the~$\phi$ mode to be in its light regime, the inductive energy~$E_L$ needs to be small. Finally, the large Josephson energy~$E_J$ provides the potential barrier between the two valleys along the~$\theta$ direction, leading to the charge-noise insensitivity and disjoint support of the qubit wavefunctions. When the circuit parameters meet all of these requirements ($E_L\approx E_C^\theta\ll E_J\approx E_C^\phi$), the ground and first excited states of the circuit are delocalized along the~$\phi$ direction while being confined to the~$\theta=0$ and~$\pi$ valleys from which the qubit takes its name.

In the hard parameter regime, it is possible to derive an effective single-mode Hamiltonian for the circuit which corresponds to a capacitively shunted tunneling element for \emph{pairs} of Cooper pairs~\cite{dipaolo2019}. In this case, the cotunneling process can be described by a~$\cos2\hat{\theta}$ term, while the single-Cooper-pair tunneling term~$\cos\hat{\theta}$ is highly suppressed. The ground or excited state of this model are even or odd charge states, respectively. The transitions between these states are forbidden because the effective Hamiltonian preserves the Cooper-pair-number parity. In analogy with the transmon, if the energy associated with the tunneling of pairs of Cooper pairs is large compared to the effective charging energy, the two qubit states are delocalized in charge space and the transition frequency is rendered insensitive to charge noise. Despite the apparent complexity of the full 0 {\textendash}~$\pi$ circuit, this simple effective model captures the exponential protection against both phase- and bit-flip errors~\cite{dipaolo2019}.

In addition to the above stringent requirements on the energy scales of the device, it is also important for the pairs of large capacitors and superinductors in the circuit to be as identical as possible. Indeed, since the 0 {\textendash}~$\pi$ circuit contains four nodes, the circuit inevitably incorporates a third degree of freedom, the~$\zeta$ mode, which can couple to the qubit modes~$\phi$ and~$\theta$ in the presence of circuit-element disorder. Because the energy of this mode is stored in the large inductors and large capacitors of the circuit [\cref{fig:zeropi}(a)], the~$\zeta$ mode is a low frequency harmonic mode. In the presence of asymmetries, thermal population of this mode can lead to dephasing of the qubit. While the 0 {\textendash}~$\pi$ is protected against this noise channel in the hard parameter regime~\cite{groszkowski2018,dipaolo2019}, the effects of~$\zeta$-mode coupling in the softer parameter regimes can be possibly mitigated with active strategies involving cooling of this mode~\cite{dipaolo2019}.

\begin{figure*}
    \centering
    \includegraphics[width = \textwidth]{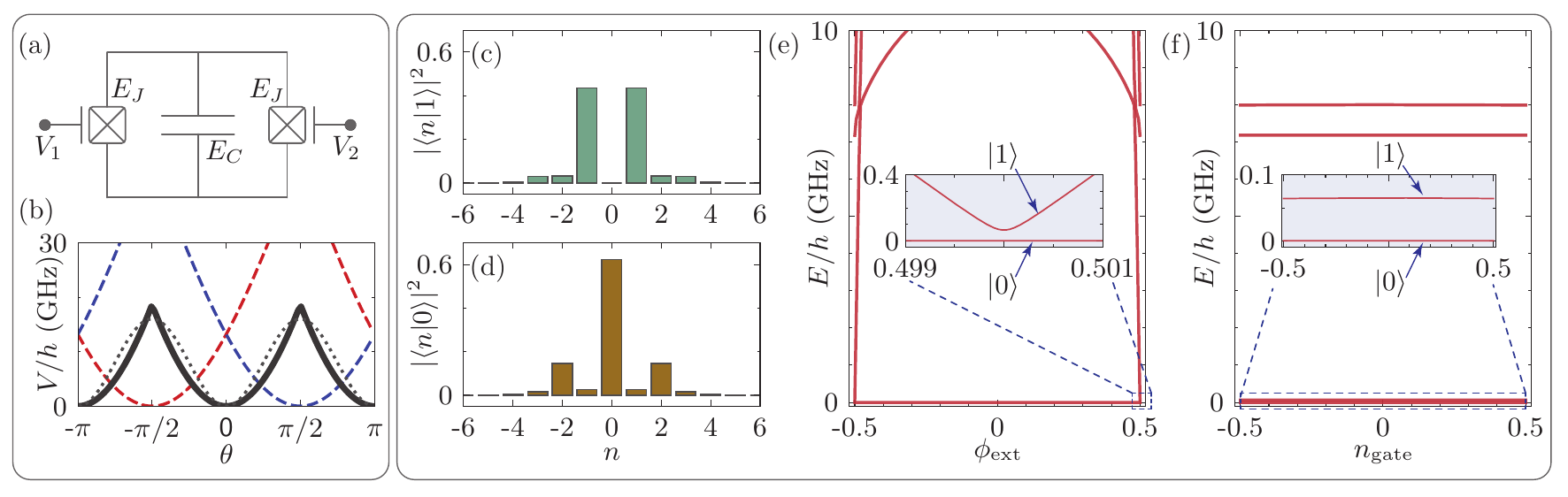}
    \caption{\textbf{Parity-protected semi-superconducting qubit.} (a) The circuit is based on two capacitively-shunted voltage-tunable junctions, which are realized as weak links in the superconducting shell of an InAs nanowire. The gate voltages~$V_1$ and~$V_2$ can be used to tune the current-phase relationship of the junctions. (b) The potential energy of the two junctions in the loop~$\hat{U}(\hat{\theta}-\phi_\mathrm{ext}/2)$ (blue dashed line),~$\hat{U}(\hat{\theta}+\phi_\mathrm{ext}/2)$ (red dashed line) and their sum (black continuous line), where~$\hat{U}(\hat{\theta}) = -\Delta\sqrt{1-T\sin^2{\hat{\theta}/2}}$,~$T=1$ and~$\Delta/2\pi$ = 45 GHz. The dotted black line indicates a~$\cos2\hat{\theta}$ potential, which approximates well the qubit potential at low energies. (c)-(d) Qubit wavefunctions in charge space are even and odd charge states at~$\phi_\mathrm{ext}=0.5$. (e)-(f) The energy level dispersion as a function of flux and offset charge. At~$\phi_\mathrm{ext}=0.5$, the qubit is first-order protected against flux noise, and exponentially protected against charge noise (insets show the energy dispersion around this optimal point). The circuit parameters and transmission coefficients are~$E_C/h=$0.284 GHz,~$\Delta/2\pi$ = 45 GHz,~$T_{J1}=[1.0, 1.0, 0.60, 0.0, 0.0]$ and~$T_{J2}=[0.99, 0.78, 0.31, 0.30]$~\cite{larsen2020}.}
    \label{fig:semiconducting_zeropi}
\end{figure*}

Experimentally realizing the hard parameter regime is challenging because of the competition between the conditions that guarantee charge- and flux-noise insensitivity. For example, simultaneously achieving the light and heavy regimes of the flux and charge modes in the same circuit requires a substantial reduction in unwanted cross capacitances. The first experimental implementation of the 0 {\textendash}~$\pi$ qubit~\cite{gyenis2019} relaxed the requirements of the light flux mode by working with smaller~$E_C^\phi$ and larger~$E_L$ than ideal, leading to considerable flux dispersion of the qubit transition. Importantly, the experimental device retained simultaneous first-order insensitivity to flux noise, exponential protection to charge noise, and the disjoint character of the qubit wavefunctions at~$\phi_\mathrm{ext}=0$ [\cref{fig:zeropi}(c)]. Therefore, this implementation demonstrates a degree of noise protection beyond that of the transmon and fluxonium qubits, constituting a blueprint for building a fully noise-protected superconducting qubit. The measured energy decay times of this soft 0 {\textendash}~$\pi$ qubit is~$T_1=1.6$ ms, which is comparable to heavy fluxonium, while the dephasing rate obtained by a spin echo measurement is~$T_\mathrm{2E}=25~\mu$s.

Single qubit gates on the soft 0 {\textendash}~$\pi$ circuit also rely on partially lifting the protection~\cite{gyenis2019}. In contrast to the bifluxon, where the system shortly leaves the protected parameter regime during the gate (namely the~$n_\mathrm{gate}=0.5$ bias point), here the system rather leaves the protected subspace (the logical~$|0\rangle$ and~$|1\rangle$ states) by temporarily occupying a higher-lying ancillary state in a Raman-type gate. Since the population of the ancillary state depends on the detuning of the drive tones, the occupation of the unprotected ancillary state can be arbitrary small at the price of increasing drive powers to avoid prohibitively long gate times. Thus, such gates require finding the balance between the amplitude of microwave drives and occupation of unprotected levels~\cite{gyenis2019}.

\section{Hybrid protected semi-superconducting qubits}\label{hybrid}

In the superconducting qubits discussed so far, the Josephson junctions use aluminum-oxide tunnel barriers between aluminum electrodes. Such superconductor-insulator-superconductor junctions result in nonlinear elements with an almost perfectly sinusoidal current-phase relationship. In this case, realizing protected qubits requires fabricating a complex combination of these Josephson junctions, large superinductors and extremely large or tiny capacitors in the same circuit. Semiconductor-based technology circumvents this obstacle by equipping Josephson junctions with non-sinusoidal current-phase behavior~\cite{ramon2020}. Such unorthodox Josephson junctions open up new possibilities to engineer exotic Hamiltonians that protect quantum information. Although the lower quality of materials currently limits this technology~\cite{casparis2018}, recent advances have enabled the implementation of the first prototypes of protected hybrid qubits. For example, a charge-insensitive Cooper pair box based on highly-transmissive Josephson junctions~\cite{kringhoj2020,bargerbos2020} and a single-mode semiconductor~$\cos2\theta$ qubit~\cite{larsen2020} illustrates the potential of this platform.

Here, we focus on the~$\cos2\theta$ hybrid qubit~\cite{larsen2020}. To do so, we briefly review the behavior of Josephson junctions from a microscopic viewpoint based on Andreev bound states. When two superconducting electrodes connect through a thin tunnel barrier, quasiparticle bound states form due to multiple Andreev reflections between the two superconducting regions. In most cases, multiple bound states mediate the coupling between the electrodes, and we can express the Josephson energy as the sum over the channels, such that~$\hat{U}(\hat{\theta}) = -\Delta\sum_i\sqrt{1-T_i\sin^2{\hat{\theta}/2}}$, where~$T_i$ is the transmission probability through the tunnel barrier for channel~$i$ and~$\Delta$ is the superconducting gap~\cite{martinis2004}. In the case of a large number of weakly transmitting channels--which applies to aluminum-oxide based tunnel junctions--we recover the well-known expression for the Josephson energy~$\hat{U}(\hat{\theta})=E_J(1-\cos\hat{\theta})$, with~$E_J=\Delta\sum_i T_i/4$. Weak-link Josephson junctions can have a few highly transmitting channels in semiconductor InAs nanowires with an epitaxially grown superconductor layer~\cite{doh2005,krogstrup2015,larsen2015,lange2015} or in proximitized two-dimensional electron gasses~\cite{casparis2018}.

\begin{table*}
\centering
\caption{Protection in various superconducting circuits. We compare the different degrees of protection in energy relaxation ($T_1$) and dephasing due to charge ($T_\varphi^{n_\mathrm{gate}}$) and flux noise ($T_\varphi^{\phi_\mathrm{ext}}$). Absent: the qubit is not protected against the error; N/A: the qubit is not coupled to the particular noise; linear: the qubit has sweet-spot protection; exponential: the qubit is insensitive to the type of error; exponential*: for the case of the fluxonium and blochnium qubits, this refers to the exponential insensitivity to offset-charge fluctuations due to coherent quantum phase slips~\cite{manucharyan2012}.}
\begin{tabular}{p{3.3cm}p{0.9cm}p{0.9cm}p{0.9cm}|p{0.9cm}p{0.9cm}p{0.9cm}}
\toprule
Circuit name &  \multicolumn{3}{c}{Ideal case} & \multicolumn{3}{c}{Realized case}\\
{}   &~$T_1$ &~$T_\varphi^{n_\mathrm{gate}}$ &~$T_\varphi^{\phi_\mathrm{ext}}$  &~$T_1$ &~$T_\varphi^{n_\mathrm{gate}}$ &~$T_\varphi^{\phi_\mathrm{ext}}$ \\
\midrule
Transmon~\cite{koch2007,schreier2008} & absent & exponential & N/A & absent & exponential & N/A \\
Fluxonium~\cite{manucharyan2009} & absent & exponential* & linear & absent & exponential* & linear \\
Blochnium~\cite{pechenezhskiy2020} & absent & exponential* & exponential & absent & exponential* & exponential \\
Heavy fluxonium~\cite{federov2011,earnest2018,lin2018} & exponential & exponential* & absent & exponential & exponential* & absent  \\
\midrule
Bifluxon~\cite{bell2016,kalashnikov2020} & exponential & linear & exponential & exponential & linear & linear  \\
0 {\textendash}~$\pi$ qubit~\cite{brooks2013,gyenis2019} & exponential & exponential & exponential & exponential & exponential & linear \\
Semiconductor~$\cos2\theta$~\cite{larsen2020} & exponential & exponential & linear & exponential & exponential & linear \\
\bottomrule
\end{tabular}\label{tab:coherence_times}
\end{table*}

In a few-channel superconductor-semiconductor junction, the~$2\pi$-periodic Josephson energy can be expressed with its Fourier components~$\hat{U}(\hat{\theta})=\sum_i A_k\cos(k\hat{\theta})$, assuming time-reversal symmetry is not broken in the junction~\cite{golubov2004}. While generally in metallic Josephson junctions the first harmonic term plays the dominant role in the dynamics ($A_1\gg A_2\gg...$), in semiconductor-based junctions the higher harmonic content can have a significant contribution as well. In a hybrid interferometer circuit formed by two identical junctions [\cref{fig:semiconducting_zeropi}(a)], the total Josephson energy can be expressed as~$\hat{U}_\mathrm{tot}(\theta)=\hat{U}(\hat{\theta}-\phi_\mathrm{ext}/2) + \hat{U}(\hat{\theta}+\phi_\mathrm{ext}/2)$. When the circuit is frustrated at half-flux quantum ($\phi_\mathrm{ext}=0.5$), the odd Fourier components of the two identical junctions cancel each other, while the even contributions constructively interfere, giving rise to a double-well potential with a~$\pi$ period that can be approximated with a~$\hat{U}_\mathrm{tot}(\hat{\theta})\approx E_{2}\cos2\hat{\theta}$ potential [\cref{fig:semiconducting_zeropi}(b)], similarly to the low-energy effective model of the 0 {\textendash}~$\pi$ qubit.

\Cref{fig:semiconducting_zeropi}(c) and (d) show the ground and first excited state of the experimentally realized circuit, which resembles odd and even charge states as expected for the disconnected parity subspaces of a~$\cos 2\theta$ element. Importantly, the qubit serves as a protected circuit only at the half-flux quantum sweet spot [\cref{fig:semiconducting_zeropi}(e)]. Regarding the charge noise of the device, as long as the mode is heavy ($E_{2}\gg E_C$), the qubit states are exponentially insensitive to the offset charge [\cref{fig:semiconducting_zeropi}(f)]. We also note that this circuit is the dual element of the bifluxon, as the offset-charge and external-flux dependence of the energy levels are reversed for the two qubits [compare~\cref{fig:bifluxon}(b) and~\cref{fig:semiconducting_zeropi}(e) and (f)]. In the first implementation of this idea, the energy relaxation rate yielded a tenfold improvement compared to the unprotected regime with~$T_1=7~\mu$s~\cite{larsen2020}.

\section{Outlook}\label{outlook}

Superconducting qubits have progressed in a truly remarkable fashion in the last two decades: Starting with coherence times of a few nanoseconds, today’s typical values are in the hundreds of microseconds. As the evolution of the transmon qubit illustrates, suitable and optimized circuits have enabled significant advancements. The fluxonium circuit is on a similar ascending trajectory, with millisecond coherence times having recently been demonstrated~\cite{Somoroff2021}. Their simplicity and partial built-in protection against dominant noise sources is an important reason for their success.

Because significant improvements in qubit performances are still necessary to realize the full potential of fault-tolerant quantum computation, it remains essential to not only continue to improve current qubits but also to explore novel qubit designs. Protected qubits play an important role in these efforts as they simultaneously suppress relaxation and dephasing. Although this is obtained at the price of increased circuit complexity, in the long run such circuits have the potential to dramatically reduce the physical qubit count needed for quantum error correction.

In addition to continue the development of protected qubits, it remains important to design single- and two-qubit logical operations on these qubits. Current approaches require leaving the protected regime~\cite{kalashnikov2020,larsen2020} or relying on non-computational states~\cite{earnest2018,gyenis2019}. Finding more fast and robust approaches to control these devices remains a theoretical and experimental challenge.

Whether protected qubits will play the same role as the ``mainstream'' qubits or be responsible for different tasks in a heterogeneous processor, such as memory or data units, still needs to be decided. Nevertheless, given the formidable challenge that it represents, superconducting quantum processors will likely benefit from the synergy of multiple strategies including quantum error correction, operation in an engineered ultra-low-noise environment, and hardware-level error protection.

\section*{Acknowledgments}
Andr\'as Gyenis thanks Charles Marcus, Pranav Mundada, Anjali Premkumar and Thomas M. Hazard for inspiring conversations about protected qubits. Agustin Di Paolo is grateful to William D. Oliver, Clemens M\"uller, Konstantin Kalashnikov, Thomas M. Hazard, Kyle Serniak and Joachim Cohen for inspiring conversations. This work was undertaken in part thanks to funding from NSERC, the Canada First Research Excellence Fund, the Minist\`ere de l'\'economie et de l'innovation du Qu\'ebec and the U.S. Army Research Office Grant No. W911NF-18-1-0411.


\begin{thebibliography}{100}%
\makeatletter
\providecommand \@ifxundefined [1]{%
 \@ifx{#1\undefined}
}%
\providecommand \@ifnum [1]{%
 \ifnum #1\expandafter \@firstoftwo
 \else \expandafter \@secondoftwo
 \fi
}%
\providecommand \@ifx [1]{%
 \ifx #1\expandafter \@firstoftwo
 \else \expandafter \@secondoftwo
 \fi
}%
\providecommand \natexlab [1]{#1}%
\providecommand \enquote  [1]{``#1''}%
\providecommand \bibnamefont  [1]{#1}%
\providecommand \bibfnamefont [1]{#1}%
\providecommand \citenamefont [1]{#1}%
\providecommand \href@noop [0]{\@secondoftwo}%
\providecommand \href [0]{\begingroup \@sanitize@url \@href}%
\providecommand \@href[1]{\@@startlink{#1}\@@href}%
\providecommand \@@href[1]{\endgroup#1\@@endlink}%
\providecommand \@sanitize@url [0]{\catcode `\\12\catcode `\$12\catcode
  `\&12\catcode `\#12\catcode `\^12\catcode `\_12\catcode `\%12\relax}%
\providecommand \@@startlink[1]{}%
\providecommand \@@endlink[0]{}%
\providecommand \url  [0]{\begingroup\@sanitize@url \@url }%
\providecommand \@url [1]{\endgroup\@href {#1}{\urlprefix }}%
\providecommand \urlprefix  [0]{URL }%
\providecommand \Eprint [0]{\href }%
\providecommand \doibase [0]{https://doi.org/}%
\providecommand \selectlanguage [0]{\@gobble}%
\providecommand \bibinfo  [0]{\@secondoftwo}%
\providecommand \bibfield  [0]{\@secondoftwo}%
\providecommand \translation [1]{[#1]}%
\providecommand \BibitemOpen [0]{}%
\providecommand \bibitemStop [0]{}%
\providecommand \bibitemNoStop [0]{.\EOS\space}%
\providecommand \EOS [0]{\spacefactor3000\relax}%
\providecommand \BibitemShut  [1]{\csname bibitem#1\endcsname}%
\let\auto@bib@innerbib\@empty
\bibitem [{\citenamefont {Martinis}\ \emph {et~al.}(2020)\citenamefont
  {Martinis}, \citenamefont {Devoret},\ and\ \citenamefont
  {Clarke}}]{martinis2020}%
  \BibitemOpen
  \bibfield  {author} {\bibinfo {author} {\bibfnamefont {J.~M.}\ \bibnamefont
  {Martinis}}, \bibinfo {author} {\bibfnamefont {M.~H.}\ \bibnamefont
  {Devoret}},\ and\ \bibinfo {author} {\bibfnamefont {J.}~\bibnamefont
  {Clarke}},\ }\bibfield  {title} {\bibinfo {title} {{Quantum Josephson
  junction circuits and the dawn of artificial atoms}},\ }\href
  {https://doi.org/10.1038/s41567-020-0829-5} {\bibfield  {journal} {\bibinfo
  {journal} {Nature Physics}\ }\textbf {\bibinfo {volume} {16}},\ \bibinfo
  {pages} {234} (\bibinfo {year} {2020})}\BibitemShut {NoStop}%
\bibitem [{\citenamefont {Kjaergaard}\ \emph
  {et~al.}(2020{\natexlab{a}})\citenamefont {Kjaergaard}, \citenamefont
  {Schwartz}, \citenamefont {Braumüller}, \citenamefont {Krantz},
  \citenamefont {Wang}, \citenamefont {Gustavsson},\ and\ \citenamefont
  {Oliver}}]{kjaergaard2020}%
  \BibitemOpen
  \bibfield  {author} {\bibinfo {author} {\bibfnamefont {M.}~\bibnamefont
  {Kjaergaard}}, \bibinfo {author} {\bibfnamefont {M.~E.}\ \bibnamefont
  {Schwartz}}, \bibinfo {author} {\bibfnamefont {J.}~\bibnamefont
  {Braumüller}}, \bibinfo {author} {\bibfnamefont {P.}~\bibnamefont {Krantz}},
  \bibinfo {author} {\bibfnamefont {J.~I.-J.}\ \bibnamefont {Wang}}, \bibinfo
  {author} {\bibfnamefont {S.}~\bibnamefont {Gustavsson}},\ and\ \bibinfo
  {author} {\bibfnamefont {W.~D.}\ \bibnamefont {Oliver}},\ }\bibfield  {title}
  {\bibinfo {title} {Superconducting qubits: {C}urrent state of play},\ }\href
  {https://doi.org/10.1146/annurev-conmatphys-031119-050605} {\bibfield
  {journal} {\bibinfo  {journal} {Annual Review of Condensed Matter Physics}\
  }\textbf {\bibinfo {volume} {11}},\ \bibinfo {pages} {369} (\bibinfo {year}
  {2020}{\natexlab{a}})}\BibitemShut {NoStop}%
\bibitem [{\citenamefont {Blais}\ \emph {et~al.}(2020)\citenamefont {Blais},
  \citenamefont {Grimsmo}, \citenamefont {Girvin},\ and\ \citenamefont
  {Wallraff}}]{blais2020}%
  \BibitemOpen
  \bibfield  {author} {\bibinfo {author} {\bibfnamefont {A.}~\bibnamefont
  {Blais}}, \bibinfo {author} {\bibfnamefont {A.~L.}\ \bibnamefont {Grimsmo}},
  \bibinfo {author} {\bibfnamefont {S.~M.}\ \bibnamefont {Girvin}},\ and\
  \bibinfo {author} {\bibfnamefont {A.}~\bibnamefont {Wallraff}},\ }\href@noop
  {} {\bibinfo {title} {Circuit quantum electrodynamics}} (\bibinfo {year}
  {2020}),\ \Eprint {https://arxiv.org/abs/2005.12667} {arXiv:2005.12667
  [quant-ph]} \BibitemShut {NoStop}%
\bibitem [{\citenamefont {Krantz}\ \emph {et~al.}(2019)\citenamefont {Krantz},
  \citenamefont {Kjaergaard}, \citenamefont {Yan}, \citenamefont {Orlando},
  \citenamefont {Gustavsson},\ and\ \citenamefont {Oliver}}]{krantz2019}%
  \BibitemOpen
  \bibfield  {author} {\bibinfo {author} {\bibfnamefont {P.}~\bibnamefont
  {Krantz}}, \bibinfo {author} {\bibfnamefont {M.}~\bibnamefont {Kjaergaard}},
  \bibinfo {author} {\bibfnamefont {F.}~\bibnamefont {Yan}}, \bibinfo {author}
  {\bibfnamefont {T.~P.}\ \bibnamefont {Orlando}}, \bibinfo {author}
  {\bibfnamefont {S.}~\bibnamefont {Gustavsson}},\ and\ \bibinfo {author}
  {\bibfnamefont {W.~D.}\ \bibnamefont {Oliver}},\ }\bibfield  {title}
  {\bibinfo {title} {A quantum engineer's guide to superconducting qubits},\
  }\href {https://doi.org/10.1063/1.5089550} {\bibfield  {journal} {\bibinfo
  {journal} {Applied Physics Reviews}\ }\textbf {\bibinfo {volume} {6}},\
  \bibinfo {pages} {021318} (\bibinfo {year} {2019})}\BibitemShut {NoStop}%
\bibitem [{\citenamefont {Wendin}(2017)}]{wendin2017}%
  \BibitemOpen
  \bibfield  {author} {\bibinfo {author} {\bibfnamefont {G.}~\bibnamefont
  {Wendin}},\ }\bibfield  {title} {\bibinfo {title} {Quantum information
  processing with superconducting circuits: a review},\ }\href
  {https://doi.org/10.1088/1361-6633/aa7e1a} {\bibfield  {journal} {\bibinfo
  {journal} {Reports on Progress in Physics}\ }\textbf {\bibinfo {volume}
  {80}},\ \bibinfo {pages} {106001} (\bibinfo {year} {2017})}\BibitemShut
  {NoStop}%
\bibitem [{\citenamefont {Clarke}\ and\ \citenamefont
  {Wilhelm}(2008)}]{clarke2008}%
  \BibitemOpen
  \bibfield  {author} {\bibinfo {author} {\bibfnamefont {J.}~\bibnamefont
  {Clarke}}\ and\ \bibinfo {author} {\bibfnamefont {F.~K.}\ \bibnamefont
  {Wilhelm}},\ }\bibfield  {title} {\bibinfo {title} {Superconducting quantum
  bits},\ }\href {https://doi.org/10.1038/nature07128} {\bibfield  {journal}
  {\bibinfo  {journal} {Nature}\ }\textbf {\bibinfo {volume} {453}},\ \bibinfo
  {pages} {1031} (\bibinfo {year} {2008})}\BibitemShut {NoStop}%
\bibitem [{\citenamefont {Schoelkopf}\ and\ \citenamefont
  {Girvin}(2008)}]{schoelkopf2008}%
  \BibitemOpen
  \bibfield  {author} {\bibinfo {author} {\bibfnamefont {R.~J.}\ \bibnamefont
  {Schoelkopf}}\ and\ \bibinfo {author} {\bibfnamefont {S.~M.}\ \bibnamefont
  {Girvin}},\ }\bibfield  {title} {\bibinfo {title} {Wiring up quantum
  systems},\ }\href {https://doi.org/10.1038/451664a} {\bibfield  {journal}
  {\bibinfo  {journal} {Nature}\ }\textbf {\bibinfo {volume} {451}},\ \bibinfo
  {pages} {664} (\bibinfo {year} {2008})}\BibitemShut {NoStop}%
\bibitem [{\citenamefont {Devoret}\ and\ \citenamefont
  {Schoelkopf}(2013)}]{devoret2013}%
  \BibitemOpen
  \bibfield  {author} {\bibinfo {author} {\bibfnamefont {M.~H.}\ \bibnamefont
  {Devoret}}\ and\ \bibinfo {author} {\bibfnamefont {R.~J.}\ \bibnamefont
  {Schoelkopf}},\ }\bibfield  {title} {\bibinfo {title} {Superconducting
  circuits for quantum information: An outlook},\ }\href
  {https://doi.org/10.1126/science.1231930} {\bibfield  {journal} {\bibinfo
  {journal} {Science}\ }\textbf {\bibinfo {volume} {339}},\ \bibinfo {pages}
  {1169} (\bibinfo {year} {2013})}\BibitemShut {NoStop}%
\bibitem [{\citenamefont {Otterbach}\ \emph {et~al.}(2017)\citenamefont
  {Otterbach}, \citenamefont {Manenti}, \citenamefont {Alidoust}, \citenamefont
  {Bestwick}, \citenamefont {Block}, \citenamefont {Bloom}, \citenamefont
  {Caldwell}, \citenamefont {Didier}, \citenamefont {Fried}, \citenamefont
  {Hong}, \citenamefont {Karalekas}, \citenamefont {Osborn}, \citenamefont
  {Papageorge}, \citenamefont {Peterson}, \citenamefont {Prawiroatmodjo},
  \citenamefont {Rubin}, \citenamefont {Ryan}, \citenamefont {Scarabelli},
  \citenamefont {Scheer}, \citenamefont {Sete}, \citenamefont {Sivarajah},
  \citenamefont {Smith}, \citenamefont {Staley}, \citenamefont {Tezak},
  \citenamefont {Zeng}, \citenamefont {Hudson}, \citenamefont {Johnson},
  \citenamefont {Reagor}, \citenamefont {da~Silva},\ and\ \citenamefont
  {Rigetti}}]{otterbach2017}%
  \BibitemOpen
  \bibfield  {author} {\bibinfo {author} {\bibfnamefont {J.~S.}\ \bibnamefont
  {Otterbach}}, \bibinfo {author} {\bibfnamefont {R.}~\bibnamefont {Manenti}},
  \bibinfo {author} {\bibfnamefont {N.}~\bibnamefont {Alidoust}}, \bibinfo
  {author} {\bibfnamefont {A.}~\bibnamefont {Bestwick}}, \bibinfo {author}
  {\bibfnamefont {M.}~\bibnamefont {Block}}, \bibinfo {author} {\bibfnamefont
  {B.}~\bibnamefont {Bloom}}, \bibinfo {author} {\bibfnamefont
  {S.}~\bibnamefont {Caldwell}}, \bibinfo {author} {\bibfnamefont
  {N.}~\bibnamefont {Didier}}, \bibinfo {author} {\bibfnamefont {E.~S.}\
  \bibnamefont {Fried}}, \bibinfo {author} {\bibfnamefont {S.}~\bibnamefont
  {Hong}}, \bibinfo {author} {\bibfnamefont {P.}~\bibnamefont {Karalekas}},
  \bibinfo {author} {\bibfnamefont {C.~B.}\ \bibnamefont {Osborn}}, \bibinfo
  {author} {\bibfnamefont {A.}~\bibnamefont {Papageorge}}, \bibinfo {author}
  {\bibfnamefont {E.~C.}\ \bibnamefont {Peterson}}, \bibinfo {author}
  {\bibfnamefont {G.}~\bibnamefont {Prawiroatmodjo}}, \bibinfo {author}
  {\bibfnamefont {N.}~\bibnamefont {Rubin}}, \bibinfo {author} {\bibfnamefont
  {C.~A.}\ \bibnamefont {Ryan}}, \bibinfo {author} {\bibfnamefont
  {D.}~\bibnamefont {Scarabelli}}, \bibinfo {author} {\bibfnamefont
  {M.}~\bibnamefont {Scheer}}, \bibinfo {author} {\bibfnamefont {E.~A.}\
  \bibnamefont {Sete}}, \bibinfo {author} {\bibfnamefont {P.}~\bibnamefont
  {Sivarajah}}, \bibinfo {author} {\bibfnamefont {R.~S.}\ \bibnamefont
  {Smith}}, \bibinfo {author} {\bibfnamefont {A.}~\bibnamefont {Staley}},
  \bibinfo {author} {\bibfnamefont {N.}~\bibnamefont {Tezak}}, \bibinfo
  {author} {\bibfnamefont {W.~J.}\ \bibnamefont {Zeng}}, \bibinfo {author}
  {\bibfnamefont {A.}~\bibnamefont {Hudson}}, \bibinfo {author} {\bibfnamefont
  {B.~R.}\ \bibnamefont {Johnson}}, \bibinfo {author} {\bibfnamefont
  {M.}~\bibnamefont {Reagor}}, \bibinfo {author} {\bibfnamefont {M.~P.}\
  \bibnamefont {da~Silva}},\ and\ \bibinfo {author} {\bibfnamefont
  {C.}~\bibnamefont {Rigetti}},\ }\href@noop {} {\bibinfo {title} {Unsupervised
  machine learning on a hybrid quantum computer}} (\bibinfo {year} {2017}),\
  \Eprint {https://arxiv.org/abs/1712.05771} {arXiv:1712.05771 [quant-ph]}
  \BibitemShut {NoStop}%
\bibitem [{\citenamefont {Kandala}\ \emph {et~al.}(2017)\citenamefont
  {Kandala}, \citenamefont {Mezzacapo}, \citenamefont {Temme}, \citenamefont
  {Takita}, \citenamefont {Brink}, \citenamefont {Chow},\ and\ \citenamefont
  {Gambetta}}]{kandala2017}%
  \BibitemOpen
  \bibfield  {author} {\bibinfo {author} {\bibfnamefont {A.}~\bibnamefont
  {Kandala}}, \bibinfo {author} {\bibfnamefont {A.}~\bibnamefont {Mezzacapo}},
  \bibinfo {author} {\bibfnamefont {K.}~\bibnamefont {Temme}}, \bibinfo
  {author} {\bibfnamefont {M.}~\bibnamefont {Takita}}, \bibinfo {author}
  {\bibfnamefont {M.}~\bibnamefont {Brink}}, \bibinfo {author} {\bibfnamefont
  {J.~M.}\ \bibnamefont {Chow}},\ and\ \bibinfo {author} {\bibfnamefont
  {J.~M.}\ \bibnamefont {Gambetta}},\ }\bibfield  {title} {\bibinfo {title}
  {Hardware-efficient variational quantum eigensolver for small molecules and
  quantum magnets},\ }\href {https://doi.org/10.1038/nature23879} {\bibfield
  {journal} {\bibinfo  {journal} {Nature}\ }\textbf {\bibinfo {volume} {549}},\
  \bibinfo {pages} {242} (\bibinfo {year} {2017})}\BibitemShut {NoStop}%
\bibitem [{\citenamefont {Neill}\ \emph {et~al.}(2018)\citenamefont {Neill},
  \citenamefont {Roushan}, \citenamefont {Kechedzhi}, \citenamefont {Boixo},
  \citenamefont {Isakov}, \citenamefont {Smelyanskiy}, \citenamefont {Megrant},
  \citenamefont {Chiaro}, \citenamefont {Dunsworth}, \citenamefont {Arya},
  \citenamefont {Barends}, \citenamefont {Burkett}, \citenamefont {Chen},
  \citenamefont {Chen}, \citenamefont {Fowler}, \citenamefont {Foxen},
  \citenamefont {Giustina}, \citenamefont {Graff}, \citenamefont {Jeffrey},
  \citenamefont {Huang}, \citenamefont {Kelly}, \citenamefont {Klimov},
  \citenamefont {Lucero}, \citenamefont {Mutus}, \citenamefont {Neeley},
  \citenamefont {Quintana}, \citenamefont {Sank}, \citenamefont {Vainsencher},
  \citenamefont {Wenner}, \citenamefont {White}, \citenamefont {Neven},\ and\
  \citenamefont {Martinis}}]{neill2018}%
  \BibitemOpen
  \bibfield  {author} {\bibinfo {author} {\bibfnamefont {C.}~\bibnamefont
  {Neill}}, \bibinfo {author} {\bibfnamefont {P.}~\bibnamefont {Roushan}},
  \bibinfo {author} {\bibfnamefont {K.}~\bibnamefont {Kechedzhi}}, \bibinfo
  {author} {\bibfnamefont {S.}~\bibnamefont {Boixo}}, \bibinfo {author}
  {\bibfnamefont {S.~V.}\ \bibnamefont {Isakov}}, \bibinfo {author}
  {\bibfnamefont {V.}~\bibnamefont {Smelyanskiy}}, \bibinfo {author}
  {\bibfnamefont {A.}~\bibnamefont {Megrant}}, \bibinfo {author} {\bibfnamefont
  {B.}~\bibnamefont {Chiaro}}, \bibinfo {author} {\bibfnamefont
  {A.}~\bibnamefont {Dunsworth}}, \bibinfo {author} {\bibfnamefont
  {K.}~\bibnamefont {Arya}}, \bibinfo {author} {\bibfnamefont {R.}~\bibnamefont
  {Barends}}, \bibinfo {author} {\bibfnamefont {B.}~\bibnamefont {Burkett}},
  \bibinfo {author} {\bibfnamefont {Y.}~\bibnamefont {Chen}}, \bibinfo {author}
  {\bibfnamefont {Z.}~\bibnamefont {Chen}}, \bibinfo {author} {\bibfnamefont
  {A.}~\bibnamefont {Fowler}}, \bibinfo {author} {\bibfnamefont
  {B.}~\bibnamefont {Foxen}}, \bibinfo {author} {\bibfnamefont
  {M.}~\bibnamefont {Giustina}}, \bibinfo {author} {\bibfnamefont
  {R.}~\bibnamefont {Graff}}, \bibinfo {author} {\bibfnamefont
  {E.}~\bibnamefont {Jeffrey}}, \bibinfo {author} {\bibfnamefont
  {T.}~\bibnamefont {Huang}}, \bibinfo {author} {\bibfnamefont
  {J.}~\bibnamefont {Kelly}}, \bibinfo {author} {\bibfnamefont
  {P.}~\bibnamefont {Klimov}}, \bibinfo {author} {\bibfnamefont
  {E.}~\bibnamefont {Lucero}}, \bibinfo {author} {\bibfnamefont
  {J.}~\bibnamefont {Mutus}}, \bibinfo {author} {\bibfnamefont
  {M.}~\bibnamefont {Neeley}}, \bibinfo {author} {\bibfnamefont
  {C.}~\bibnamefont {Quintana}}, \bibinfo {author} {\bibfnamefont
  {D.}~\bibnamefont {Sank}}, \bibinfo {author} {\bibfnamefont {A.}~\bibnamefont
  {Vainsencher}}, \bibinfo {author} {\bibfnamefont {J.}~\bibnamefont {Wenner}},
  \bibinfo {author} {\bibfnamefont {T.~C.}\ \bibnamefont {White}}, \bibinfo
  {author} {\bibfnamefont {H.}~\bibnamefont {Neven}},\ and\ \bibinfo {author}
  {\bibfnamefont {J.~M.}\ \bibnamefont {Martinis}},\ }\bibfield  {title}
  {\bibinfo {title} {A blueprint for demonstrating quantum supremacy with
  superconducting qubits},\ }\href {https://doi.org/10.1126/science.aao4309}
  {\bibfield  {journal} {\bibinfo  {journal} {Science}\ }\textbf {\bibinfo
  {volume} {360}},\ \bibinfo {pages} {195} (\bibinfo {year}
  {2018})}\BibitemShut {NoStop}%
\bibitem [{\citenamefont {Havl{\'i}{\v{c}}ek}\ \emph
  {et~al.}(2019)\citenamefont {Havl{\'i}{\v{c}}ek}, \citenamefont
  {C{\'o}rcoles}, \citenamefont {Temme}, \citenamefont {Harrow}, \citenamefont
  {Kandala}, \citenamefont {Chow},\ and\ \citenamefont
  {Gambetta}}]{havlicek2019}%
  \BibitemOpen
  \bibfield  {author} {\bibinfo {author} {\bibfnamefont {V.}~\bibnamefont
  {Havl{\'i}{\v{c}}ek}}, \bibinfo {author} {\bibfnamefont {A.~D.}\ \bibnamefont
  {C{\'o}rcoles}}, \bibinfo {author} {\bibfnamefont {K.}~\bibnamefont {Temme}},
  \bibinfo {author} {\bibfnamefont {A.~W.}\ \bibnamefont {Harrow}}, \bibinfo
  {author} {\bibfnamefont {A.}~\bibnamefont {Kandala}}, \bibinfo {author}
  {\bibfnamefont {J.~M.}\ \bibnamefont {Chow}},\ and\ \bibinfo {author}
  {\bibfnamefont {J.~M.}\ \bibnamefont {Gambetta}},\ }\bibfield  {title}
  {\bibinfo {title} {Supervised learning with quantum-enhanced feature
  spaces},\ }\href {https://doi.org/10.1038/s41586-019-0980-2} {\bibfield
  {journal} {\bibinfo  {journal} {Nature}\ }\textbf {\bibinfo {volume} {567}},\
  \bibinfo {pages} {209} (\bibinfo {year} {2019})}\BibitemShut {NoStop}%
\bibitem [{\citenamefont {Arute}\ \emph {et~al.}(2019)\citenamefont {Arute},
  \citenamefont {Arya}, \citenamefont {Babbush}, \citenamefont {Bacon},
  \citenamefont {Bardin}, \citenamefont {Barends}, \citenamefont {Biswas},
  \citenamefont {Boixo}, \citenamefont {Brandao}, \citenamefont {Buell},
  \citenamefont {Burkett}, \citenamefont {Chen}, \citenamefont {Chen},
  \citenamefont {Chiaro}, \citenamefont {Collins}, \citenamefont {Courtney},
  \citenamefont {Dunsworth}, \citenamefont {Farhi}, \citenamefont {Foxen},
  \citenamefont {Fowler}, \citenamefont {Gidney}, \citenamefont {Giustina},
  \citenamefont {Graff}, \citenamefont {Guerin}, \citenamefont {Habegger},
  \citenamefont {Harrigan}, \citenamefont {Hartmann}, \citenamefont {Ho},
  \citenamefont {Hoffmann}, \citenamefont {Huang}, \citenamefont {Humble},
  \citenamefont {Isakov}, \citenamefont {Jeffrey}, \citenamefont {Jiang},
  \citenamefont {Kafri}, \citenamefont {Kechedzhi}, \citenamefont {Kelly},
  \citenamefont {Klimov}, \citenamefont {Knysh}, \citenamefont {Korotkov},
  \citenamefont {Kostritsa}, \citenamefont {Landhuis}, \citenamefont
  {Lindmark}, \citenamefont {Lucero}, \citenamefont {Lyakh}, \citenamefont
  {Mandr{\`a}}, \citenamefont {McClean}, \citenamefont {McEwen}, \citenamefont
  {Megrant}, \citenamefont {Mi}, \citenamefont {Michielsen}, \citenamefont
  {Mohseni}, \citenamefont {Mutus}, \citenamefont {Naaman}, \citenamefont
  {Neeley}, \citenamefont {Neill}, \citenamefont {Niu}, \citenamefont {Ostby},
  \citenamefont {Petukhov}, \citenamefont {Platt}, \citenamefont {Quintana},
  \citenamefont {Rieffel}, \citenamefont {Roushan}, \citenamefont {Rubin},
  \citenamefont {Sank}, \citenamefont {Satzinger}, \citenamefont {Smelyanskiy},
  \citenamefont {Sung}, \citenamefont {Trevithick}, \citenamefont
  {Vainsencher}, \citenamefont {Villalonga}, \citenamefont {White},
  \citenamefont {Yao}, \citenamefont {Yeh}, \citenamefont {Zalcman},
  \citenamefont {Neven},\ and\ \citenamefont {Martinis}}]{arute2019}%
  \BibitemOpen
  \bibfield  {author} {\bibinfo {author} {\bibfnamefont {F.}~\bibnamefont
  {Arute}}, \bibinfo {author} {\bibfnamefont {K.}~\bibnamefont {Arya}},
  \bibinfo {author} {\bibfnamefont {R.}~\bibnamefont {Babbush}}, \bibinfo
  {author} {\bibfnamefont {D.}~\bibnamefont {Bacon}}, \bibinfo {author}
  {\bibfnamefont {J.~C.}\ \bibnamefont {Bardin}}, \bibinfo {author}
  {\bibfnamefont {R.}~\bibnamefont {Barends}}, \bibinfo {author} {\bibfnamefont
  {R.}~\bibnamefont {Biswas}}, \bibinfo {author} {\bibfnamefont
  {S.}~\bibnamefont {Boixo}}, \bibinfo {author} {\bibfnamefont {F.~G. S.~L.}\
  \bibnamefont {Brandao}}, \bibinfo {author} {\bibfnamefont {D.~A.}\
  \bibnamefont {Buell}}, \bibinfo {author} {\bibfnamefont {B.}~\bibnamefont
  {Burkett}}, \bibinfo {author} {\bibfnamefont {Y.}~\bibnamefont {Chen}},
  \bibinfo {author} {\bibfnamefont {Z.}~\bibnamefont {Chen}}, \bibinfo {author}
  {\bibfnamefont {B.}~\bibnamefont {Chiaro}}, \bibinfo {author} {\bibfnamefont
  {R.}~\bibnamefont {Collins}}, \bibinfo {author} {\bibfnamefont
  {W.}~\bibnamefont {Courtney}}, \bibinfo {author} {\bibfnamefont
  {A.}~\bibnamefont {Dunsworth}}, \bibinfo {author} {\bibfnamefont
  {E.}~\bibnamefont {Farhi}}, \bibinfo {author} {\bibfnamefont
  {B.}~\bibnamefont {Foxen}}, \bibinfo {author} {\bibfnamefont
  {A.}~\bibnamefont {Fowler}}, \bibinfo {author} {\bibfnamefont
  {C.}~\bibnamefont {Gidney}}, \bibinfo {author} {\bibfnamefont
  {M.}~\bibnamefont {Giustina}}, \bibinfo {author} {\bibfnamefont
  {R.}~\bibnamefont {Graff}}, \bibinfo {author} {\bibfnamefont
  {K.}~\bibnamefont {Guerin}}, \bibinfo {author} {\bibfnamefont
  {S.}~\bibnamefont {Habegger}}, \bibinfo {author} {\bibfnamefont {M.~P.}\
  \bibnamefont {Harrigan}}, \bibinfo {author} {\bibfnamefont {M.~J.}\
  \bibnamefont {Hartmann}}, \bibinfo {author} {\bibfnamefont {A.}~\bibnamefont
  {Ho}}, \bibinfo {author} {\bibfnamefont {M.}~\bibnamefont {Hoffmann}},
  \bibinfo {author} {\bibfnamefont {T.}~\bibnamefont {Huang}}, \bibinfo
  {author} {\bibfnamefont {T.~S.}\ \bibnamefont {Humble}}, \bibinfo {author}
  {\bibfnamefont {S.~V.}\ \bibnamefont {Isakov}}, \bibinfo {author}
  {\bibfnamefont {E.}~\bibnamefont {Jeffrey}}, \bibinfo {author} {\bibfnamefont
  {Z.}~\bibnamefont {Jiang}}, \bibinfo {author} {\bibfnamefont
  {D.}~\bibnamefont {Kafri}}, \bibinfo {author} {\bibfnamefont
  {K.}~\bibnamefont {Kechedzhi}}, \bibinfo {author} {\bibfnamefont
  {J.}~\bibnamefont {Kelly}}, \bibinfo {author} {\bibfnamefont {P.~V.}\
  \bibnamefont {Klimov}}, \bibinfo {author} {\bibfnamefont {S.}~\bibnamefont
  {Knysh}}, \bibinfo {author} {\bibfnamefont {A.}~\bibnamefont {Korotkov}},
  \bibinfo {author} {\bibfnamefont {F.}~\bibnamefont {Kostritsa}}, \bibinfo
  {author} {\bibfnamefont {D.}~\bibnamefont {Landhuis}}, \bibinfo {author}
  {\bibfnamefont {M.}~\bibnamefont {Lindmark}}, \bibinfo {author}
  {\bibfnamefont {E.}~\bibnamefont {Lucero}}, \bibinfo {author} {\bibfnamefont
  {D.}~\bibnamefont {Lyakh}}, \bibinfo {author} {\bibfnamefont
  {S.}~\bibnamefont {Mandr{\`a}}}, \bibinfo {author} {\bibfnamefont {J.~R.}\
  \bibnamefont {McClean}}, \bibinfo {author} {\bibfnamefont {M.}~\bibnamefont
  {McEwen}}, \bibinfo {author} {\bibfnamefont {A.}~\bibnamefont {Megrant}},
  \bibinfo {author} {\bibfnamefont {X.}~\bibnamefont {Mi}}, \bibinfo {author}
  {\bibfnamefont {K.}~\bibnamefont {Michielsen}}, \bibinfo {author}
  {\bibfnamefont {M.}~\bibnamefont {Mohseni}}, \bibinfo {author} {\bibfnamefont
  {J.}~\bibnamefont {Mutus}}, \bibinfo {author} {\bibfnamefont
  {O.}~\bibnamefont {Naaman}}, \bibinfo {author} {\bibfnamefont
  {M.}~\bibnamefont {Neeley}}, \bibinfo {author} {\bibfnamefont
  {C.}~\bibnamefont {Neill}}, \bibinfo {author} {\bibfnamefont {M.~Y.}\
  \bibnamefont {Niu}}, \bibinfo {author} {\bibfnamefont {E.}~\bibnamefont
  {Ostby}}, \bibinfo {author} {\bibfnamefont {A.}~\bibnamefont {Petukhov}},
  \bibinfo {author} {\bibfnamefont {J.~C.}\ \bibnamefont {Platt}}, \bibinfo
  {author} {\bibfnamefont {C.}~\bibnamefont {Quintana}}, \bibinfo {author}
  {\bibfnamefont {E.~G.}\ \bibnamefont {Rieffel}}, \bibinfo {author}
  {\bibfnamefont {P.}~\bibnamefont {Roushan}}, \bibinfo {author} {\bibfnamefont
  {N.~C.}\ \bibnamefont {Rubin}}, \bibinfo {author} {\bibfnamefont
  {D.}~\bibnamefont {Sank}}, \bibinfo {author} {\bibfnamefont {K.~J.}\
  \bibnamefont {Satzinger}}, \bibinfo {author} {\bibfnamefont {V.}~\bibnamefont
  {Smelyanskiy}}, \bibinfo {author} {\bibfnamefont {K.~J.}\ \bibnamefont
  {Sung}}, \bibinfo {author} {\bibfnamefont {M.~D.}\ \bibnamefont
  {Trevithick}}, \bibinfo {author} {\bibfnamefont {A.}~\bibnamefont
  {Vainsencher}}, \bibinfo {author} {\bibfnamefont {B.}~\bibnamefont
  {Villalonga}}, \bibinfo {author} {\bibfnamefont {T.}~\bibnamefont {White}},
  \bibinfo {author} {\bibfnamefont {Z.~J.}\ \bibnamefont {Yao}}, \bibinfo
  {author} {\bibfnamefont {P.}~\bibnamefont {Yeh}}, \bibinfo {author}
  {\bibfnamefont {A.}~\bibnamefont {Zalcman}}, \bibinfo {author} {\bibfnamefont
  {H.}~\bibnamefont {Neven}},\ and\ \bibinfo {author} {\bibfnamefont {J.~M.}\
  \bibnamefont {Martinis}},\ }\bibfield  {title} {\bibinfo {title} {Quantum
  supremacy using a programmable superconducting processor},\ }\href
  {https://doi.org/10.1038/s41586-019-1666-5} {\bibfield  {journal} {\bibinfo
  {journal} {Nature}\ }\textbf {\bibinfo {volume} {574}},\ \bibinfo {pages}
  {505} (\bibinfo {year} {2019})}\BibitemShut {NoStop}%
\bibitem [{\citenamefont {Kjaergaard}\ \emph
  {et~al.}(2020{\natexlab{b}})\citenamefont {Kjaergaard}, \citenamefont
  {Schwartz}, \citenamefont {Greene}, \citenamefont {Samach}, \citenamefont
  {Bengtsson}, \citenamefont {O'Keeffe}, \citenamefont {McNally}, \citenamefont
  {Braumüller}, \citenamefont {Kim}, \citenamefont {Krantz}, \citenamefont
  {Marvian}, \citenamefont {Melville}, \citenamefont {Niedzielski},
  \citenamefont {Sung}, \citenamefont {Winik}, \citenamefont {Yoder},
  \citenamefont {Rosenberg}, \citenamefont {Obenland}, \citenamefont {Lloyd},
  \citenamefont {Orlando}, \citenamefont {Marvian}, \citenamefont
  {Gustavsson},\ and\ \citenamefont {Oliver}}]{kjaergaard2020b}%
  \BibitemOpen
  \bibfield  {author} {\bibinfo {author} {\bibfnamefont {M.}~\bibnamefont
  {Kjaergaard}}, \bibinfo {author} {\bibfnamefont {M.~E.}\ \bibnamefont
  {Schwartz}}, \bibinfo {author} {\bibfnamefont {A.}~\bibnamefont {Greene}},
  \bibinfo {author} {\bibfnamefont {G.~O.}\ \bibnamefont {Samach}}, \bibinfo
  {author} {\bibfnamefont {A.}~\bibnamefont {Bengtsson}}, \bibinfo {author}
  {\bibfnamefont {M.}~\bibnamefont {O'Keeffe}}, \bibinfo {author}
  {\bibfnamefont {C.~M.}\ \bibnamefont {McNally}}, \bibinfo {author}
  {\bibfnamefont {J.}~\bibnamefont {Braumüller}}, \bibinfo {author}
  {\bibfnamefont {D.~K.}\ \bibnamefont {Kim}}, \bibinfo {author} {\bibfnamefont
  {P.}~\bibnamefont {Krantz}}, \bibinfo {author} {\bibfnamefont
  {M.}~\bibnamefont {Marvian}}, \bibinfo {author} {\bibfnamefont
  {A.}~\bibnamefont {Melville}}, \bibinfo {author} {\bibfnamefont {B.~M.}\
  \bibnamefont {Niedzielski}}, \bibinfo {author} {\bibfnamefont
  {Y.}~\bibnamefont {Sung}}, \bibinfo {author} {\bibfnamefont {R.}~\bibnamefont
  {Winik}}, \bibinfo {author} {\bibfnamefont {J.}~\bibnamefont {Yoder}},
  \bibinfo {author} {\bibfnamefont {D.}~\bibnamefont {Rosenberg}}, \bibinfo
  {author} {\bibfnamefont {K.}~\bibnamefont {Obenland}}, \bibinfo {author}
  {\bibfnamefont {S.}~\bibnamefont {Lloyd}}, \bibinfo {author} {\bibfnamefont
  {T.~P.}\ \bibnamefont {Orlando}}, \bibinfo {author} {\bibfnamefont
  {I.}~\bibnamefont {Marvian}}, \bibinfo {author} {\bibfnamefont
  {S.}~\bibnamefont {Gustavsson}},\ and\ \bibinfo {author} {\bibfnamefont
  {W.~D.}\ \bibnamefont {Oliver}},\ }\href@noop {} {\bibinfo {title} {A quantum
  instruction set implemented on a superconducting quantum processor}}
  (\bibinfo {year} {2020}{\natexlab{b}}),\ \Eprint
  {https://arxiv.org/abs/2001.08838} {arXiv:2001.08838 [quant-ph]} \BibitemShut
  {NoStop}%
\bibitem [{\citenamefont {Koch}\ \emph {et~al.}(2007)\citenamefont {Koch},
  \citenamefont {Yu}, \citenamefont {Gambetta}, \citenamefont {Houck},
  \citenamefont {Schuster}, \citenamefont {Majer}, \citenamefont {Blais},
  \citenamefont {Devoret}, \citenamefont {Girvin},\ and\ \citenamefont
  {Schoelkopf}}]{koch2007}%
  \BibitemOpen
  \bibfield  {author} {\bibinfo {author} {\bibfnamefont {J.}~\bibnamefont
  {Koch}}, \bibinfo {author} {\bibfnamefont {T.~M.}\ \bibnamefont {Yu}},
  \bibinfo {author} {\bibfnamefont {J.}~\bibnamefont {Gambetta}}, \bibinfo
  {author} {\bibfnamefont {A.~A.}\ \bibnamefont {Houck}}, \bibinfo {author}
  {\bibfnamefont {D.~I.}\ \bibnamefont {Schuster}}, \bibinfo {author}
  {\bibfnamefont {J.}~\bibnamefont {Majer}}, \bibinfo {author} {\bibfnamefont
  {A.}~\bibnamefont {Blais}}, \bibinfo {author} {\bibfnamefont {M.~H.}\
  \bibnamefont {Devoret}}, \bibinfo {author} {\bibfnamefont {S.~M.}\
  \bibnamefont {Girvin}},\ and\ \bibinfo {author} {\bibfnamefont {R.~J.}\
  \bibnamefont {Schoelkopf}},\ }\bibfield  {title} {\bibinfo {title}
  {{Charge-insensitive qubit design derived from the Cooper pair box}},\ }\href
  {https://doi.org/10.1103/PhysRevA.76.042319} {\bibfield  {journal} {\bibinfo
  {journal} {Phys. Rev. A}\ }\textbf {\bibinfo {volume} {76}},\ \bibinfo
  {pages} {042319} (\bibinfo {year} {2007})}\BibitemShut {NoStop}%
\bibitem [{\citenamefont {Houck}\ \emph {et~al.}(2008)\citenamefont {Houck},
  \citenamefont {Schreier}, \citenamefont {Johnson}, \citenamefont {Chow},
  \citenamefont {Koch}, \citenamefont {Gambetta}, \citenamefont {Schuster},
  \citenamefont {Frunzio}, \citenamefont {Devoret}, \citenamefont {Girvin},\
  and\ \citenamefont {Schoelkopf}}]{houck2008}%
  \BibitemOpen
  \bibfield  {author} {\bibinfo {author} {\bibfnamefont {A.~A.}\ \bibnamefont
  {Houck}}, \bibinfo {author} {\bibfnamefont {J.~A.}\ \bibnamefont {Schreier}},
  \bibinfo {author} {\bibfnamefont {B.~R.}\ \bibnamefont {Johnson}}, \bibinfo
  {author} {\bibfnamefont {J.~M.}\ \bibnamefont {Chow}}, \bibinfo {author}
  {\bibfnamefont {J.}~\bibnamefont {Koch}}, \bibinfo {author} {\bibfnamefont
  {J.~M.}\ \bibnamefont {Gambetta}}, \bibinfo {author} {\bibfnamefont {D.~I.}\
  \bibnamefont {Schuster}}, \bibinfo {author} {\bibfnamefont {L.}~\bibnamefont
  {Frunzio}}, \bibinfo {author} {\bibfnamefont {M.~H.}\ \bibnamefont
  {Devoret}}, \bibinfo {author} {\bibfnamefont {S.~M.}\ \bibnamefont
  {Girvin}},\ and\ \bibinfo {author} {\bibfnamefont {R.~J.}\ \bibnamefont
  {Schoelkopf}},\ }\bibfield  {title} {\bibinfo {title} {Controlling the
  spontaneous emission of a superconducting transmon qubit},\ }\href
  {https://doi.org/10.1103/PhysRevLett.101.080502} {\bibfield  {journal}
  {\bibinfo  {journal} {Phys. Rev. Lett.}\ }\textbf {\bibinfo {volume} {101}},\
  \bibinfo {pages} {080502} (\bibinfo {year} {2008})}\BibitemShut {NoStop}%
\bibitem [{\citenamefont {Oliver}\ and\ \citenamefont
  {Welander}(2013)}]{oliver2013}%
  \BibitemOpen
  \bibfield  {author} {\bibinfo {author} {\bibfnamefont {W.~D.}\ \bibnamefont
  {Oliver}}\ and\ \bibinfo {author} {\bibfnamefont {P.~B.}\ \bibnamefont
  {Welander}},\ }\bibfield  {title} {\bibinfo {title} {Materials in
  superconducting quantum bits},\ }\href {https://doi.org/10.1557/mrs.2013.229}
  {\bibfield  {journal} {\bibinfo  {journal} {MRS Bulletin}\ }\textbf {\bibinfo
  {volume} {38}},\ \bibinfo {pages} {816–825} (\bibinfo {year}
  {2013})}\BibitemShut {NoStop}%
\bibitem [{\citenamefont {Quintana}\ \emph {et~al.}(2014)\citenamefont
  {Quintana}, \citenamefont {Megrant}, \citenamefont {Chen}, \citenamefont
  {Dunsworth}, \citenamefont {Chiaro}, \citenamefont {Barends}, \citenamefont
  {Campbell}, \citenamefont {Chen}, \citenamefont {Hoi}, \citenamefont
  {Jeffrey}, \citenamefont {Kelly}, \citenamefont {Mutus}, \citenamefont
  {O'Malley}, \citenamefont {Neill}, \citenamefont {Roushan}, \citenamefont
  {Sank}, \citenamefont {Vainsencher}, \citenamefont {Wenner}, \citenamefont
  {White}, \citenamefont {Cleland},\ and\ \citenamefont
  {Martinis}}]{quintana2014}%
  \BibitemOpen
  \bibfield  {author} {\bibinfo {author} {\bibfnamefont {C.~M.}\ \bibnamefont
  {Quintana}}, \bibinfo {author} {\bibfnamefont {A.}~\bibnamefont {Megrant}},
  \bibinfo {author} {\bibfnamefont {Z.}~\bibnamefont {Chen}}, \bibinfo {author}
  {\bibfnamefont {A.}~\bibnamefont {Dunsworth}}, \bibinfo {author}
  {\bibfnamefont {B.}~\bibnamefont {Chiaro}}, \bibinfo {author} {\bibfnamefont
  {R.}~\bibnamefont {Barends}}, \bibinfo {author} {\bibfnamefont
  {B.}~\bibnamefont {Campbell}}, \bibinfo {author} {\bibfnamefont
  {Y.}~\bibnamefont {Chen}}, \bibinfo {author} {\bibfnamefont {I.-C.}\
  \bibnamefont {Hoi}}, \bibinfo {author} {\bibfnamefont {E.}~\bibnamefont
  {Jeffrey}}, \bibinfo {author} {\bibfnamefont {J.}~\bibnamefont {Kelly}},
  \bibinfo {author} {\bibfnamefont {J.~Y.}\ \bibnamefont {Mutus}}, \bibinfo
  {author} {\bibfnamefont {P.~J.~J.}\ \bibnamefont {O'Malley}}, \bibinfo
  {author} {\bibfnamefont {C.}~\bibnamefont {Neill}}, \bibinfo {author}
  {\bibfnamefont {P.}~\bibnamefont {Roushan}}, \bibinfo {author} {\bibfnamefont
  {D.}~\bibnamefont {Sank}}, \bibinfo {author} {\bibfnamefont {A.}~\bibnamefont
  {Vainsencher}}, \bibinfo {author} {\bibfnamefont {J.}~\bibnamefont {Wenner}},
  \bibinfo {author} {\bibfnamefont {T.~C.}\ \bibnamefont {White}}, \bibinfo
  {author} {\bibfnamefont {A.~N.}\ \bibnamefont {Cleland}},\ and\ \bibinfo
  {author} {\bibfnamefont {J.~M.}\ \bibnamefont {Martinis}},\ }\bibfield
  {title} {\bibinfo {title} {Characterization and reduction of
  microfabrication-induced decoherence in superconducting quantum circuits},\
  }\href {https://doi.org/10.1063/1.4893297} {\bibfield  {journal} {\bibinfo
  {journal} {Applied Physics Letters}\ }\textbf {\bibinfo {volume} {105}},\
  \bibinfo {pages} {062601} (\bibinfo {year} {2014})},\ \Eprint
  {https://arxiv.org/abs/https://doi.org/10.1063/1.4893297}
  {https://doi.org/10.1063/1.4893297} \BibitemShut {NoStop}%
\bibitem [{\citenamefont {Braum\"uller}\ \emph {et~al.}(2020)\citenamefont
  {Braum\"uller}, \citenamefont {Ding}, \citenamefont {Veps\"al\"ainen},
  \citenamefont {Sung}, \citenamefont {Kjaergaard}, \citenamefont {Menke},
  \citenamefont {Winik}, \citenamefont {Kim}, \citenamefont {Niedzielski},
  \citenamefont {Melville}, \citenamefont {Yoder}, \citenamefont
  {Hirjibehedin}, \citenamefont {Orlando}, \citenamefont {Gustavsson},\ and\
  \citenamefont {Oliver}}]{braumuller2020}%
  \BibitemOpen
  \bibfield  {author} {\bibinfo {author} {\bibfnamefont {J.}~\bibnamefont
  {Braum\"uller}}, \bibinfo {author} {\bibfnamefont {L.}~\bibnamefont {Ding}},
  \bibinfo {author} {\bibfnamefont {A.~P.}\ \bibnamefont {Veps\"al\"ainen}},
  \bibinfo {author} {\bibfnamefont {Y.}~\bibnamefont {Sung}}, \bibinfo {author}
  {\bibfnamefont {M.}~\bibnamefont {Kjaergaard}}, \bibinfo {author}
  {\bibfnamefont {T.}~\bibnamefont {Menke}}, \bibinfo {author} {\bibfnamefont
  {R.}~\bibnamefont {Winik}}, \bibinfo {author} {\bibfnamefont
  {D.}~\bibnamefont {Kim}}, \bibinfo {author} {\bibfnamefont {B.~M.}\
  \bibnamefont {Niedzielski}}, \bibinfo {author} {\bibfnamefont
  {A.}~\bibnamefont {Melville}}, \bibinfo {author} {\bibfnamefont {J.~L.}\
  \bibnamefont {Yoder}}, \bibinfo {author} {\bibfnamefont {C.~F.}\ \bibnamefont
  {Hirjibehedin}}, \bibinfo {author} {\bibfnamefont {T.~P.}\ \bibnamefont
  {Orlando}}, \bibinfo {author} {\bibfnamefont {S.}~\bibnamefont
  {Gustavsson}},\ and\ \bibinfo {author} {\bibfnamefont {W.~D.}\ \bibnamefont
  {Oliver}},\ }\bibfield  {title} {\bibinfo {title} {Characterizing and
  optimizing qubit coherence based on squid geometry},\ }\href
  {https://doi.org/10.1103/PhysRevApplied.13.054079} {\bibfield  {journal}
  {\bibinfo  {journal} {Phys. Rev. Applied}\ }\textbf {\bibinfo {volume}
  {13}},\ \bibinfo {pages} {054079} (\bibinfo {year} {2020})}\BibitemShut
  {NoStop}%
\bibitem [{\citenamefont {Place}\ \emph {et~al.}(2021)\citenamefont {Place},
  \citenamefont {Rodgers}, \citenamefont {Mundada}, \citenamefont {Smitham},
  \citenamefont {Fitzpatrick}, \citenamefont {Leng}, \citenamefont {Premkumar},
  \citenamefont {Bryon}, \citenamefont {Vrajitoarea}, \citenamefont {Sussman},
  \citenamefont {Cheng}, \citenamefont {Madhavan}, \citenamefont {Babla},
  \citenamefont {Le}, \citenamefont {Gang}, \citenamefont {J{\"a}ck},
  \citenamefont {Gyenis}, \citenamefont {Yao}, \citenamefont {Cava},
  \citenamefont {de~Leon},\ and\ \citenamefont {Houck}}]{place2020}%
  \BibitemOpen
  \bibfield  {author} {\bibinfo {author} {\bibfnamefont {A.~P.~M.}\
  \bibnamefont {Place}}, \bibinfo {author} {\bibfnamefont {L.~V.~H.}\
  \bibnamefont {Rodgers}}, \bibinfo {author} {\bibfnamefont {P.}~\bibnamefont
  {Mundada}}, \bibinfo {author} {\bibfnamefont {B.~M.}\ \bibnamefont
  {Smitham}}, \bibinfo {author} {\bibfnamefont {M.}~\bibnamefont
  {Fitzpatrick}}, \bibinfo {author} {\bibfnamefont {Z.}~\bibnamefont {Leng}},
  \bibinfo {author} {\bibfnamefont {A.}~\bibnamefont {Premkumar}}, \bibinfo
  {author} {\bibfnamefont {J.}~\bibnamefont {Bryon}}, \bibinfo {author}
  {\bibfnamefont {A.}~\bibnamefont {Vrajitoarea}}, \bibinfo {author}
  {\bibfnamefont {S.}~\bibnamefont {Sussman}}, \bibinfo {author} {\bibfnamefont
  {G.}~\bibnamefont {Cheng}}, \bibinfo {author} {\bibfnamefont
  {T.}~\bibnamefont {Madhavan}}, \bibinfo {author} {\bibfnamefont {H.~K.}\
  \bibnamefont {Babla}}, \bibinfo {author} {\bibfnamefont {X.~H.}\ \bibnamefont
  {Le}}, \bibinfo {author} {\bibfnamefont {Y.}~\bibnamefont {Gang}}, \bibinfo
  {author} {\bibfnamefont {B.}~\bibnamefont {J{\"a}ck}}, \bibinfo {author}
  {\bibfnamefont {A.}~\bibnamefont {Gyenis}}, \bibinfo {author} {\bibfnamefont
  {N.}~\bibnamefont {Yao}}, \bibinfo {author} {\bibfnamefont {R.~J.}\
  \bibnamefont {Cava}}, \bibinfo {author} {\bibfnamefont {N.~P.}\ \bibnamefont
  {de~Leon}},\ and\ \bibinfo {author} {\bibfnamefont {A.~A.}\ \bibnamefont
  {Houck}},\ }\bibfield  {title} {\bibinfo {title} {New material platform for
  superconducting transmon qubits with coherence times exceeding 0.3
  milliseconds},\ }\href {https://doi.org/10.1038/s41467-021-22030-5}
  {\bibfield  {journal} {\bibinfo  {journal} {Nature Communications}\ }\textbf
  {\bibinfo {volume} {12}},\ \bibinfo {pages} {1779} (\bibinfo {year}
  {2021})}\BibitemShut {NoStop}%
\bibitem [{\citenamefont {{Rosenberg}}\ \emph {et~al.}(2020)\citenamefont
  {{Rosenberg}}, \citenamefont {{Weber}}, \citenamefont {{Conway}},
  \citenamefont {{Yost}}, \citenamefont {{Mallek}}, \citenamefont {{Calusine}},
  \citenamefont {{Das}}, \citenamefont {{Kim}}, \citenamefont {{Schwartz}},
  \citenamefont {{Woods}}, \citenamefont {{Yoder}},\ and\ \citenamefont
  {{Oliver}}}]{rosenberg2020}%
  \BibitemOpen
  \bibfield  {author} {\bibinfo {author} {\bibfnamefont {D.}~\bibnamefont
  {{Rosenberg}}}, \bibinfo {author} {\bibfnamefont {S.~J.}\ \bibnamefont
  {{Weber}}}, \bibinfo {author} {\bibfnamefont {D.}~\bibnamefont {{Conway}}},
  \bibinfo {author} {\bibfnamefont {D.~W.}\ \bibnamefont {{Yost}}}, \bibinfo
  {author} {\bibfnamefont {J.}~\bibnamefont {{Mallek}}}, \bibinfo {author}
  {\bibfnamefont {G.}~\bibnamefont {{Calusine}}}, \bibinfo {author}
  {\bibfnamefont {R.}~\bibnamefont {{Das}}}, \bibinfo {author} {\bibfnamefont
  {D.}~\bibnamefont {{Kim}}}, \bibinfo {author} {\bibfnamefont {M.~E.}\
  \bibnamefont {{Schwartz}}}, \bibinfo {author} {\bibfnamefont
  {W.}~\bibnamefont {{Woods}}}, \bibinfo {author} {\bibfnamefont {J.~L.}\
  \bibnamefont {{Yoder}}},\ and\ \bibinfo {author} {\bibfnamefont {W.~D.}\
  \bibnamefont {{Oliver}}},\ }\bibfield  {title} {\bibinfo {title} {Solid-state
  qubits: {3D} integration and packaging},\ }\href
  {https://doi.org/10.1109/MMM.2020.2993478} {\bibfield  {journal} {\bibinfo
  {journal} {IEEE Microwave Magazine}\ }\textbf {\bibinfo {volume} {21}},\
  \bibinfo {pages} {72} (\bibinfo {year} {2020})}\BibitemShut {NoStop}%
\bibitem [{\citenamefont {Kitaev}(2003)}]{kitaev2003}%
  \BibitemOpen
  \bibfield  {author} {\bibinfo {author} {\bibfnamefont {A.}~\bibnamefont
  {Kitaev}},\ }\bibfield  {title} {\bibinfo {title} {Fault-tolerant quantum
  computation by anyons},\ }\href
  {https://doi.org/https://doi.org/10.1016/S0003-4916(02)00018-0} {\bibfield
  {journal} {\bibinfo  {journal} {Annals of Physics}\ }\textbf {\bibinfo
  {volume} {303}},\ \bibinfo {pages} {2 } (\bibinfo {year} {2003})}\BibitemShut
  {NoStop}%
\bibitem [{\citenamefont {Ioffe}\ \emph {et~al.}(2002)\citenamefont {Ioffe},
  \citenamefont {Feigel'man}, \citenamefont {Ioselevich}, \citenamefont
  {Ivanov}, \citenamefont {Troyer},\ and\ \citenamefont
  {Blatter}}]{ioffe2002a}%
  \BibitemOpen
  \bibfield  {author} {\bibinfo {author} {\bibfnamefont {L.~B.}\ \bibnamefont
  {Ioffe}}, \bibinfo {author} {\bibfnamefont {M.~V.}\ \bibnamefont
  {Feigel'man}}, \bibinfo {author} {\bibfnamefont {A.}~\bibnamefont
  {Ioselevich}}, \bibinfo {author} {\bibfnamefont {D.}~\bibnamefont {Ivanov}},
  \bibinfo {author} {\bibfnamefont {M.}~\bibnamefont {Troyer}},\ and\ \bibinfo
  {author} {\bibfnamefont {G.}~\bibnamefont {Blatter}},\ }\bibfield  {title}
  {\bibinfo {title} {{Topologically protected quantum bits using Josephson
  junction arrays}},\ }\href {https://doi.org/10.1038/415503a} {\bibfield
  {journal} {\bibinfo  {journal} {Nature}\ }\textbf {\bibinfo {volume} {415}},\
  \bibinfo {pages} {503} (\bibinfo {year} {2002})}\BibitemShut {NoStop}%
\bibitem [{\citenamefont {Ioffe}\ and\ \citenamefont
  {Feigel'man}(2002)}]{ioffe2002b}%
  \BibitemOpen
  \bibfield  {author} {\bibinfo {author} {\bibfnamefont {L.~B.}\ \bibnamefont
  {Ioffe}}\ and\ \bibinfo {author} {\bibfnamefont {M.~V.}\ \bibnamefont
  {Feigel'man}},\ }\bibfield  {title} {\bibinfo {title} {{Possible realization
  of an ideal quantum computer in Josephson junction array}},\ }\href
  {https://doi.org/10.1103/PhysRevB.66.224503} {\bibfield  {journal} {\bibinfo
  {journal} {Phys. Rev. B}\ }\textbf {\bibinfo {volume} {66}},\ \bibinfo
  {pages} {224503} (\bibinfo {year} {2002})}\BibitemShut {NoStop}%
\bibitem [{\citenamefont {Dou\ifmmode~\mbox{\c{c}}\else \c{c}\fi{}ot}\ and\
  \citenamefont {Vidal}(2002)}]{doucot2002}%
  \BibitemOpen
  \bibfield  {author} {\bibinfo {author} {\bibfnamefont {B.}~\bibnamefont
  {Dou\ifmmode~\mbox{\c{c}}\else \c{c}\fi{}ot}}\ and\ \bibinfo {author}
  {\bibfnamefont {J.}~\bibnamefont {Vidal}},\ }\bibfield  {title} {\bibinfo
  {title} {Pairing of {Cooper} pairs in a fully frustrated {Josephson}-junction
  chain},\ }\href {https://doi.org/10.1103/PhysRevLett.88.227005} {\bibfield
  {journal} {\bibinfo  {journal} {Phys. Rev. Lett.}\ }\textbf {\bibinfo
  {volume} {88}},\ \bibinfo {pages} {227005} (\bibinfo {year}
  {2002})}\BibitemShut {NoStop}%
\bibitem [{\citenamefont {Dou\ifmmode~\mbox{\c{c}}\else \c{c}\fi{}ot}\ \emph
  {et~al.}(2005)\citenamefont {Dou\ifmmode~\mbox{\c{c}}\else \c{c}\fi{}ot},
  \citenamefont {Feigel'man}, \citenamefont {Ioffe},\ and\ \citenamefont
  {Ioselevich}}]{doucot2005}%
  \BibitemOpen
  \bibfield  {author} {\bibinfo {author} {\bibfnamefont {B.}~\bibnamefont
  {Dou\ifmmode~\mbox{\c{c}}\else \c{c}\fi{}ot}}, \bibinfo {author}
  {\bibfnamefont {M.~V.}\ \bibnamefont {Feigel'man}}, \bibinfo {author}
  {\bibfnamefont {L.~B.}\ \bibnamefont {Ioffe}},\ and\ \bibinfo {author}
  {\bibfnamefont {A.~S.}\ \bibnamefont {Ioselevich}},\ }\bibfield  {title}
  {\bibinfo {title} {{Protected qubits and Chern-Simons theories in Josephson
  junction arrays}},\ }\href {https://doi.org/10.1103/PhysRevB.71.024505}
  {\bibfield  {journal} {\bibinfo  {journal} {Phys. Rev. B}\ }\textbf {\bibinfo
  {volume} {71}},\ \bibinfo {pages} {024505} (\bibinfo {year}
  {2005})}\BibitemShut {NoStop}%
\bibitem [{\citenamefont {Gladchenko}\ \emph {et~al.}(2008)\citenamefont
  {Gladchenko}, \citenamefont {Olaya}, \citenamefont {Dupont-Ferrier},
  \citenamefont {Dou{\c{c}}ot}, \citenamefont {Ioffe},\ and\ \citenamefont
  {Gershenson}}]{gladchenko2008}%
  \BibitemOpen
  \bibfield  {author} {\bibinfo {author} {\bibfnamefont {S.}~\bibnamefont
  {Gladchenko}}, \bibinfo {author} {\bibfnamefont {D.}~\bibnamefont {Olaya}},
  \bibinfo {author} {\bibfnamefont {E.}~\bibnamefont {Dupont-Ferrier}},
  \bibinfo {author} {\bibfnamefont {B.}~\bibnamefont {Dou{\c{c}}ot}}, \bibinfo
  {author} {\bibfnamefont {L.~B.}\ \bibnamefont {Ioffe}},\ and\ \bibinfo
  {author} {\bibfnamefont {M.~E.}\ \bibnamefont {Gershenson}},\ }\bibfield
  {title} {\bibinfo {title} {Superconducting nanocircuits for topologically
  protected qubits},\ }\href {https://doi.org/10.1038/nphys1151} {\bibfield
  {journal} {\bibinfo  {journal} {Nature Physics}\ }\textbf {\bibinfo {volume}
  {5}},\ \bibinfo {pages} {48} (\bibinfo {year} {2008})}\BibitemShut {NoStop}%
\bibitem [{\citenamefont {Dou{\c{c}}ot}\ and\ \citenamefont
  {Ioffe}(2012)}]{doucot2012}%
  \BibitemOpen
  \bibfield  {author} {\bibinfo {author} {\bibfnamefont {B.}~\bibnamefont
  {Dou{\c{c}}ot}}\ and\ \bibinfo {author} {\bibfnamefont {L.~B.}\ \bibnamefont
  {Ioffe}},\ }\bibfield  {title} {\bibinfo {title} {Physical implementation of
  protected qubits},\ }\href {https://doi.org/10.1088/0034-4885/75/7/072001}
  {\bibfield  {journal} {\bibinfo  {journal} {Reports on Progress in Physics}\
  }\textbf {\bibinfo {volume} {75}},\ \bibinfo {pages} {072001} (\bibinfo
  {year} {2012})}\BibitemShut {NoStop}%
\bibitem [{\citenamefont {Bell}\ \emph {et~al.}(2014)\citenamefont {Bell},
  \citenamefont {Paramanandam}, \citenamefont {Ioffe},\ and\ \citenamefont
  {Gershenson}}]{bell2014}%
  \BibitemOpen
  \bibfield  {author} {\bibinfo {author} {\bibfnamefont {M.~T.}\ \bibnamefont
  {Bell}}, \bibinfo {author} {\bibfnamefont {J.}~\bibnamefont {Paramanandam}},
  \bibinfo {author} {\bibfnamefont {L.~B.}\ \bibnamefont {Ioffe}},\ and\
  \bibinfo {author} {\bibfnamefont {M.~E.}\ \bibnamefont {Gershenson}},\
  }\bibfield  {title} {\bibinfo {title} {Protected {J}osephson rhombus
  chains},\ }\href {https://doi.org/10.1103/PhysRevLett.112.167001} {\bibfield
  {journal} {\bibinfo  {journal} {Phys. Rev. Lett.}\ }\textbf {\bibinfo
  {volume} {112}},\ \bibinfo {pages} {167001} (\bibinfo {year}
  {2014})}\BibitemShut {NoStop}%
\bibitem [{\citenamefont {Shor}(1995)}]{shor1995}%
  \BibitemOpen
  \bibfield  {author} {\bibinfo {author} {\bibfnamefont {P.~W.}\ \bibnamefont
  {Shor}},\ }\bibfield  {title} {\bibinfo {title} {Scheme for reducing
  decoherence in quantum computer memory},\ }\href
  {https://doi.org/10.1103/PhysRevA.52.R2493} {\bibfield  {journal} {\bibinfo
  {journal} {Phys. Rev. A}\ }\textbf {\bibinfo {volume} {52}},\ \bibinfo
  {pages} {R2493} (\bibinfo {year} {1995})}\BibitemShut {NoStop}%
\bibitem [{\citenamefont {Fowler}\ \emph {et~al.}(2012)\citenamefont {Fowler},
  \citenamefont {Mariantoni}, \citenamefont {Martinis},\ and\ \citenamefont
  {Cleland}}]{fowler2012}%
  \BibitemOpen
  \bibfield  {author} {\bibinfo {author} {\bibfnamefont {A.~G.}\ \bibnamefont
  {Fowler}}, \bibinfo {author} {\bibfnamefont {M.}~\bibnamefont {Mariantoni}},
  \bibinfo {author} {\bibfnamefont {J.~M.}\ \bibnamefont {Martinis}},\ and\
  \bibinfo {author} {\bibfnamefont {A.~N.}\ \bibnamefont {Cleland}},\
  }\bibfield  {title} {\bibinfo {title} {Surface codes: Towards practical
  large-scale quantum computation},\ }\href
  {https://doi.org/10.1103/PhysRevA.86.032324} {\bibfield  {journal} {\bibinfo
  {journal} {Phys. Rev. A}\ }\textbf {\bibinfo {volume} {86}},\ \bibinfo
  {pages} {032324} (\bibinfo {year} {2012})}\BibitemShut {NoStop}%
\bibitem [{\citenamefont {Reed}\ \emph {et~al.}(2012)\citenamefont {Reed},
  \citenamefont {DiCarlo}, \citenamefont {Nigg}, \citenamefont {Sun},
  \citenamefont {Frunzio}, \citenamefont {Girvin},\ and\ \citenamefont
  {Schoelkopf}}]{reed2012}%
  \BibitemOpen
  \bibfield  {author} {\bibinfo {author} {\bibfnamefont {M.~D.}\ \bibnamefont
  {Reed}}, \bibinfo {author} {\bibfnamefont {L.}~\bibnamefont {DiCarlo}},
  \bibinfo {author} {\bibfnamefont {S.~E.}\ \bibnamefont {Nigg}}, \bibinfo
  {author} {\bibfnamefont {L.}~\bibnamefont {Sun}}, \bibinfo {author}
  {\bibfnamefont {L.}~\bibnamefont {Frunzio}}, \bibinfo {author} {\bibfnamefont
  {S.~M.}\ \bibnamefont {Girvin}},\ and\ \bibinfo {author} {\bibfnamefont
  {R.~J.}\ \bibnamefont {Schoelkopf}},\ }\bibfield  {title} {\bibinfo {title}
  {Realization of three-qubit quantum error correction with superconducting
  circuits},\ }\href {https://www.nature.com/articles/nature10786} {\bibfield
  {journal} {\bibinfo  {journal} {Nature}\ }\textbf {\bibinfo {volume} {482}},\
  \bibinfo {pages} {382} (\bibinfo {year} {2012})}\BibitemShut {NoStop}%
\bibitem [{\citenamefont {Chow}\ \emph {et~al.}(2014)\citenamefont {Chow},
  \citenamefont {Gambetta}, \citenamefont {Magesan}, \citenamefont {Abraham},
  \citenamefont {Cross}, \citenamefont {Johnson}, \citenamefont {Masluk},
  \citenamefont {Ryan}, \citenamefont {Smolin}, \citenamefont {Srinivasan},\
  and\ \citenamefont {Steffen}}]{chow2014}%
  \BibitemOpen
  \bibfield  {author} {\bibinfo {author} {\bibfnamefont {J.~M.}\ \bibnamefont
  {Chow}}, \bibinfo {author} {\bibfnamefont {J.~M.}\ \bibnamefont {Gambetta}},
  \bibinfo {author} {\bibfnamefont {E.}~\bibnamefont {Magesan}}, \bibinfo
  {author} {\bibfnamefont {D.~W.}\ \bibnamefont {Abraham}}, \bibinfo {author}
  {\bibfnamefont {A.~W.}\ \bibnamefont {Cross}}, \bibinfo {author}
  {\bibfnamefont {B.~R.}\ \bibnamefont {Johnson}}, \bibinfo {author}
  {\bibfnamefont {N.~A.}\ \bibnamefont {Masluk}}, \bibinfo {author}
  {\bibfnamefont {C.~A.}\ \bibnamefont {Ryan}}, \bibinfo {author}
  {\bibfnamefont {J.~A.}\ \bibnamefont {Smolin}}, \bibinfo {author}
  {\bibfnamefont {S.~J.}\ \bibnamefont {Srinivasan}},\ and\ \bibinfo {author}
  {\bibfnamefont {M.}~\bibnamefont {Steffen}},\ }\bibfield  {title} {\bibinfo
  {title} {Implementing a strand of a scalable fault-tolerant quantum computing
  fabric},\ }\href {https://doi.org/10.1038/ncomms5015} {\bibfield  {journal}
  {\bibinfo  {journal} {Nature Communications}\ }\textbf {\bibinfo {volume}
  {5}},\ \bibinfo {pages} {4015} (\bibinfo {year} {2014})}\BibitemShut
  {NoStop}%
\bibitem [{\citenamefont {Saira}\ \emph {et~al.}(2014)\citenamefont {Saira},
  \citenamefont {Groen}, \citenamefont {Cramer}, \citenamefont {Meretska},
  \citenamefont {de~Lange},\ and\ \citenamefont {DiCarlo}}]{saira2014}%
  \BibitemOpen
  \bibfield  {author} {\bibinfo {author} {\bibfnamefont {O.-P.}\ \bibnamefont
  {Saira}}, \bibinfo {author} {\bibfnamefont {J.~P.}\ \bibnamefont {Groen}},
  \bibinfo {author} {\bibfnamefont {J.}~\bibnamefont {Cramer}}, \bibinfo
  {author} {\bibfnamefont {M.}~\bibnamefont {Meretska}}, \bibinfo {author}
  {\bibfnamefont {G.}~\bibnamefont {de~Lange}},\ and\ \bibinfo {author}
  {\bibfnamefont {L.}~\bibnamefont {DiCarlo}},\ }\bibfield  {title} {\bibinfo
  {title} {{Entanglement genesis by ancilla-based parity measurement in 2D
  circuit QED}},\ }\href {https://doi.org/10.1103/PhysRevLett.112.070502}
  {\bibfield  {journal} {\bibinfo  {journal} {Phys. Rev. Lett.}\ }\textbf
  {\bibinfo {volume} {112}},\ \bibinfo {pages} {070502} (\bibinfo {year}
  {2014})}\BibitemShut {NoStop}%
\bibitem [{\citenamefont {Kelly}\ \emph {et~al.}(2015)\citenamefont {Kelly},
  \citenamefont {Barends}, \citenamefont {Fowler}, \citenamefont {Megrant},
  \citenamefont {Jeffrey}, \citenamefont {White}, \citenamefont {Sank},
  \citenamefont {Mutus}, \citenamefont {Campbell}, \citenamefont {Chen},
  \citenamefont {Chen}, \citenamefont {Chiaro}, \citenamefont {Dunsworth},
  \citenamefont {Hoi}, \citenamefont {Neill}, \citenamefont {O'Malley},
  \citenamefont {Quintana}, \citenamefont {Roushan}, \citenamefont
  {Vainsencher}, \citenamefont {Wenner}, \citenamefont {Cleland},\ and\
  \citenamefont {Martinis}}]{kelly2015}%
  \BibitemOpen
  \bibfield  {author} {\bibinfo {author} {\bibfnamefont {J.}~\bibnamefont
  {Kelly}}, \bibinfo {author} {\bibfnamefont {R.}~\bibnamefont {Barends}},
  \bibinfo {author} {\bibfnamefont {A.~G.}\ \bibnamefont {Fowler}}, \bibinfo
  {author} {\bibfnamefont {A.}~\bibnamefont {Megrant}}, \bibinfo {author}
  {\bibfnamefont {E.}~\bibnamefont {Jeffrey}}, \bibinfo {author} {\bibfnamefont
  {T.~C.}\ \bibnamefont {White}}, \bibinfo {author} {\bibfnamefont
  {D.}~\bibnamefont {Sank}}, \bibinfo {author} {\bibfnamefont {J.~Y.}\
  \bibnamefont {Mutus}}, \bibinfo {author} {\bibfnamefont {B.}~\bibnamefont
  {Campbell}}, \bibinfo {author} {\bibfnamefont {Y.}~\bibnamefont {Chen}},
  \bibinfo {author} {\bibfnamefont {Z.}~\bibnamefont {Chen}}, \bibinfo {author}
  {\bibfnamefont {B.}~\bibnamefont {Chiaro}}, \bibinfo {author} {\bibfnamefont
  {A.}~\bibnamefont {Dunsworth}}, \bibinfo {author} {\bibfnamefont {I.-C.}\
  \bibnamefont {Hoi}}, \bibinfo {author} {\bibfnamefont {C.}~\bibnamefont
  {Neill}}, \bibinfo {author} {\bibfnamefont {P.~J.~J.}\ \bibnamefont
  {O'Malley}}, \bibinfo {author} {\bibfnamefont {C.}~\bibnamefont {Quintana}},
  \bibinfo {author} {\bibfnamefont {P.}~\bibnamefont {Roushan}}, \bibinfo
  {author} {\bibfnamefont {A.}~\bibnamefont {Vainsencher}}, \bibinfo {author}
  {\bibfnamefont {J.}~\bibnamefont {Wenner}}, \bibinfo {author} {\bibfnamefont
  {A.~N.}\ \bibnamefont {Cleland}},\ and\ \bibinfo {author} {\bibfnamefont
  {J.~M.}\ \bibnamefont {Martinis}},\ }\bibfield  {title} {\bibinfo {title}
  {State preservation by repetitive error detection in a superconducting
  quantum circuit},\ }\href {https://doi.org/10.1038/nature14270} {\bibfield
  {journal} {\bibinfo  {journal} {Nature}\ }\textbf {\bibinfo {volume} {519}},\
  \bibinfo {pages} {66} (\bibinfo {year} {2015})}\BibitemShut {NoStop}%
\bibitem [{\citenamefont {Ofek}\ \emph {et~al.}(2016)\citenamefont {Ofek},
  \citenamefont {Petrenko}, \citenamefont {Heeres}, \citenamefont {Reinhold},
  \citenamefont {Leghtas}, \citenamefont {Vlastakis}, \citenamefont {Liu},
  \citenamefont {Frunzio}, \citenamefont {Girvin}, \citenamefont {Jiang},
  \citenamefont {Mirrahimi}, \citenamefont {Devoret},\ and\ \citenamefont
  {Schoelkopf}}]{ofek2016}%
  \BibitemOpen
  \bibfield  {author} {\bibinfo {author} {\bibfnamefont {N.}~\bibnamefont
  {Ofek}}, \bibinfo {author} {\bibfnamefont {A.}~\bibnamefont {Petrenko}},
  \bibinfo {author} {\bibfnamefont {R.}~\bibnamefont {Heeres}}, \bibinfo
  {author} {\bibfnamefont {P.}~\bibnamefont {Reinhold}}, \bibinfo {author}
  {\bibfnamefont {Z.}~\bibnamefont {Leghtas}}, \bibinfo {author} {\bibfnamefont
  {B.}~\bibnamefont {Vlastakis}}, \bibinfo {author} {\bibfnamefont
  {Y.}~\bibnamefont {Liu}}, \bibinfo {author} {\bibfnamefont {L.}~\bibnamefont
  {Frunzio}}, \bibinfo {author} {\bibfnamefont {S.~M.}\ \bibnamefont {Girvin}},
  \bibinfo {author} {\bibfnamefont {L.}~\bibnamefont {Jiang}}, \bibinfo
  {author} {\bibfnamefont {M.}~\bibnamefont {Mirrahimi}}, \bibinfo {author}
  {\bibfnamefont {M.~H.}\ \bibnamefont {Devoret}},\ and\ \bibinfo {author}
  {\bibfnamefont {R.~J.}\ \bibnamefont {Schoelkopf}},\ }\bibfield  {title}
  {\bibinfo {title} {Extending the lifetime of a quantum bit with error
  correction in superconducting circuits},\ }\href
  {https://doi.org/10.1038/nature18949} {\bibfield  {journal} {\bibinfo
  {journal} {Nature}\ }\textbf {\bibinfo {volume} {536}},\ \bibinfo {pages}
  {441} (\bibinfo {year} {2016})}\BibitemShut {NoStop}%
\bibitem [{\citenamefont {Gambetta}\ \emph {et~al.}(2017)\citenamefont
  {Gambetta}, \citenamefont {Chow},\ and\ \citenamefont
  {Steffen}}]{gambetta2017}%
  \BibitemOpen
  \bibfield  {author} {\bibinfo {author} {\bibfnamefont {J.~M.}\ \bibnamefont
  {Gambetta}}, \bibinfo {author} {\bibfnamefont {J.~M.}\ \bibnamefont {Chow}},\
  and\ \bibinfo {author} {\bibfnamefont {M.}~\bibnamefont {Steffen}},\
  }\bibfield  {title} {\bibinfo {title} {Building logical qubits in a
  superconducting quantum computing system},\ }\href
  {https://doi.org/10.1038/s41534-016-0004-0} {\bibfield  {journal} {\bibinfo
  {journal} {npj Quantum Information}\ }\textbf {\bibinfo {volume} {3}},\
  \bibinfo {pages} {2} (\bibinfo {year} {2017})}\BibitemShut {NoStop}%
\bibitem [{\citenamefont {Gottesman}\ \emph {et~al.}(2001)\citenamefont
  {Gottesman}, \citenamefont {Kitaev},\ and\ \citenamefont
  {Preskill}}]{gottesman2001}%
  \BibitemOpen
  \bibfield  {author} {\bibinfo {author} {\bibfnamefont {D.}~\bibnamefont
  {Gottesman}}, \bibinfo {author} {\bibfnamefont {A.}~\bibnamefont {Kitaev}},\
  and\ \bibinfo {author} {\bibfnamefont {J.}~\bibnamefont {Preskill}},\
  }\bibfield  {title} {\bibinfo {title} {Encoding a qubit in an oscillator},\
  }\href {https://doi.org/10.1103/PhysRevA.64.012310} {\bibfield  {journal}
  {\bibinfo  {journal} {Phys. Rev. A}\ }\textbf {\bibinfo {volume} {64}},\
  \bibinfo {pages} {012310} (\bibinfo {year} {2001})}\BibitemShut {NoStop}%
\bibitem [{\citenamefont {Campagne-Ibarcq}\ \emph {et~al.}(2020)\citenamefont
  {Campagne-Ibarcq}, \citenamefont {Eickbusch}, \citenamefont {Touzard},
  \citenamefont {Zalys-Geller}, \citenamefont {Frattini}, \citenamefont
  {Sivak}, \citenamefont {Reinhold}, \citenamefont {Puri}, \citenamefont
  {Shankar}, \citenamefont {Schoelkopf}, \citenamefont {Frunzio}, \citenamefont
  {Mirrahimi},\ and\ \citenamefont {Devoret}}]{campagne2020}%
  \BibitemOpen
  \bibfield  {author} {\bibinfo {author} {\bibfnamefont {P.}~\bibnamefont
  {Campagne-Ibarcq}}, \bibinfo {author} {\bibfnamefont {A.}~\bibnamefont
  {Eickbusch}}, \bibinfo {author} {\bibfnamefont {S.}~\bibnamefont {Touzard}},
  \bibinfo {author} {\bibfnamefont {E.}~\bibnamefont {Zalys-Geller}}, \bibinfo
  {author} {\bibfnamefont {N.~E.}\ \bibnamefont {Frattini}}, \bibinfo {author}
  {\bibfnamefont {V.~V.}\ \bibnamefont {Sivak}}, \bibinfo {author}
  {\bibfnamefont {P.}~\bibnamefont {Reinhold}}, \bibinfo {author}
  {\bibfnamefont {S.}~\bibnamefont {Puri}}, \bibinfo {author} {\bibfnamefont
  {S.}~\bibnamefont {Shankar}}, \bibinfo {author} {\bibfnamefont {R.~J.}\
  \bibnamefont {Schoelkopf}}, \bibinfo {author} {\bibfnamefont
  {L.}~\bibnamefont {Frunzio}}, \bibinfo {author} {\bibfnamefont
  {M.}~\bibnamefont {Mirrahimi}},\ and\ \bibinfo {author} {\bibfnamefont
  {M.~H.}\ \bibnamefont {Devoret}},\ }\bibfield  {title} {\bibinfo {title}
  {Quantum error correction of a qubit encoded in grid states of an
  oscillator},\ }\href {https://doi.org/10.1038/s41586-020-2603-3} {\bibfield
  {journal} {\bibinfo  {journal} {Nature}\ }\textbf {\bibinfo {volume} {584}},\
  \bibinfo {pages} {368} (\bibinfo {year} {2020})}\BibitemShut {NoStop}%
\bibitem [{\citenamefont {Shnirman}\ \emph {et~al.}(1997)\citenamefont
  {Shnirman}, \citenamefont {Sch\"on},\ and\ \citenamefont
  {Hermon}}]{shnirman1997}%
  \BibitemOpen
  \bibfield  {author} {\bibinfo {author} {\bibfnamefont {A.}~\bibnamefont
  {Shnirman}}, \bibinfo {author} {\bibfnamefont {G.}~\bibnamefont {Sch\"on}},\
  and\ \bibinfo {author} {\bibfnamefont {Z.}~\bibnamefont {Hermon}},\
  }\bibfield  {title} {\bibinfo {title} {Quantum manipulations of small
  {J}osephson junctions},\ }\href {https://doi.org/10.1103/PhysRevLett.79.2371}
  {\bibfield  {journal} {\bibinfo  {journal} {Phys. Rev. Lett.}\ }\textbf
  {\bibinfo {volume} {79}},\ \bibinfo {pages} {2371} (\bibinfo {year}
  {1997})}\BibitemShut {NoStop}%
\bibitem [{\citenamefont {Bouchiat}\ \emph {et~al.}(1998)\citenamefont
  {Bouchiat}, \citenamefont {Vion}, \citenamefont {Joyez}, \citenamefont
  {Esteve},\ and\ \citenamefont {Devoret}}]{bouchiat1998}%
  \BibitemOpen
  \bibfield  {author} {\bibinfo {author} {\bibfnamefont {V.}~\bibnamefont
  {Bouchiat}}, \bibinfo {author} {\bibfnamefont {D.}~\bibnamefont {Vion}},
  \bibinfo {author} {\bibfnamefont {P.}~\bibnamefont {Joyez}}, \bibinfo
  {author} {\bibfnamefont {D.}~\bibnamefont {Esteve}},\ and\ \bibinfo {author}
  {\bibfnamefont {M.~H.}\ \bibnamefont {Devoret}},\ }\bibfield  {title}
  {\bibinfo {title} {Quantum coherence with a single {C}ooper pair},\ }\href
  {https://doi.org/10.1238/physica.topical.076a00165} {\bibfield  {journal}
  {\bibinfo  {journal} {Physica Scripta}\ }\textbf {\bibinfo {volume} {T76}},\
  \bibinfo {pages} {165} (\bibinfo {year} {1998})}\BibitemShut {NoStop}%
\bibitem [{\citenamefont {Nakamura}\ \emph {et~al.}(1999)\citenamefont
  {Nakamura}, \citenamefont {Pashkin},\ and\ \citenamefont
  {Tsai}}]{nakamura1999}%
  \BibitemOpen
  \bibfield  {author} {\bibinfo {author} {\bibfnamefont {Y.}~\bibnamefont
  {Nakamura}}, \bibinfo {author} {\bibfnamefont {Y.~A.}\ \bibnamefont
  {Pashkin}},\ and\ \bibinfo {author} {\bibfnamefont {J.~S.}\ \bibnamefont
  {Tsai}},\ }\bibfield  {title} {\bibinfo {title} {Coherent control of
  macroscopic quantum states in a single-{C}ooper-pair box},\ }\href
  {https://doi.org/10.1038/19718} {\bibfield  {journal} {\bibinfo  {journal}
  {Nature}\ }\textbf {\bibinfo {volume} {398}},\ \bibinfo {pages} {786}
  (\bibinfo {year} {1999})}\BibitemShut {NoStop}%
\bibitem [{\citenamefont {Vion}\ \emph {et~al.}(2002)\citenamefont {Vion},
  \citenamefont {Aassime}, \citenamefont {Cottet}, \citenamefont {Joyez},
  \citenamefont {Pothier}, \citenamefont {Urbina}, \citenamefont {Esteve},\
  and\ \citenamefont {Devoret}}]{vion2002}%
  \BibitemOpen
  \bibfield  {author} {\bibinfo {author} {\bibfnamefont {D.}~\bibnamefont
  {Vion}}, \bibinfo {author} {\bibfnamefont {A.}~\bibnamefont {Aassime}},
  \bibinfo {author} {\bibfnamefont {A.}~\bibnamefont {Cottet}}, \bibinfo
  {author} {\bibfnamefont {P.}~\bibnamefont {Joyez}}, \bibinfo {author}
  {\bibfnamefont {H.}~\bibnamefont {Pothier}}, \bibinfo {author} {\bibfnamefont
  {C.}~\bibnamefont {Urbina}}, \bibinfo {author} {\bibfnamefont
  {D.}~\bibnamefont {Esteve}},\ and\ \bibinfo {author} {\bibfnamefont {M.~H.}\
  \bibnamefont {Devoret}},\ }\bibfield  {title} {\bibinfo {title} {Manipulating
  the quantum state of an electrical circuit},\ }\href
  {https://doi.org/10.1126/science.1069372} {\bibfield  {journal} {\bibinfo
  {journal} {Science}\ }\textbf {\bibinfo {volume} {296}},\ \bibinfo {pages}
  {886} (\bibinfo {year} {2002})}\BibitemShut {NoStop}%
\bibitem [{\citenamefont {Fedorov}\ \emph {et~al.}(2011)\citenamefont
  {Fedorov}, \citenamefont {Macha}, \citenamefont {Feofanov}, \citenamefont
  {Harmans},\ and\ \citenamefont {Mooij}}]{federov2011}%
  \BibitemOpen
  \bibfield  {author} {\bibinfo {author} {\bibfnamefont {A.}~\bibnamefont
  {Fedorov}}, \bibinfo {author} {\bibfnamefont {P.}~\bibnamefont {Macha}},
  \bibinfo {author} {\bibfnamefont {A.~K.}\ \bibnamefont {Feofanov}}, \bibinfo
  {author} {\bibfnamefont {C.~J. P.~M.}\ \bibnamefont {Harmans}},\ and\
  \bibinfo {author} {\bibfnamefont {J.~E.}\ \bibnamefont {Mooij}},\ }\bibfield
  {title} {\bibinfo {title} {Tuned transition from quantum to classical for
  macroscopic quantum states},\ }\href
  {https://doi.org/10.1103/PhysRevLett.106.170404} {\bibfield  {journal}
  {\bibinfo  {journal} {Phys. Rev. Lett.}\ }\textbf {\bibinfo {volume} {106}},\
  \bibinfo {pages} {170404} (\bibinfo {year} {2011})}\BibitemShut {NoStop}%
\bibitem [{\citenamefont {Earnest}\ \emph {et~al.}(2018)\citenamefont
  {Earnest}, \citenamefont {Chakram}, \citenamefont {Lu}, \citenamefont
  {Irons}, \citenamefont {Naik}, \citenamefont {Leung}, \citenamefont {Ocola},
  \citenamefont {Czaplewski}, \citenamefont {Baker}, \citenamefont {Lawrence},
  \citenamefont {Koch},\ and\ \citenamefont {Schuster}}]{earnest2018}%
  \BibitemOpen
  \bibfield  {author} {\bibinfo {author} {\bibfnamefont {N.}~\bibnamefont
  {Earnest}}, \bibinfo {author} {\bibfnamefont {S.}~\bibnamefont {Chakram}},
  \bibinfo {author} {\bibfnamefont {Y.}~\bibnamefont {Lu}}, \bibinfo {author}
  {\bibfnamefont {N.}~\bibnamefont {Irons}}, \bibinfo {author} {\bibfnamefont
  {R.~K.}\ \bibnamefont {Naik}}, \bibinfo {author} {\bibfnamefont
  {N.}~\bibnamefont {Leung}}, \bibinfo {author} {\bibfnamefont
  {L.}~\bibnamefont {Ocola}}, \bibinfo {author} {\bibfnamefont {D.~A.}\
  \bibnamefont {Czaplewski}}, \bibinfo {author} {\bibfnamefont
  {B.}~\bibnamefont {Baker}}, \bibinfo {author} {\bibfnamefont
  {J.}~\bibnamefont {Lawrence}}, \bibinfo {author} {\bibfnamefont
  {J.}~\bibnamefont {Koch}},\ and\ \bibinfo {author} {\bibfnamefont {D.~I.}\
  \bibnamefont {Schuster}},\ }\bibfield  {title} {\bibinfo {title} {Realization
  of a $\mathrm{\ensuremath{\Lambda}}$ system with metastable states of a
  capacitively shunted fluxonium},\ }\href
  {https://doi.org/10.1103/PhysRevLett.120.150504} {\bibfield  {journal}
  {\bibinfo  {journal} {Phys. Rev. Lett.}\ }\textbf {\bibinfo {volume} {120}},\
  \bibinfo {pages} {150504} (\bibinfo {year} {2018})}\BibitemShut {NoStop}%
\bibitem [{\citenamefont {Lin}\ \emph {et~al.}(2018)\citenamefont {Lin},
  \citenamefont {Nguyen}, \citenamefont {Grabon}, \citenamefont {San~Miguel},
  \citenamefont {Pankratova},\ and\ \citenamefont {Manucharyan}}]{lin2018}%
  \BibitemOpen
  \bibfield  {author} {\bibinfo {author} {\bibfnamefont {Y.-H.}\ \bibnamefont
  {Lin}}, \bibinfo {author} {\bibfnamefont {L.~B.}\ \bibnamefont {Nguyen}},
  \bibinfo {author} {\bibfnamefont {N.}~\bibnamefont {Grabon}}, \bibinfo
  {author} {\bibfnamefont {J.}~\bibnamefont {San~Miguel}}, \bibinfo {author}
  {\bibfnamefont {N.}~\bibnamefont {Pankratova}},\ and\ \bibinfo {author}
  {\bibfnamefont {V.~E.}\ \bibnamefont {Manucharyan}},\ }\bibfield  {title}
  {\bibinfo {title} {Demonstration of protection of a superconducting qubit
  from energy decay},\ }\href {https://doi.org/10.1103/PhysRevLett.120.150503}
  {\bibfield  {journal} {\bibinfo  {journal} {Phys. Rev. Lett.}\ }\textbf
  {\bibinfo {volume} {120}},\ \bibinfo {pages} {150503} (\bibinfo {year}
  {2018})}\BibitemShut {NoStop}%
\bibitem [{\citenamefont {Ithier}\ \emph {et~al.}(2005)\citenamefont {Ithier},
  \citenamefont {Collin}, \citenamefont {Joyez}, \citenamefont {Meeson},
  \citenamefont {Vion}, \citenamefont {Esteve}, \citenamefont {Chiarello},
  \citenamefont {Shnirman}, \citenamefont {Makhlin}, \citenamefont {Schriefl},\
  and\ \citenamefont {Sch\"on}}]{ithier2005}%
  \BibitemOpen
  \bibfield  {author} {\bibinfo {author} {\bibfnamefont {G.}~\bibnamefont
  {Ithier}}, \bibinfo {author} {\bibfnamefont {E.}~\bibnamefont {Collin}},
  \bibinfo {author} {\bibfnamefont {P.}~\bibnamefont {Joyez}}, \bibinfo
  {author} {\bibfnamefont {P.~J.}\ \bibnamefont {Meeson}}, \bibinfo {author}
  {\bibfnamefont {D.}~\bibnamefont {Vion}}, \bibinfo {author} {\bibfnamefont
  {D.}~\bibnamefont {Esteve}}, \bibinfo {author} {\bibfnamefont
  {F.}~\bibnamefont {Chiarello}}, \bibinfo {author} {\bibfnamefont
  {A.}~\bibnamefont {Shnirman}}, \bibinfo {author} {\bibfnamefont
  {Y.}~\bibnamefont {Makhlin}}, \bibinfo {author} {\bibfnamefont
  {J.}~\bibnamefont {Schriefl}},\ and\ \bibinfo {author} {\bibfnamefont
  {G.}~\bibnamefont {Sch\"on}},\ }\bibfield  {title} {\bibinfo {title}
  {Decoherence in a superconducting quantum bit circuit},\ }\href
  {https://doi.org/10.1103/PhysRevB.72.134519} {\bibfield  {journal} {\bibinfo
  {journal} {Phys. Rev. B}\ }\textbf {\bibinfo {volume} {72}},\ \bibinfo
  {pages} {134519} (\bibinfo {year} {2005})}\BibitemShut {NoStop}%
\bibitem [{\citenamefont {Dempster}\ \emph {et~al.}(2014)\citenamefont
  {Dempster}, \citenamefont {Fu}, \citenamefont {Ferguson}, \citenamefont
  {Schuster},\ and\ \citenamefont {Koch}}]{dempster2014}%
  \BibitemOpen
  \bibfield  {author} {\bibinfo {author} {\bibfnamefont {J.~M.}\ \bibnamefont
  {Dempster}}, \bibinfo {author} {\bibfnamefont {B.}~\bibnamefont {Fu}},
  \bibinfo {author} {\bibfnamefont {D.~G.}\ \bibnamefont {Ferguson}}, \bibinfo
  {author} {\bibfnamefont {D.~I.}\ \bibnamefont {Schuster}},\ and\ \bibinfo
  {author} {\bibfnamefont {J.}~\bibnamefont {Koch}},\ }\bibfield  {title}
  {\bibinfo {title} {{Understanding degenerate ground states of a protected
  quantum circuit in the presence of disorder}},\ }\href
  {https://doi.org/10.1103/PhysRevB.90.094518} {\bibfield  {journal} {\bibinfo
  {journal} {Phys. Rev. B}\ }\textbf {\bibinfo {volume} {90}},\ \bibinfo
  {pages} {094518} (\bibinfo {year} {2014})}\BibitemShut {NoStop}%
\bibitem [{\citenamefont {Brooks}\ \emph {et~al.}(2013)\citenamefont {Brooks},
  \citenamefont {Kitaev},\ and\ \citenamefont {Preskill}}]{brooks2013}%
  \BibitemOpen
  \bibfield  {author} {\bibinfo {author} {\bibfnamefont {P.}~\bibnamefont
  {Brooks}}, \bibinfo {author} {\bibfnamefont {A.}~\bibnamefont {Kitaev}},\
  and\ \bibinfo {author} {\bibfnamefont {J.}~\bibnamefont {Preskill}},\
  }\bibfield  {title} {\bibinfo {title} {Protected gates for superconducting
  qubits},\ }\href {https://doi.org/10.1103/PhysRevA.87.052306} {\bibfield
  {journal} {\bibinfo  {journal} {Phys. Rev. A}\ }\textbf {\bibinfo {volume}
  {87}},\ \bibinfo {pages} {052306} (\bibinfo {year} {2013})}\BibitemShut
  {NoStop}%
\bibitem [{\citenamefont {Smith}\ \emph {et~al.}(2020)\citenamefont {Smith},
  \citenamefont {Kou}, \citenamefont {Xiao}, \citenamefont {Vool},\ and\
  \citenamefont {Devoret}}]{smith2020}%
  \BibitemOpen
  \bibfield  {author} {\bibinfo {author} {\bibfnamefont {W.~C.}\ \bibnamefont
  {Smith}}, \bibinfo {author} {\bibfnamefont {A.}~\bibnamefont {Kou}}, \bibinfo
  {author} {\bibfnamefont {X.}~\bibnamefont {Xiao}}, \bibinfo {author}
  {\bibfnamefont {U.}~\bibnamefont {Vool}},\ and\ \bibinfo {author}
  {\bibfnamefont {M.~H.}\ \bibnamefont {Devoret}},\ }\bibfield  {title}
  {\bibinfo {title} {Superconducting circuit protected by two-{C}ooper-pair
  tunneling},\ }\href {https://doi.org/10.1038/s41534-019-0231-2} {\bibfield
  {journal} {\bibinfo  {journal} {npj Quantum Information}\ }\textbf {\bibinfo
  {volume} {6}},\ \bibinfo {pages} {8} (\bibinfo {year} {2020})}\BibitemShut
  {NoStop}%
\bibitem [{\citenamefont {Kalashnikov}\ \emph {et~al.}(2020)\citenamefont
  {Kalashnikov}, \citenamefont {Hsieh}, \citenamefont {Zhang}, \citenamefont
  {Lu}, \citenamefont {Kamenov}, \citenamefont {Di~Paolo}, \citenamefont
  {Blais}, \citenamefont {Gershenson},\ and\ \citenamefont
  {Bell}}]{kalashnikov2020}%
  \BibitemOpen
  \bibfield  {author} {\bibinfo {author} {\bibfnamefont {K.}~\bibnamefont
  {Kalashnikov}}, \bibinfo {author} {\bibfnamefont {W.~T.}\ \bibnamefont
  {Hsieh}}, \bibinfo {author} {\bibfnamefont {W.}~\bibnamefont {Zhang}},
  \bibinfo {author} {\bibfnamefont {W.-S.}\ \bibnamefont {Lu}}, \bibinfo
  {author} {\bibfnamefont {P.}~\bibnamefont {Kamenov}}, \bibinfo {author}
  {\bibfnamefont {A.}~\bibnamefont {Di~Paolo}}, \bibinfo {author}
  {\bibfnamefont {A.}~\bibnamefont {Blais}}, \bibinfo {author} {\bibfnamefont
  {M.~E.}\ \bibnamefont {Gershenson}},\ and\ \bibinfo {author} {\bibfnamefont
  {M.}~\bibnamefont {Bell}},\ }\bibfield  {title} {\bibinfo {title} {Bifluxon:
  Fluxon-parity-protected superconducting qubit},\ }\href
  {https://doi.org/10.1103/PRXQuantum.1.010307} {\bibfield  {journal} {\bibinfo
   {journal} {PRX Quantum}\ }\textbf {\bibinfo {volume} {1}},\ \bibinfo {pages}
  {010307} (\bibinfo {year} {2020})}\BibitemShut {NoStop}%
\bibitem [{\citenamefont {Mirrahimi}\ \emph {et~al.}(2014)\citenamefont
  {Mirrahimi}, \citenamefont {Leghtas}, \citenamefont {Albert}, \citenamefont
  {Touzard}, \citenamefont {Schoelkopf}, \citenamefont {Jiang},\ and\
  \citenamefont {Devoret}}]{mirrahimi2014}%
  \BibitemOpen
  \bibfield  {author} {\bibinfo {author} {\bibfnamefont {M.}~\bibnamefont
  {Mirrahimi}}, \bibinfo {author} {\bibfnamefont {Z.}~\bibnamefont {Leghtas}},
  \bibinfo {author} {\bibfnamefont {V.~V.}\ \bibnamefont {Albert}}, \bibinfo
  {author} {\bibfnamefont {S.}~\bibnamefont {Touzard}}, \bibinfo {author}
  {\bibfnamefont {R.~J.}\ \bibnamefont {Schoelkopf}}, \bibinfo {author}
  {\bibfnamefont {L.}~\bibnamefont {Jiang}},\ and\ \bibinfo {author}
  {\bibfnamefont {M.~H.}\ \bibnamefont {Devoret}},\ }\bibfield  {title}
  {\bibinfo {title} {Dynamically protected cat-qubits: a new paradigm for
  universal quantum computation},\ }\href
  {https://doi.org/10.1088/1367-2630/16/4/045014} {\bibfield  {journal}
  {\bibinfo  {journal} {New Journal of Physics}\ }\textbf {\bibinfo {volume}
  {16}},\ \bibinfo {pages} {045014} (\bibinfo {year} {2014})}\BibitemShut
  {NoStop}%
\bibitem [{\citenamefont {Puri}\ \emph {et~al.}(2017)\citenamefont {Puri},
  \citenamefont {Boutin},\ and\ \citenamefont {Blais}}]{puri2017}%
  \BibitemOpen
  \bibfield  {author} {\bibinfo {author} {\bibfnamefont {S.}~\bibnamefont
  {Puri}}, \bibinfo {author} {\bibfnamefont {S.}~\bibnamefont {Boutin}},\ and\
  \bibinfo {author} {\bibfnamefont {A.}~\bibnamefont {Blais}},\ }\bibfield
  {title} {\bibinfo {title} {Engineering the quantum states of light in a
  {Kerr}-nonlinear resonator by two-photon driving},\ }\href
  {https://doi.org/10.1038/s41534-017-0019-1} {\bibfield  {journal} {\bibinfo
  {journal} {npj Quantum Information}\ }\textbf {\bibinfo {volume} {3}},\
  \bibinfo {pages} {18} (\bibinfo {year} {2017})}\BibitemShut {NoStop}%
\bibitem [{\citenamefont {Grimm}\ \emph {et~al.}(2020)\citenamefont {Grimm},
  \citenamefont {Frattini}, \citenamefont {Puri}, \citenamefont {Mundhada},
  \citenamefont {Touzard}, \citenamefont {Mirrahimi}, \citenamefont {Girvin},
  \citenamefont {Shankar},\ and\ \citenamefont {Devoret}}]{grimm2020}%
  \BibitemOpen
  \bibfield  {author} {\bibinfo {author} {\bibfnamefont {A.}~\bibnamefont
  {Grimm}}, \bibinfo {author} {\bibfnamefont {N.~E.}\ \bibnamefont {Frattini}},
  \bibinfo {author} {\bibfnamefont {S.}~\bibnamefont {Puri}}, \bibinfo {author}
  {\bibfnamefont {S.~O.}\ \bibnamefont {Mundhada}}, \bibinfo {author}
  {\bibfnamefont {S.}~\bibnamefont {Touzard}}, \bibinfo {author} {\bibfnamefont
  {M.}~\bibnamefont {Mirrahimi}}, \bibinfo {author} {\bibfnamefont {S.~M.}\
  \bibnamefont {Girvin}}, \bibinfo {author} {\bibfnamefont {S.}~\bibnamefont
  {Shankar}},\ and\ \bibinfo {author} {\bibfnamefont {M.~H.}\ \bibnamefont
  {Devoret}},\ }\bibfield  {title} {\bibinfo {title} {Stabilization and
  operation of a {K}err-cat qubit},\ }\href
  {https://doi.org/10.1038/s41586-020-2587-z} {\bibfield  {journal} {\bibinfo
  {journal} {Nature}\ }\textbf {\bibinfo {volume} {584}},\ \bibinfo {pages}
  {205} (\bibinfo {year} {2020})}\BibitemShut {NoStop}%
\bibitem [{\citenamefont {Mundada}\ \emph {et~al.}(2020)\citenamefont
  {Mundada}, \citenamefont {Gyenis}, \citenamefont {Huang}, \citenamefont
  {Koch},\ and\ \citenamefont {Houck}}]{mundada2020}%
  \BibitemOpen
  \bibfield  {author} {\bibinfo {author} {\bibfnamefont {P.~S.}\ \bibnamefont
  {Mundada}}, \bibinfo {author} {\bibfnamefont {A.}~\bibnamefont {Gyenis}},
  \bibinfo {author} {\bibfnamefont {Z.}~\bibnamefont {Huang}}, \bibinfo
  {author} {\bibfnamefont {J.}~\bibnamefont {Koch}},\ and\ \bibinfo {author}
  {\bibfnamefont {A.~A.}\ \bibnamefont {Houck}},\ }\bibfield  {title} {\bibinfo
  {title} {Floquet-engineered enhancement of coherence times in a driven
  fluxonium qubit},\ }\href {https://doi.org/10.1103/PhysRevApplied.14.054033}
  {\bibfield  {journal} {\bibinfo  {journal} {Phys. Rev. Applied}\ }\textbf
  {\bibinfo {volume} {14}},\ \bibinfo {pages} {054033} (\bibinfo {year}
  {2020})}\BibitemShut {NoStop}%
\bibitem [{\citenamefont {Huang}\ \emph {et~al.}(2021)\citenamefont {Huang},
  \citenamefont {Mundada}, \citenamefont {Gyenis}, \citenamefont {Schuster},
  \citenamefont {Houck},\ and\ \citenamefont {Koch}}]{huang2020}%
  \BibitemOpen
  \bibfield  {author} {\bibinfo {author} {\bibfnamefont {Z.}~\bibnamefont
  {Huang}}, \bibinfo {author} {\bibfnamefont {P.~S.}\ \bibnamefont {Mundada}},
  \bibinfo {author} {\bibfnamefont {A.}~\bibnamefont {Gyenis}}, \bibinfo
  {author} {\bibfnamefont {D.~I.}\ \bibnamefont {Schuster}}, \bibinfo {author}
  {\bibfnamefont {A.~A.}\ \bibnamefont {Houck}},\ and\ \bibinfo {author}
  {\bibfnamefont {J.}~\bibnamefont {Koch}},\ }\bibfield  {title} {\bibinfo
  {title} {Engineering dynamical sweet spots to protect qubits from $1/f$
  noise},\ }\href {https://doi.org/10.1103/PhysRevApplied.15.034065} {\bibfield
   {journal} {\bibinfo  {journal} {Phys. Rev. Applied}\ }\textbf {\bibinfo
  {volume} {15}},\ \bibinfo {pages} {034065} (\bibinfo {year}
  {2021})}\BibitemShut {NoStop}%
\bibitem [{\citenamefont {Kitaev}(2006)}]{kitaev2006}%
  \BibitemOpen
  \bibfield  {author} {\bibinfo {author} {\bibfnamefont {A.}~\bibnamefont
  {Kitaev}},\ }\href@noop {} {\bibinfo {title} {Protected qubit based on a
  superconducting current mirror}} (\bibinfo {year} {2006}),\ \Eprint
  {https://arxiv.org/abs/cond-mat/0609441} {arXiv:cond-mat/0609441
  [cond-mat.mes-hall]} \BibitemShut {NoStop}%
\bibitem [{\citenamefont {Di~Paolo}(2020)}]{dipaolo2020}%
  \BibitemOpen
  \bibfield  {author} {\bibinfo {author} {\bibfnamefont {A.}~\bibnamefont
  {Di~Paolo}},\ }\emph {\bibinfo {title} {Qubits supraconducteurs protégés
  basés sur des modes à haute impédancet}},\ \href@noop {} {Ph.D. thesis}
  (\bibinfo {year} {2020})\BibitemShut {NoStop}%
\bibitem [{\citenamefont {Clerk}\ \emph {et~al.}(2010)\citenamefont {Clerk},
  \citenamefont {Devoret}, \citenamefont {Girvin}, \citenamefont {Marquardt},\
  and\ \citenamefont {Schoelkopf}}]{clerk2010}%
  \BibitemOpen
  \bibfield  {author} {\bibinfo {author} {\bibfnamefont {A.~A.}\ \bibnamefont
  {Clerk}}, \bibinfo {author} {\bibfnamefont {M.~H.}\ \bibnamefont {Devoret}},
  \bibinfo {author} {\bibfnamefont {S.~M.}\ \bibnamefont {Girvin}}, \bibinfo
  {author} {\bibfnamefont {F.}~\bibnamefont {Marquardt}},\ and\ \bibinfo
  {author} {\bibfnamefont {R.~J.}\ \bibnamefont {Schoelkopf}},\ }\bibfield
  {title} {\bibinfo {title} {Introduction to quantum noise, measurement, and
  amplification},\ }\href {https://doi.org/10.1103/RevModPhys.82.1155}
  {\bibfield  {journal} {\bibinfo  {journal} {Rev. Mod. Phys.}\ }\textbf
  {\bibinfo {volume} {82}},\ \bibinfo {pages} {1155} (\bibinfo {year}
  {2010})}\BibitemShut {NoStop}%
\bibitem [{\citenamefont {Manucharyan}\ \emph {et~al.}(2012)\citenamefont
  {Manucharyan}, \citenamefont {Masluk}, \citenamefont {Kamal}, \citenamefont
  {Koch}, \citenamefont {Glazman},\ and\ \citenamefont
  {Devoret}}]{manucharyan2012}%
  \BibitemOpen
  \bibfield  {author} {\bibinfo {author} {\bibfnamefont {V.~E.}\ \bibnamefont
  {Manucharyan}}, \bibinfo {author} {\bibfnamefont {N.~A.}\ \bibnamefont
  {Masluk}}, \bibinfo {author} {\bibfnamefont {A.}~\bibnamefont {Kamal}},
  \bibinfo {author} {\bibfnamefont {J.}~\bibnamefont {Koch}}, \bibinfo {author}
  {\bibfnamefont {L.~I.}\ \bibnamefont {Glazman}},\ and\ \bibinfo {author}
  {\bibfnamefont {M.~H.}\ \bibnamefont {Devoret}},\ }\bibfield  {title}
  {\bibinfo {title} {Evidence for coherent quantum phase slips across a
  {J}osephson junction array},\ }\href
  {https://doi.org/10.1103/PhysRevB.85.024521} {\bibfield  {journal} {\bibinfo
  {journal} {Phys. Rev. B}\ }\textbf {\bibinfo {volume} {85}},\ \bibinfo
  {pages} {024521} (\bibinfo {year} {2012})}\BibitemShut {NoStop}%
\bibitem [{\citenamefont {Kou}\ \emph {et~al.}(2017)\citenamefont {Kou},
  \citenamefont {Smith}, \citenamefont {Vool}, \citenamefont {Brierley},
  \citenamefont {Meier}, \citenamefont {Frunzio}, \citenamefont {Girvin},
  \citenamefont {Glazman},\ and\ \citenamefont {Devoret}}]{kou2017}%
  \BibitemOpen
  \bibfield  {author} {\bibinfo {author} {\bibfnamefont {A.}~\bibnamefont
  {Kou}}, \bibinfo {author} {\bibfnamefont {W.~C.}\ \bibnamefont {Smith}},
  \bibinfo {author} {\bibfnamefont {U.}~\bibnamefont {Vool}}, \bibinfo {author}
  {\bibfnamefont {R.~T.}\ \bibnamefont {Brierley}}, \bibinfo {author}
  {\bibfnamefont {H.}~\bibnamefont {Meier}}, \bibinfo {author} {\bibfnamefont
  {L.}~\bibnamefont {Frunzio}}, \bibinfo {author} {\bibfnamefont {S.~M.}\
  \bibnamefont {Girvin}}, \bibinfo {author} {\bibfnamefont {L.~I.}\
  \bibnamefont {Glazman}},\ and\ \bibinfo {author} {\bibfnamefont {M.~H.}\
  \bibnamefont {Devoret}},\ }\bibfield  {title} {\bibinfo {title}
  {Fluxonium-based artificial molecule with a tunable magnetic moment},\ }\href
  {https://doi.org/10.1103/PhysRevX.7.031037} {\bibfield  {journal} {\bibinfo
  {journal} {Phys. Rev. X}\ }\textbf {\bibinfo {volume} {7}},\ \bibinfo {pages}
  {031037} (\bibinfo {year} {2017})}\BibitemShut {NoStop}%
\bibitem [{\citenamefont {Slichter}\ \emph {et~al.}(2012)\citenamefont
  {Slichter}, \citenamefont {Vijay}, \citenamefont {Weber}, \citenamefont
  {Boutin}, \citenamefont {Boissonneault}, \citenamefont {Gambetta},
  \citenamefont {Blais},\ and\ \citenamefont {Siddiqi}}]{slichter2012}%
  \BibitemOpen
  \bibfield  {author} {\bibinfo {author} {\bibfnamefont {D.~H.}\ \bibnamefont
  {Slichter}}, \bibinfo {author} {\bibfnamefont {R.}~\bibnamefont {Vijay}},
  \bibinfo {author} {\bibfnamefont {S.~J.}\ \bibnamefont {Weber}}, \bibinfo
  {author} {\bibfnamefont {S.}~\bibnamefont {Boutin}}, \bibinfo {author}
  {\bibfnamefont {M.}~\bibnamefont {Boissonneault}}, \bibinfo {author}
  {\bibfnamefont {J.~M.}\ \bibnamefont {Gambetta}}, \bibinfo {author}
  {\bibfnamefont {A.}~\bibnamefont {Blais}},\ and\ \bibinfo {author}
  {\bibfnamefont {I.}~\bibnamefont {Siddiqi}},\ }\bibfield  {title} {\bibinfo
  {title} {Measurement-induced qubit state mixing in circuit {QED} from
  up-converted dephasing noise},\ }\href
  {https://doi.org/10.1103/PhysRevLett.109.153601} {\bibfield  {journal}
  {\bibinfo  {journal} {Phys. Rev. Lett.}\ }\textbf {\bibinfo {volume} {109}},\
  \bibinfo {pages} {153601} (\bibinfo {year} {2012})}\BibitemShut {NoStop}%
\bibitem [{\citenamefont {Yan}\ \emph {et~al.}(2016)\citenamefont {Yan},
  \citenamefont {Gustavsson}, \citenamefont {Kamal}, \citenamefont {Birenbaum},
  \citenamefont {Sears}, \citenamefont {Hover}, \citenamefont {Gudmundsen},
  \citenamefont {Rosenberg}, \citenamefont {Samach}, \citenamefont {Weber},
  \citenamefont {Yoder}, \citenamefont {Orlando}, \citenamefont {Clarke},
  \citenamefont {Kerman},\ and\ \citenamefont {Oliver}}]{yan2016}%
  \BibitemOpen
  \bibfield  {author} {\bibinfo {author} {\bibfnamefont {F.}~\bibnamefont
  {Yan}}, \bibinfo {author} {\bibfnamefont {S.}~\bibnamefont {Gustavsson}},
  \bibinfo {author} {\bibfnamefont {A.}~\bibnamefont {Kamal}}, \bibinfo
  {author} {\bibfnamefont {J.}~\bibnamefont {Birenbaum}}, \bibinfo {author}
  {\bibfnamefont {A.~P.}\ \bibnamefont {Sears}}, \bibinfo {author}
  {\bibfnamefont {D.}~\bibnamefont {Hover}}, \bibinfo {author} {\bibfnamefont
  {T.~J.}\ \bibnamefont {Gudmundsen}}, \bibinfo {author} {\bibfnamefont
  {D.}~\bibnamefont {Rosenberg}}, \bibinfo {author} {\bibfnamefont
  {G.}~\bibnamefont {Samach}}, \bibinfo {author} {\bibfnamefont
  {S.}~\bibnamefont {Weber}}, \bibinfo {author} {\bibfnamefont {J.~L.}\
  \bibnamefont {Yoder}}, \bibinfo {author} {\bibfnamefont {T.~P.}\ \bibnamefont
  {Orlando}}, \bibinfo {author} {\bibfnamefont {J.}~\bibnamefont {Clarke}},
  \bibinfo {author} {\bibfnamefont {A.~J.}\ \bibnamefont {Kerman}},\ and\
  \bibinfo {author} {\bibfnamefont {W.~D.}\ \bibnamefont {Oliver}},\ }\bibfield
   {title} {\bibinfo {title} {The flux qubit revisited to enhance coherence and
  reproducibility},\ }\href {https://doi.org/10.1038/ncomms12964} {\bibfield
  {journal} {\bibinfo  {journal} {Nature Communications}\ }\textbf {\bibinfo
  {volume} {7}},\ \bibinfo {pages} {12964} (\bibinfo {year}
  {2016})}\BibitemShut {NoStop}%
\bibitem [{\citenamefont {Quintana}\ \emph {et~al.}(2017)\citenamefont
  {Quintana}, \citenamefont {Chen}, \citenamefont {Sank}, \citenamefont
  {Petukhov}, \citenamefont {White}, \citenamefont {Kafri}, \citenamefont
  {Chiaro}, \citenamefont {Megrant}, \citenamefont {Barends}, \citenamefont
  {Campbell}, \citenamefont {Chen}, \citenamefont {Dunsworth}, \citenamefont
  {Fowler}, \citenamefont {Graff}, \citenamefont {Jeffrey}, \citenamefont
  {Kelly}, \citenamefont {Lucero}, \citenamefont {Mutus}, \citenamefont
  {Neeley}, \citenamefont {Neill}, \citenamefont {O'Malley}, \citenamefont
  {Roushan}, \citenamefont {Shabani}, \citenamefont {Smelyanskiy},
  \citenamefont {Vainsencher}, \citenamefont {Wenner}, \citenamefont {Neven},\
  and\ \citenamefont {Martinis}}]{quintana2017}%
  \BibitemOpen
  \bibfield  {author} {\bibinfo {author} {\bibfnamefont {C.~M.}\ \bibnamefont
  {Quintana}}, \bibinfo {author} {\bibfnamefont {Y.}~\bibnamefont {Chen}},
  \bibinfo {author} {\bibfnamefont {D.}~\bibnamefont {Sank}}, \bibinfo {author}
  {\bibfnamefont {A.~G.}\ \bibnamefont {Petukhov}}, \bibinfo {author}
  {\bibfnamefont {T.~C.}\ \bibnamefont {White}}, \bibinfo {author}
  {\bibfnamefont {D.}~\bibnamefont {Kafri}}, \bibinfo {author} {\bibfnamefont
  {B.}~\bibnamefont {Chiaro}}, \bibinfo {author} {\bibfnamefont
  {A.}~\bibnamefont {Megrant}}, \bibinfo {author} {\bibfnamefont
  {R.}~\bibnamefont {Barends}}, \bibinfo {author} {\bibfnamefont
  {B.}~\bibnamefont {Campbell}}, \bibinfo {author} {\bibfnamefont
  {Z.}~\bibnamefont {Chen}}, \bibinfo {author} {\bibfnamefont {A.}~\bibnamefont
  {Dunsworth}}, \bibinfo {author} {\bibfnamefont {A.~G.}\ \bibnamefont
  {Fowler}}, \bibinfo {author} {\bibfnamefont {R.}~\bibnamefont {Graff}},
  \bibinfo {author} {\bibfnamefont {E.}~\bibnamefont {Jeffrey}}, \bibinfo
  {author} {\bibfnamefont {J.}~\bibnamefont {Kelly}}, \bibinfo {author}
  {\bibfnamefont {E.}~\bibnamefont {Lucero}}, \bibinfo {author} {\bibfnamefont
  {J.~Y.}\ \bibnamefont {Mutus}}, \bibinfo {author} {\bibfnamefont
  {M.}~\bibnamefont {Neeley}}, \bibinfo {author} {\bibfnamefont
  {C.}~\bibnamefont {Neill}}, \bibinfo {author} {\bibfnamefont {P.~J.~J.}\
  \bibnamefont {O'Malley}}, \bibinfo {author} {\bibfnamefont {P.}~\bibnamefont
  {Roushan}}, \bibinfo {author} {\bibfnamefont {A.}~\bibnamefont {Shabani}},
  \bibinfo {author} {\bibfnamefont {V.~N.}\ \bibnamefont {Smelyanskiy}},
  \bibinfo {author} {\bibfnamefont {A.}~\bibnamefont {Vainsencher}}, \bibinfo
  {author} {\bibfnamefont {J.}~\bibnamefont {Wenner}}, \bibinfo {author}
  {\bibfnamefont {H.}~\bibnamefont {Neven}},\ and\ \bibinfo {author}
  {\bibfnamefont {J.~M.}\ \bibnamefont {Martinis}},\ }\bibfield  {title}
  {\bibinfo {title} {Observation of classical-quantum crossover of $1/f$ flux
  noise and its paramagnetic temperature dependence},\ }\href
  {https://doi.org/10.1103/PhysRevLett.118.057702} {\bibfield  {journal}
  {\bibinfo  {journal} {Phys. Rev. Lett.}\ }\textbf {\bibinfo {volume} {118}},\
  \bibinfo {pages} {057702} (\bibinfo {year} {2017})}\BibitemShut {NoStop}%
\bibitem [{\citenamefont {Astafiev}\ \emph {et~al.}(2004)\citenamefont
  {Astafiev}, \citenamefont {Pashkin}, \citenamefont {Nakamura}, \citenamefont
  {Yamamoto},\ and\ \citenamefont {Tsai}}]{astafiev2004}%
  \BibitemOpen
  \bibfield  {author} {\bibinfo {author} {\bibfnamefont {O.}~\bibnamefont
  {Astafiev}}, \bibinfo {author} {\bibfnamefont {Y.~A.}\ \bibnamefont
  {Pashkin}}, \bibinfo {author} {\bibfnamefont {Y.}~\bibnamefont {Nakamura}},
  \bibinfo {author} {\bibfnamefont {T.}~\bibnamefont {Yamamoto}},\ and\
  \bibinfo {author} {\bibfnamefont {J.~S.}\ \bibnamefont {Tsai}},\ }\bibfield
  {title} {\bibinfo {title} {Quantum noise in the {Josephson} charge qubit},\
  }\href {https://doi.org/10.1103/PhysRevLett.93.267007} {\bibfield  {journal}
  {\bibinfo  {journal} {Phys. Rev. Lett.}\ }\textbf {\bibinfo {volume} {93}},\
  \bibinfo {pages} {267007} (\bibinfo {year} {2004})}\BibitemShut {NoStop}%
\bibitem [{\citenamefont {Shnirman}\ \emph {et~al.}(2005)\citenamefont
  {Shnirman}, \citenamefont {Sch\"on}, \citenamefont {Martin},\ and\
  \citenamefont {Makhlin}}]{shnirman2005}%
  \BibitemOpen
  \bibfield  {author} {\bibinfo {author} {\bibfnamefont {A.}~\bibnamefont
  {Shnirman}}, \bibinfo {author} {\bibfnamefont {G.}~\bibnamefont {Sch\"on}},
  \bibinfo {author} {\bibfnamefont {I.}~\bibnamefont {Martin}},\ and\ \bibinfo
  {author} {\bibfnamefont {Y.}~\bibnamefont {Makhlin}},\ }\bibfield  {title}
  {\bibinfo {title} {Low- and high-frequency noise from coherent two-level
  systems},\ }\href {https://doi.org/10.1103/PhysRevLett.94.127002} {\bibfield
  {journal} {\bibinfo  {journal} {Phys. Rev. Lett.}\ }\textbf {\bibinfo
  {volume} {94}},\ \bibinfo {pages} {127002} (\bibinfo {year}
  {2005})}\BibitemShut {NoStop}%
\bibitem [{\citenamefont {Christensen}\ \emph {et~al.}(2019)\citenamefont
  {Christensen}, \citenamefont {Wilen}, \citenamefont {Opremcak}, \citenamefont
  {Nelson}, \citenamefont {Schlenker}, \citenamefont {Zimonick}, \citenamefont
  {Faoro}, \citenamefont {Ioffe}, \citenamefont {Rosen}, \citenamefont
  {DuBois}, \citenamefont {Plourde},\ and\ \citenamefont
  {McDermott}}]{christensen2019}%
  \BibitemOpen
  \bibfield  {author} {\bibinfo {author} {\bibfnamefont {B.~G.}\ \bibnamefont
  {Christensen}}, \bibinfo {author} {\bibfnamefont {C.~D.}\ \bibnamefont
  {Wilen}}, \bibinfo {author} {\bibfnamefont {A.}~\bibnamefont {Opremcak}},
  \bibinfo {author} {\bibfnamefont {J.}~\bibnamefont {Nelson}}, \bibinfo
  {author} {\bibfnamefont {F.}~\bibnamefont {Schlenker}}, \bibinfo {author}
  {\bibfnamefont {C.~H.}\ \bibnamefont {Zimonick}}, \bibinfo {author}
  {\bibfnamefont {L.}~\bibnamefont {Faoro}}, \bibinfo {author} {\bibfnamefont
  {L.~B.}\ \bibnamefont {Ioffe}}, \bibinfo {author} {\bibfnamefont {Y.~J.}\
  \bibnamefont {Rosen}}, \bibinfo {author} {\bibfnamefont {J.~L.}\ \bibnamefont
  {DuBois}}, \bibinfo {author} {\bibfnamefont {B.~L.~T.}\ \bibnamefont
  {Plourde}},\ and\ \bibinfo {author} {\bibfnamefont {R.}~\bibnamefont
  {McDermott}},\ }\bibfield  {title} {\bibinfo {title} {Anomalous charge noise
  in superconducting qubits},\ }\href
  {https://doi.org/10.1103/PhysRevB.100.140503} {\bibfield  {journal} {\bibinfo
   {journal} {Phys. Rev. B}\ }\textbf {\bibinfo {volume} {100}},\ \bibinfo
  {pages} {140503} (\bibinfo {year} {2019})}\BibitemShut {NoStop}%
\bibitem [{\citenamefont {You}\ \emph {et~al.}(2021)\citenamefont {You},
  \citenamefont {Clerk},\ and\ \citenamefont {Koch}}]{you2021}%
  \BibitemOpen
  \bibfield  {author} {\bibinfo {author} {\bibfnamefont {X.}~\bibnamefont
  {You}}, \bibinfo {author} {\bibfnamefont {A.~A.}\ \bibnamefont {Clerk}},\
  and\ \bibinfo {author} {\bibfnamefont {J.}~\bibnamefont {Koch}},\ }\bibfield
  {title} {\bibinfo {title} {Positive- and negative-frequency noise from an
  ensemble of two-level fluctuators},\ }\href
  {https://doi.org/10.1103/PhysRevResearch.3.013045} {\bibfield  {journal}
  {\bibinfo  {journal} {Phys. Rev. Research}\ }\textbf {\bibinfo {volume}
  {3}},\ \bibinfo {pages} {013045} (\bibinfo {year} {2021})}\BibitemShut
  {NoStop}%
\bibitem [{\citenamefont {Pechenezhskiy}\ \emph {et~al.}(2020)\citenamefont
  {Pechenezhskiy}, \citenamefont {Mencia}, \citenamefont {Nguyen},
  \citenamefont {Lin},\ and\ \citenamefont {Manucharyan}}]{pechenezhskiy2020}%
  \BibitemOpen
  \bibfield  {author} {\bibinfo {author} {\bibfnamefont {I.~V.}\ \bibnamefont
  {Pechenezhskiy}}, \bibinfo {author} {\bibfnamefont {R.~A.}\ \bibnamefont
  {Mencia}}, \bibinfo {author} {\bibfnamefont {L.~B.}\ \bibnamefont {Nguyen}},
  \bibinfo {author} {\bibfnamefont {Y.-H.}\ \bibnamefont {Lin}},\ and\ \bibinfo
  {author} {\bibfnamefont {V.~E.}\ \bibnamefont {Manucharyan}},\ }\bibfield
  {title} {\bibinfo {title} {The superconducting quasicharge qubit},\ }\href
  {https://doi.org/10.1038/s41586-020-2687-9} {\bibfield  {journal} {\bibinfo
  {journal} {Nature}\ }\textbf {\bibinfo {volume} {585}},\ \bibinfo {pages}
  {368} (\bibinfo {year} {2020})}\BibitemShut {NoStop}%
\bibitem [{\citenamefont {Schreier}\ \emph {et~al.}(2008)\citenamefont
  {Schreier}, \citenamefont {Houck}, \citenamefont {Koch}, \citenamefont
  {Schuster}, \citenamefont {Johnson}, \citenamefont {Chow}, \citenamefont
  {Gambetta}, \citenamefont {Majer}, \citenamefont {Frunzio}, \citenamefont
  {Devoret}, \citenamefont {Girvin},\ and\ \citenamefont
  {Schoelkopf}}]{schreier2008}%
  \BibitemOpen
  \bibfield  {author} {\bibinfo {author} {\bibfnamefont {J.~A.}\ \bibnamefont
  {Schreier}}, \bibinfo {author} {\bibfnamefont {A.~A.}\ \bibnamefont {Houck}},
  \bibinfo {author} {\bibfnamefont {J.}~\bibnamefont {Koch}}, \bibinfo {author}
  {\bibfnamefont {D.~I.}\ \bibnamefont {Schuster}}, \bibinfo {author}
  {\bibfnamefont {B.~R.}\ \bibnamefont {Johnson}}, \bibinfo {author}
  {\bibfnamefont {J.~M.}\ \bibnamefont {Chow}}, \bibinfo {author}
  {\bibfnamefont {J.~M.}\ \bibnamefont {Gambetta}}, \bibinfo {author}
  {\bibfnamefont {J.}~\bibnamefont {Majer}}, \bibinfo {author} {\bibfnamefont
  {L.}~\bibnamefont {Frunzio}}, \bibinfo {author} {\bibfnamefont {M.~H.}\
  \bibnamefont {Devoret}}, \bibinfo {author} {\bibfnamefont {S.~M.}\
  \bibnamefont {Girvin}},\ and\ \bibinfo {author} {\bibfnamefont {R.~J.}\
  \bibnamefont {Schoelkopf}},\ }\bibfield  {title} {\bibinfo {title}
  {Suppressing charge noise decoherence in superconducting charge qubits},\
  }\href {https://doi.org/10.1103/PhysRevB.77.180502} {\bibfield  {journal}
  {\bibinfo  {journal} {Phys. Rev. B}\ }\textbf {\bibinfo {volume} {77}},\
  \bibinfo {pages} {180502} (\bibinfo {year} {2008})}\BibitemShut {NoStop}%
\bibitem [{\citenamefont {Wang}\ \emph {et~al.}(2021)\citenamefont {Wang},
  \citenamefont {Li}, \citenamefont {Xu}, \citenamefont {Li}, \citenamefont
  {Wang}, \citenamefont {Yang}, \citenamefont {Mi}, \citenamefont {Liang},
  \citenamefont {Su}, \citenamefont {Yang}, \citenamefont {Wang}, \citenamefont
  {Wang}, \citenamefont {Li}, \citenamefont {Chen}, \citenamefont {Li},
  \citenamefont {Linghu}, \citenamefont {Han}, \citenamefont {Zhang},
  \citenamefont {Feng}, \citenamefont {Song}, \citenamefont {Ma}, \citenamefont
  {Zhang}, \citenamefont {Wang}, \citenamefont {Zhao}, \citenamefont {Liu},
  \citenamefont {Xue}, \citenamefont {Jin},\ and\ \citenamefont
  {Yu}}]{wang2021}%
  \BibitemOpen
  \bibfield  {author} {\bibinfo {author} {\bibfnamefont {C.}~\bibnamefont
  {Wang}}, \bibinfo {author} {\bibfnamefont {X.}~\bibnamefont {Li}}, \bibinfo
  {author} {\bibfnamefont {H.}~\bibnamefont {Xu}}, \bibinfo {author}
  {\bibfnamefont {Z.}~\bibnamefont {Li}}, \bibinfo {author} {\bibfnamefont
  {J.}~\bibnamefont {Wang}}, \bibinfo {author} {\bibfnamefont {Z.}~\bibnamefont
  {Yang}}, \bibinfo {author} {\bibfnamefont {Z.}~\bibnamefont {Mi}}, \bibinfo
  {author} {\bibfnamefont {X.}~\bibnamefont {Liang}}, \bibinfo {author}
  {\bibfnamefont {T.}~\bibnamefont {Su}}, \bibinfo {author} {\bibfnamefont
  {C.}~\bibnamefont {Yang}}, \bibinfo {author} {\bibfnamefont {G.}~\bibnamefont
  {Wang}}, \bibinfo {author} {\bibfnamefont {W.}~\bibnamefont {Wang}}, \bibinfo
  {author} {\bibfnamefont {Y.}~\bibnamefont {Li}}, \bibinfo {author}
  {\bibfnamefont {M.}~\bibnamefont {Chen}}, \bibinfo {author} {\bibfnamefont
  {C.}~\bibnamefont {Li}}, \bibinfo {author} {\bibfnamefont {K.}~\bibnamefont
  {Linghu}}, \bibinfo {author} {\bibfnamefont {J.}~\bibnamefont {Han}},
  \bibinfo {author} {\bibfnamefont {Y.}~\bibnamefont {Zhang}}, \bibinfo
  {author} {\bibfnamefont {Y.}~\bibnamefont {Feng}}, \bibinfo {author}
  {\bibfnamefont {Y.}~\bibnamefont {Song}}, \bibinfo {author} {\bibfnamefont
  {T.}~\bibnamefont {Ma}}, \bibinfo {author} {\bibfnamefont {J.}~\bibnamefont
  {Zhang}}, \bibinfo {author} {\bibfnamefont {R.}~\bibnamefont {Wang}},
  \bibinfo {author} {\bibfnamefont {P.}~\bibnamefont {Zhao}}, \bibinfo {author}
  {\bibfnamefont {W.}~\bibnamefont {Liu}}, \bibinfo {author} {\bibfnamefont
  {G.}~\bibnamefont {Xue}}, \bibinfo {author} {\bibfnamefont {Y.}~\bibnamefont
  {Jin}},\ and\ \bibinfo {author} {\bibfnamefont {H.}~\bibnamefont {Yu}},\
  }\href@noop {} {\bibinfo {title} {Transmon qubit with relaxation time
  exceeding 0.5 milliseconds}} (\bibinfo {year} {2021}),\ \Eprint
  {https://arxiv.org/abs/2105.09890} {arXiv:2105.09890 [quant-ph]} \BibitemShut
  {NoStop}%
\bibitem [{\citenamefont {Koch}\ \emph {et~al.}(2009)\citenamefont {Koch},
  \citenamefont {Manucharyan}, \citenamefont {Devoret},\ and\ \citenamefont
  {Glazman}}]{koch2009}%
  \BibitemOpen
  \bibfield  {author} {\bibinfo {author} {\bibfnamefont {J.}~\bibnamefont
  {Koch}}, \bibinfo {author} {\bibfnamefont {V.}~\bibnamefont {Manucharyan}},
  \bibinfo {author} {\bibfnamefont {M.~H.}\ \bibnamefont {Devoret}},\ and\
  \bibinfo {author} {\bibfnamefont {L.~I.}\ \bibnamefont {Glazman}},\
  }\bibfield  {title} {\bibinfo {title} {Charging effects in the inductively
  shunted {J}osephson junction},\ }\href
  {https://doi.org/10.1103/PhysRevLett.103.217004} {\bibfield  {journal}
  {\bibinfo  {journal} {Phys. Rev. Lett.}\ }\textbf {\bibinfo {volume} {103}},\
  \bibinfo {pages} {217004} (\bibinfo {year} {2009})}\BibitemShut {NoStop}%
\bibitem [{\citenamefont {Manucharyan}\ \emph {et~al.}(2009)\citenamefont
  {Manucharyan}, \citenamefont {Koch}, \citenamefont {Glazman},\ and\
  \citenamefont {Devoret}}]{manucharyan2009}%
  \BibitemOpen
  \bibfield  {author} {\bibinfo {author} {\bibfnamefont {V.~E.}\ \bibnamefont
  {Manucharyan}}, \bibinfo {author} {\bibfnamefont {J.}~\bibnamefont {Koch}},
  \bibinfo {author} {\bibfnamefont {L.~I.}\ \bibnamefont {Glazman}},\ and\
  \bibinfo {author} {\bibfnamefont {M.~H.}\ \bibnamefont {Devoret}},\
  }\bibfield  {title} {\bibinfo {title} {Fluxonium: Single {C}ooper-pair
  circuit free of charge offsets},\ }\href
  {https://doi.org/10.1126/science.1175552} {\bibfield  {journal} {\bibinfo
  {journal} {Science}\ }\textbf {\bibinfo {volume} {326}},\ \bibinfo {pages}
  {113} (\bibinfo {year} {2009})}\BibitemShut {NoStop}%
\bibitem [{\citenamefont {Peruzzo}\ \emph {et~al.}(2020)\citenamefont
  {Peruzzo}, \citenamefont {Trioni}, \citenamefont {Hassani}, \citenamefont
  {Zemlicka},\ and\ \citenamefont {Fink}}]{peruzzo2020}%
  \BibitemOpen
  \bibfield  {author} {\bibinfo {author} {\bibfnamefont {M.}~\bibnamefont
  {Peruzzo}}, \bibinfo {author} {\bibfnamefont {A.}~\bibnamefont {Trioni}},
  \bibinfo {author} {\bibfnamefont {F.}~\bibnamefont {Hassani}}, \bibinfo
  {author} {\bibfnamefont {M.}~\bibnamefont {Zemlicka}},\ and\ \bibinfo
  {author} {\bibfnamefont {J.~M.}\ \bibnamefont {Fink}},\ }\bibfield  {title}
  {\bibinfo {title} {Surpassing the resistance quantum with a geometric
  superinductor},\ }\href {https://doi.org/10.1103/PhysRevApplied.14.044055}
  {\bibfield  {journal} {\bibinfo  {journal} {Phys. Rev. Applied}\ }\textbf
  {\bibinfo {volume} {14}},\ \bibinfo {pages} {044055} (\bibinfo {year}
  {2020})}\BibitemShut {NoStop}%
\bibitem [{\citenamefont {Masluk}\ \emph {et~al.}(2012)\citenamefont {Masluk},
  \citenamefont {Pop}, \citenamefont {Kamal}, \citenamefont {Minev},\ and\
  \citenamefont {Devoret}}]{masluk2012}%
  \BibitemOpen
  \bibfield  {author} {\bibinfo {author} {\bibfnamefont {N.~A.}\ \bibnamefont
  {Masluk}}, \bibinfo {author} {\bibfnamefont {I.~M.}\ \bibnamefont {Pop}},
  \bibinfo {author} {\bibfnamefont {A.}~\bibnamefont {Kamal}}, \bibinfo
  {author} {\bibfnamefont {Z.~K.}\ \bibnamefont {Minev}},\ and\ \bibinfo
  {author} {\bibfnamefont {M.~H.}\ \bibnamefont {Devoret}},\ }\bibfield
  {title} {\bibinfo {title} {Microwave characterization of josephson junction
  arrays: Implementing a low loss superinductance},\ }\href@noop {} {\bibfield
  {journal} {\bibinfo  {journal} {Physical review letters}\ }\textbf {\bibinfo
  {volume} {109}},\ \bibinfo {pages} {137002} (\bibinfo {year}
  {2012})}\BibitemShut {NoStop}%
\bibitem [{\citenamefont {Hazard}\ \emph {et~al.}(2019)\citenamefont {Hazard},
  \citenamefont {Gyenis}, \citenamefont {Di~Paolo}, \citenamefont {Asfaw},
  \citenamefont {Lyon}, \citenamefont {Blais},\ and\ \citenamefont
  {Houck}}]{hazard2019}%
  \BibitemOpen
  \bibfield  {author} {\bibinfo {author} {\bibfnamefont {T.~M.}\ \bibnamefont
  {Hazard}}, \bibinfo {author} {\bibfnamefont {A.}~\bibnamefont {Gyenis}},
  \bibinfo {author} {\bibfnamefont {A.}~\bibnamefont {Di~Paolo}}, \bibinfo
  {author} {\bibfnamefont {A.~T.}\ \bibnamefont {Asfaw}}, \bibinfo {author}
  {\bibfnamefont {S.~A.}\ \bibnamefont {Lyon}}, \bibinfo {author}
  {\bibfnamefont {A.}~\bibnamefont {Blais}},\ and\ \bibinfo {author}
  {\bibfnamefont {A.~A.}\ \bibnamefont {Houck}},\ }\bibfield  {title} {\bibinfo
  {title} {Nanowire superinductance fluxonium qubit},\ }\href
  {https://doi.org/10.1103/PhysRevLett.122.010504} {\bibfield  {journal}
  {\bibinfo  {journal} {Phys. Rev. Lett.}\ }\textbf {\bibinfo {volume} {122}},\
  \bibinfo {pages} {010504} (\bibinfo {year} {2019})}\BibitemShut {NoStop}%
\bibitem [{\citenamefont {Di~Paolo}\ \emph {et~al.}(2021)\citenamefont
  {Di~Paolo}, \citenamefont {Baker}, \citenamefont {Foley}, \citenamefont
  {S{\'e}n{\'e}chal},\ and\ \citenamefont {Blais}}]{di2021}%
  \BibitemOpen
  \bibfield  {author} {\bibinfo {author} {\bibfnamefont {A.}~\bibnamefont
  {Di~Paolo}}, \bibinfo {author} {\bibfnamefont {T.~E.}\ \bibnamefont {Baker}},
  \bibinfo {author} {\bibfnamefont {A.}~\bibnamefont {Foley}}, \bibinfo
  {author} {\bibfnamefont {D.}~\bibnamefont {S{\'e}n{\'e}chal}},\ and\ \bibinfo
  {author} {\bibfnamefont {A.}~\bibnamefont {Blais}},\ }\bibfield  {title}
  {\bibinfo {title} {Efficient modeling of superconducting quantum circuits
  with tensor networks},\ }\href@noop {} {\bibfield  {journal} {\bibinfo
  {journal} {npj Quantum Information}\ }\textbf {\bibinfo {volume} {7}},\
  \bibinfo {pages} {1} (\bibinfo {year} {2021})}\BibitemShut {NoStop}%
\bibitem [{\citenamefont {Manucharyan}(2012)}]{manucharyan2012superinductance}%
  \BibitemOpen
  \bibfield  {author} {\bibinfo {author} {\bibfnamefont {V.~E.}\ \bibnamefont
  {Manucharyan}},\ }\emph {\bibinfo {title} {Superinductance}},\ \href@noop {}
  {Ph.D. thesis},\ \bibinfo  {school} {Yale University} (\bibinfo {year}
  {2012})\BibitemShut {NoStop}%
\bibitem [{\citenamefont {Nguyen}\ \emph {et~al.}(2019)\citenamefont {Nguyen},
  \citenamefont {Lin}, \citenamefont {Somoroff}, \citenamefont {Mencia},
  \citenamefont {Grabon},\ and\ \citenamefont {Manucharyan}}]{nguyen2019}%
  \BibitemOpen
  \bibfield  {author} {\bibinfo {author} {\bibfnamefont {L.~B.}\ \bibnamefont
  {Nguyen}}, \bibinfo {author} {\bibfnamefont {Y.-H.}\ \bibnamefont {Lin}},
  \bibinfo {author} {\bibfnamefont {A.}~\bibnamefont {Somoroff}}, \bibinfo
  {author} {\bibfnamefont {R.}~\bibnamefont {Mencia}}, \bibinfo {author}
  {\bibfnamefont {N.}~\bibnamefont {Grabon}},\ and\ \bibinfo {author}
  {\bibfnamefont {V.~E.}\ \bibnamefont {Manucharyan}},\ }\bibfield  {title}
  {\bibinfo {title} {High-coherence fluxonium qubit},\ }\href
  {https://doi.org/10.1103/PhysRevX.9.041041} {\bibfield  {journal} {\bibinfo
  {journal} {Phys. Rev. X}\ }\textbf {\bibinfo {volume} {9}},\ \bibinfo {pages}
  {041041} (\bibinfo {year} {2019})}\BibitemShut {NoStop}%
\bibitem [{\citenamefont {Zhang}\ \emph {et~al.}(2021)\citenamefont {Zhang},
  \citenamefont {Chakram}, \citenamefont {Roy}, \citenamefont {Earnest},
  \citenamefont {Lu}, \citenamefont {Huang}, \citenamefont {Weiss},
  \citenamefont {Koch},\ and\ \citenamefont {Schuster}}]{zhang2020}%
  \BibitemOpen
  \bibfield  {author} {\bibinfo {author} {\bibfnamefont {H.}~\bibnamefont
  {Zhang}}, \bibinfo {author} {\bibfnamefont {S.}~\bibnamefont {Chakram}},
  \bibinfo {author} {\bibfnamefont {T.}~\bibnamefont {Roy}}, \bibinfo {author}
  {\bibfnamefont {N.}~\bibnamefont {Earnest}}, \bibinfo {author} {\bibfnamefont
  {Y.}~\bibnamefont {Lu}}, \bibinfo {author} {\bibfnamefont {Z.}~\bibnamefont
  {Huang}}, \bibinfo {author} {\bibfnamefont {D.~K.}\ \bibnamefont {Weiss}},
  \bibinfo {author} {\bibfnamefont {J.}~\bibnamefont {Koch}},\ and\ \bibinfo
  {author} {\bibfnamefont {D.~I.}\ \bibnamefont {Schuster}},\ }\bibfield
  {title} {\bibinfo {title} {Universal fast-flux control of a coherent,
  low-frequency qubit},\ }\href {https://doi.org/10.1103/PhysRevX.11.011010}
  {\bibfield  {journal} {\bibinfo  {journal} {Phys. Rev. X}\ }\textbf {\bibinfo
  {volume} {11}},\ \bibinfo {pages} {011010} (\bibinfo {year}
  {2021})}\BibitemShut {NoStop}%
\bibitem [{\citenamefont {{Somoroff}}\ \emph {et~al.}(2021)\citenamefont
  {{Somoroff}}, \citenamefont {{Ficheux}}, \citenamefont {{Mencia}},
  \citenamefont {{Xiong}}, \citenamefont {{Kuzmin}},\ and\ \citenamefont
  {{Manucharyan}}}]{Somoroff2021}%
  \BibitemOpen
  \bibfield  {author} {\bibinfo {author} {\bibfnamefont {A.}~\bibnamefont
  {{Somoroff}}}, \bibinfo {author} {\bibfnamefont {Q.}~\bibnamefont
  {{Ficheux}}}, \bibinfo {author} {\bibfnamefont {R.~A.}\ \bibnamefont
  {{Mencia}}}, \bibinfo {author} {\bibfnamefont {H.}~\bibnamefont {{Xiong}}},
  \bibinfo {author} {\bibfnamefont {R.~V.}\ \bibnamefont {{Kuzmin}}},\ and\
  \bibinfo {author} {\bibfnamefont {V.~E.}\ \bibnamefont {{Manucharyan}}},\
  }\bibfield  {title} {\bibinfo {title} {{Millisecond coherence in a
  superconducting qubit}},\ }\href@noop {} {\bibfield  {journal} {\bibinfo
  {journal} {arXiv e-prints}\ ,\ \bibinfo {eid} {arXiv:2103.08578}} (\bibinfo
  {year} {2021})},\ \Eprint {https://arxiv.org/abs/2103.08578}
  {arXiv:2103.08578 [quant-ph]} \BibitemShut {NoStop}%
\bibitem [{\citenamefont {Groszkowski}\ \emph {et~al.}(2018)\citenamefont
  {Groszkowski}, \citenamefont {Paolo}, \citenamefont {Grimsmo}, \citenamefont
  {Blais}, \citenamefont {Schuster}, \citenamefont {Houck},\ and\ \citenamefont
  {Koch}}]{groszkowski2018}%
  \BibitemOpen
  \bibfield  {author} {\bibinfo {author} {\bibfnamefont {P.}~\bibnamefont
  {Groszkowski}}, \bibinfo {author} {\bibfnamefont {A.~D.}\ \bibnamefont
  {Paolo}}, \bibinfo {author} {\bibfnamefont {A.~L.}\ \bibnamefont {Grimsmo}},
  \bibinfo {author} {\bibfnamefont {A.}~\bibnamefont {Blais}}, \bibinfo
  {author} {\bibfnamefont {D.~I.}\ \bibnamefont {Schuster}}, \bibinfo {author}
  {\bibfnamefont {A.~A.}\ \bibnamefont {Houck}},\ and\ \bibinfo {author}
  {\bibfnamefont {J.}~\bibnamefont {Koch}},\ }\bibfield  {title} {\bibinfo
  {title} {{Coherence properties of the 0-$\pi$ qubit}},\ }\href
  {https://doi.org/10.1088/1367-2630/aab7cd} {\bibfield  {journal} {\bibinfo
  {journal} {New J. Phys.}\ }\textbf {\bibinfo {volume} {20}},\ \bibinfo
  {pages} {043053} (\bibinfo {year} {2018})}\BibitemShut {NoStop}%
\bibitem [{\citenamefont {Paolo}\ \emph {et~al.}(2019)\citenamefont {Paolo},
  \citenamefont {Grimsmo}, \citenamefont {Groszkowski}, \citenamefont {Koch},\
  and\ \citenamefont {Blais}}]{dipaolo2019}%
  \BibitemOpen
  \bibfield  {author} {\bibinfo {author} {\bibfnamefont {A.~D.}\ \bibnamefont
  {Paolo}}, \bibinfo {author} {\bibfnamefont {A.~L.}\ \bibnamefont {Grimsmo}},
  \bibinfo {author} {\bibfnamefont {P.}~\bibnamefont {Groszkowski}}, \bibinfo
  {author} {\bibfnamefont {J.}~\bibnamefont {Koch}},\ and\ \bibinfo {author}
  {\bibfnamefont {A.}~\bibnamefont {Blais}},\ }\bibfield  {title} {\bibinfo
  {title} {Control and coherence time enhancement of the 0{\textendash}$\pi$
  qubit},\ }\href {https://doi.org/10.1088/1367-2630/ab09b0} {\bibfield
  {journal} {\bibinfo  {journal} {New J. Phys.}\ }\textbf {\bibinfo {volume}
  {21}},\ \bibinfo {pages} {043002} (\bibinfo {year} {2019})}\BibitemShut
  {NoStop}%
\bibitem [{\citenamefont {Gyenis}\ \emph {et~al.}(2021)\citenamefont {Gyenis},
  \citenamefont {Mundada}, \citenamefont {Di~Paolo}, \citenamefont {Hazard},
  \citenamefont {You}, \citenamefont {Schuster}, \citenamefont {Koch},
  \citenamefont {Blais},\ and\ \citenamefont {Houck}}]{gyenis2019}%
  \BibitemOpen
  \bibfield  {author} {\bibinfo {author} {\bibfnamefont {A.}~\bibnamefont
  {Gyenis}}, \bibinfo {author} {\bibfnamefont {P.~S.}\ \bibnamefont {Mundada}},
  \bibinfo {author} {\bibfnamefont {A.}~\bibnamefont {Di~Paolo}}, \bibinfo
  {author} {\bibfnamefont {T.~M.}\ \bibnamefont {Hazard}}, \bibinfo {author}
  {\bibfnamefont {X.}~\bibnamefont {You}}, \bibinfo {author} {\bibfnamefont
  {D.~I.}\ \bibnamefont {Schuster}}, \bibinfo {author} {\bibfnamefont
  {J.}~\bibnamefont {Koch}}, \bibinfo {author} {\bibfnamefont {A.}~\bibnamefont
  {Blais}},\ and\ \bibinfo {author} {\bibfnamefont {A.~A.}\ \bibnamefont
  {Houck}},\ }\bibfield  {title} {\bibinfo {title} {Experimental realization of
  a protected superconducting circuit derived from the $0$--$\ensuremath{\pi}$
  qubit},\ }\href {https://doi.org/10.1103/PRXQuantum.2.010339} {\bibfield
  {journal} {\bibinfo  {journal} {PRX Quantum}\ }\textbf {\bibinfo {volume}
  {2}},\ \bibinfo {pages} {010339} (\bibinfo {year} {2021})}\BibitemShut
  {NoStop}%
\bibitem [{\citenamefont {Bell}\ \emph {et~al.}(2016)\citenamefont {Bell},
  \citenamefont {Zhang}, \citenamefont {Ioffe},\ and\ \citenamefont
  {Gershenson}}]{bell2016}%
  \BibitemOpen
  \bibfield  {author} {\bibinfo {author} {\bibfnamefont {M.~T.}\ \bibnamefont
  {Bell}}, \bibinfo {author} {\bibfnamefont {W.}~\bibnamefont {Zhang}},
  \bibinfo {author} {\bibfnamefont {L.~B.}\ \bibnamefont {Ioffe}},\ and\
  \bibinfo {author} {\bibfnamefont {M.~E.}\ \bibnamefont {Gershenson}},\
  }\bibfield  {title} {\bibinfo {title} {Spectroscopic evidence of the
  {Aharonov-Casher} effect in a {C}ooper pair box},\ }\href
  {https://doi.org/10.1103/PhysRevLett.116.107002} {\bibfield  {journal}
  {\bibinfo  {journal} {Phys. Rev. Lett.}\ }\textbf {\bibinfo {volume} {116}},\
  \bibinfo {pages} {107002} (\bibinfo {year} {2016})}\BibitemShut {NoStop}%
\bibitem [{\citenamefont {Protopopov}\ and\ \citenamefont
  {Feigel'man}(2004)}]{protopopov2004}%
  \BibitemOpen
  \bibfield  {author} {\bibinfo {author} {\bibfnamefont {I.~V.}\ \bibnamefont
  {Protopopov}}\ and\ \bibinfo {author} {\bibfnamefont {M.~V.}\ \bibnamefont
  {Feigel'man}},\ }\bibfield  {title} {\bibinfo {title} {Anomalous periodicity
  of supercurrent in long frustrated {Josephson-junction} rhombi chains},\
  }\href {https://doi.org/10.1103/PhysRevB.70.184519} {\bibfield  {journal}
  {\bibinfo  {journal} {Phys. Rev. B}\ }\textbf {\bibinfo {volume} {70}},\
  \bibinfo {pages} {184519} (\bibinfo {year} {2004})}\BibitemShut {NoStop}%
\bibitem [{\citenamefont {Protopopov}\ and\ \citenamefont
  {Feigel'man}(2006)}]{protopopov2006}%
  \BibitemOpen
  \bibfield  {author} {\bibinfo {author} {\bibfnamefont {I.~V.}\ \bibnamefont
  {Protopopov}}\ and\ \bibinfo {author} {\bibfnamefont {M.~V.}\ \bibnamefont
  {Feigel'man}},\ }\bibfield  {title} {\bibinfo {title} {Coherent transport in
  {Josephson-junction} rhombi chain with quenched disorder},\ }\href
  {https://doi.org/10.1103/PhysRevB.74.064516} {\bibfield  {journal} {\bibinfo
  {journal} {Phys. Rev. B}\ }\textbf {\bibinfo {volume} {74}},\ \bibinfo
  {pages} {064516} (\bibinfo {year} {2006})}\BibitemShut {NoStop}%
\bibitem [{\citenamefont {Friedman}\ and\ \citenamefont
  {Averin}(2002)}]{Friedman2002}%
  \BibitemOpen
  \bibfield  {author} {\bibinfo {author} {\bibfnamefont {J.~R.}\ \bibnamefont
  {Friedman}}\ and\ \bibinfo {author} {\bibfnamefont {D.~V.}\ \bibnamefont
  {Averin}},\ }\bibfield  {title} {\bibinfo {title} {Aharonov-casher-effect
  suppression of macroscopic tunneling of magnetic flux},\ }\href
  {https://doi.org/10.1103/PhysRevLett.88.050403} {\bibfield  {journal}
  {\bibinfo  {journal} {Phys. Rev. Lett.}\ }\textbf {\bibinfo {volume} {88}},\
  \bibinfo {pages} {050403} (\bibinfo {year} {2002})}\BibitemShut {NoStop}%
\bibitem [{\citenamefont {Weiss}\ \emph {et~al.}(2019)\citenamefont {Weiss},
  \citenamefont {Li}, \citenamefont {Ferguson},\ and\ \citenamefont
  {Koch}}]{weiss2019}%
  \BibitemOpen
  \bibfield  {author} {\bibinfo {author} {\bibfnamefont {D.~K.}\ \bibnamefont
  {Weiss}}, \bibinfo {author} {\bibfnamefont {A.~C.~Y.}\ \bibnamefont {Li}},
  \bibinfo {author} {\bibfnamefont {D.~G.}\ \bibnamefont {Ferguson}},\ and\
  \bibinfo {author} {\bibfnamefont {J.}~\bibnamefont {Koch}},\ }\bibfield
  {title} {\bibinfo {title} {Spectrum and coherence properties of the
  current-mirror qubit},\ }\href {https://doi.org/10.1103/PhysRevB.100.224507}
  {\bibfield  {journal} {\bibinfo  {journal} {Phys. Rev. B}\ }\textbf {\bibinfo
  {volume} {100}},\ \bibinfo {pages} {224507} (\bibinfo {year}
  {2019})}\BibitemShut {NoStop}%
\bibitem [{\citenamefont {Larsen}\ \emph {et~al.}(2020)\citenamefont {Larsen},
  \citenamefont {Gershenson}, \citenamefont {Casparis}, \citenamefont
  {Kringh\o{}j}, \citenamefont {Pearson}, \citenamefont {McNeil}, \citenamefont
  {Kuemmeth}, \citenamefont {Krogstrup}, \citenamefont {Petersson},\ and\
  \citenamefont {Marcus}}]{larsen2020}%
  \BibitemOpen
  \bibfield  {author} {\bibinfo {author} {\bibfnamefont {T.~W.}\ \bibnamefont
  {Larsen}}, \bibinfo {author} {\bibfnamefont {M.~E.}\ \bibnamefont
  {Gershenson}}, \bibinfo {author} {\bibfnamefont {L.}~\bibnamefont
  {Casparis}}, \bibinfo {author} {\bibfnamefont {A.}~\bibnamefont
  {Kringh\o{}j}}, \bibinfo {author} {\bibfnamefont {N.~J.}\ \bibnamefont
  {Pearson}}, \bibinfo {author} {\bibfnamefont {R.~P.~G.}\ \bibnamefont
  {McNeil}}, \bibinfo {author} {\bibfnamefont {F.}~\bibnamefont {Kuemmeth}},
  \bibinfo {author} {\bibfnamefont {P.}~\bibnamefont {Krogstrup}}, \bibinfo
  {author} {\bibfnamefont {K.~D.}\ \bibnamefont {Petersson}},\ and\ \bibinfo
  {author} {\bibfnamefont {C.~M.}\ \bibnamefont {Marcus}},\ }\bibfield  {title}
  {\bibinfo {title} {Parity-protected superconductor-semiconductor qubit},\
  }\href {https://doi.org/10.1103/PhysRevLett.125.056801} {\bibfield  {journal}
  {\bibinfo  {journal} {Phys. Rev. Lett.}\ }\textbf {\bibinfo {volume} {125}},\
  \bibinfo {pages} {056801} (\bibinfo {year} {2020})}\BibitemShut {NoStop}%
\bibitem [{\citenamefont {Aguado}(2020)}]{ramon2020}%
  \BibitemOpen
  \bibfield  {author} {\bibinfo {author} {\bibfnamefont {R.}~\bibnamefont
  {Aguado}},\ }\bibfield  {title} {\bibinfo {title} {A perspective on
  semiconductor-based superconducting qubits},\ }\href
  {https://doi.org/10.1063/5.0024124} {\bibfield  {journal} {\bibinfo
  {journal} {Applied Physics Letters}\ }\textbf {\bibinfo {volume} {117}},\
  \bibinfo {pages} {240501} (\bibinfo {year} {2020})},\ \Eprint
  {https://arxiv.org/abs/https://doi.org/10.1063/5.0024124}
  {https://doi.org/10.1063/5.0024124} \BibitemShut {NoStop}%
\bibitem [{\citenamefont {Casparis}\ \emph {et~al.}(2018)\citenamefont
  {Casparis}, \citenamefont {Connolly}, \citenamefont {Kjaergaard},
  \citenamefont {Pearson}, \citenamefont {Kringh{\o}j}, \citenamefont {Larsen},
  \citenamefont {Kuemmeth}, \citenamefont {Wang}, \citenamefont {Thomas},
  \citenamefont {Gronin}, \citenamefont {Gardner}, \citenamefont {Manfra},
  \citenamefont {Marcus},\ and\ \citenamefont {Petersson}}]{casparis2018}%
  \BibitemOpen
  \bibfield  {author} {\bibinfo {author} {\bibfnamefont {L.}~\bibnamefont
  {Casparis}}, \bibinfo {author} {\bibfnamefont {M.~R.}\ \bibnamefont
  {Connolly}}, \bibinfo {author} {\bibfnamefont {M.}~\bibnamefont
  {Kjaergaard}}, \bibinfo {author} {\bibfnamefont {N.~J.}\ \bibnamefont
  {Pearson}}, \bibinfo {author} {\bibfnamefont {A.}~\bibnamefont
  {Kringh{\o}j}}, \bibinfo {author} {\bibfnamefont {T.~W.}\ \bibnamefont
  {Larsen}}, \bibinfo {author} {\bibfnamefont {F.}~\bibnamefont {Kuemmeth}},
  \bibinfo {author} {\bibfnamefont {T.}~\bibnamefont {Wang}}, \bibinfo {author}
  {\bibfnamefont {C.}~\bibnamefont {Thomas}}, \bibinfo {author} {\bibfnamefont
  {S.}~\bibnamefont {Gronin}}, \bibinfo {author} {\bibfnamefont {G.~C.}\
  \bibnamefont {Gardner}}, \bibinfo {author} {\bibfnamefont {M.~J.}\
  \bibnamefont {Manfra}}, \bibinfo {author} {\bibfnamefont {C.~M.}\
  \bibnamefont {Marcus}},\ and\ \bibinfo {author} {\bibfnamefont {K.~D.}\
  \bibnamefont {Petersson}},\ }\bibfield  {title} {\bibinfo {title}
  {Superconducting gatemon qubit based on a proximitized two-dimensional
  electron gas},\ }\href {https://doi.org/10.1038/s41565-018-0207-y} {\bibfield
   {journal} {\bibinfo  {journal} {Nature Nanotechnology}\ }\textbf {\bibinfo
  {volume} {13}},\ \bibinfo {pages} {915} (\bibinfo {year} {2018})}\BibitemShut
  {NoStop}%
\bibitem [{\citenamefont {Kringh\o{}j}\ \emph {et~al.}(2020)\citenamefont
  {Kringh\o{}j}, \citenamefont {van Heck}, \citenamefont {Larsen},
  \citenamefont {Erlandsson}, \citenamefont {Sabonis}, \citenamefont
  {Krogstrup}, \citenamefont {Casparis}, \citenamefont {Petersson},\ and\
  \citenamefont {Marcus}}]{kringhoj2020}%
  \BibitemOpen
  \bibfield  {author} {\bibinfo {author} {\bibfnamefont {A.}~\bibnamefont
  {Kringh\o{}j}}, \bibinfo {author} {\bibfnamefont {B.}~\bibnamefont {van
  Heck}}, \bibinfo {author} {\bibfnamefont {T.~W.}\ \bibnamefont {Larsen}},
  \bibinfo {author} {\bibfnamefont {O.}~\bibnamefont {Erlandsson}}, \bibinfo
  {author} {\bibfnamefont {D.}~\bibnamefont {Sabonis}}, \bibinfo {author}
  {\bibfnamefont {P.}~\bibnamefont {Krogstrup}}, \bibinfo {author}
  {\bibfnamefont {L.}~\bibnamefont {Casparis}}, \bibinfo {author}
  {\bibfnamefont {K.~D.}\ \bibnamefont {Petersson}},\ and\ \bibinfo {author}
  {\bibfnamefont {C.~M.}\ \bibnamefont {Marcus}},\ }\bibfield  {title}
  {\bibinfo {title} {Suppressed charge dispersion via resonant tunneling in a
  single-channel transmon},\ }\href
  {https://doi.org/10.1103/PhysRevLett.124.246803} {\bibfield  {journal}
  {\bibinfo  {journal} {Phys. Rev. Lett.}\ }\textbf {\bibinfo {volume} {124}},\
  \bibinfo {pages} {246803} (\bibinfo {year} {2020})}\BibitemShut {NoStop}%
\bibitem [{\citenamefont {Bargerbos}\ \emph {et~al.}(2020)\citenamefont
  {Bargerbos}, \citenamefont {Uilhoorn}, \citenamefont {Yang}, \citenamefont
  {Krogstrup}, \citenamefont {Kouwenhoven}, \citenamefont {de~Lange},
  \citenamefont {van Heck},\ and\ \citenamefont {Kou}}]{bargerbos2020}%
  \BibitemOpen
  \bibfield  {author} {\bibinfo {author} {\bibfnamefont {A.}~\bibnamefont
  {Bargerbos}}, \bibinfo {author} {\bibfnamefont {W.}~\bibnamefont {Uilhoorn}},
  \bibinfo {author} {\bibfnamefont {C.-K.}\ \bibnamefont {Yang}}, \bibinfo
  {author} {\bibfnamefont {P.}~\bibnamefont {Krogstrup}}, \bibinfo {author}
  {\bibfnamefont {L.~P.}\ \bibnamefont {Kouwenhoven}}, \bibinfo {author}
  {\bibfnamefont {G.}~\bibnamefont {de~Lange}}, \bibinfo {author}
  {\bibfnamefont {B.}~\bibnamefont {van Heck}},\ and\ \bibinfo {author}
  {\bibfnamefont {A.}~\bibnamefont {Kou}},\ }\bibfield  {title} {\bibinfo
  {title} {Observation of vanishing charge dispersion of a nearly open
  superconducting island},\ }\href
  {https://doi.org/10.1103/PhysRevLett.124.246802} {\bibfield  {journal}
  {\bibinfo  {journal} {Phys. Rev. Lett.}\ }\textbf {\bibinfo {volume} {124}},\
  \bibinfo {pages} {246802} (\bibinfo {year} {2020})}\BibitemShut {NoStop}%
\bibitem [{\citenamefont {Martinis}\ and\ \citenamefont
  {Osborne}(2004)}]{martinis2004}%
  \BibitemOpen
  \bibfield  {author} {\bibinfo {author} {\bibfnamefont {J.~M.}\ \bibnamefont
  {Martinis}}\ and\ \bibinfo {author} {\bibfnamefont {K.}~\bibnamefont
  {Osborne}},\ }\href@noop {} {\bibinfo {title} {Superconducting qubits and the
  physics of {Josephson} junctions}} (\bibinfo {year} {2004}),\ \Eprint
  {https://arxiv.org/abs/cond-mat/0402415} {arXiv:cond-mat/0402415
  [cond-mat.supr-con]} \BibitemShut {NoStop}%
\bibitem [{\citenamefont {Doh}\ \emph {et~al.}(2005)\citenamefont {Doh},
  \citenamefont {van Dam}, \citenamefont {Roest}, \citenamefont {Bakkers},
  \citenamefont {Kouwenhoven},\ and\ \citenamefont {De~Franceschi}}]{doh2005}%
  \BibitemOpen
  \bibfield  {author} {\bibinfo {author} {\bibfnamefont {Y.-J.}\ \bibnamefont
  {Doh}}, \bibinfo {author} {\bibfnamefont {J.~A.}\ \bibnamefont {van Dam}},
  \bibinfo {author} {\bibfnamefont {A.~L.}\ \bibnamefont {Roest}}, \bibinfo
  {author} {\bibfnamefont {E.~P. A.~M.}\ \bibnamefont {Bakkers}}, \bibinfo
  {author} {\bibfnamefont {L.~P.}\ \bibnamefont {Kouwenhoven}},\ and\ \bibinfo
  {author} {\bibfnamefont {S.}~\bibnamefont {De~Franceschi}},\ }\bibfield
  {title} {\bibinfo {title} {Tunable supercurrent through semiconductor
  nanowires},\ }\href {https://doi.org/10.1126/science.1113523} {\bibfield
  {journal} {\bibinfo  {journal} {Science}\ }\textbf {\bibinfo {volume}
  {309}},\ \bibinfo {pages} {272} (\bibinfo {year} {2005})}\BibitemShut
  {NoStop}%
\bibitem [{\citenamefont {Krogstrup}\ \emph {et~al.}(2015)\citenamefont
  {Krogstrup}, \citenamefont {Ziino}, \citenamefont {Chang}, \citenamefont
  {Albrecht}, \citenamefont {Madsen}, \citenamefont {Johnson}, \citenamefont
  {Nyg{\aa}rd}, \citenamefont {Marcus},\ and\ \citenamefont
  {Jespersen}}]{krogstrup2015}%
  \BibitemOpen
  \bibfield  {author} {\bibinfo {author} {\bibfnamefont {P.}~\bibnamefont
  {Krogstrup}}, \bibinfo {author} {\bibfnamefont {N.~L.~B.}\ \bibnamefont
  {Ziino}}, \bibinfo {author} {\bibfnamefont {W.}~\bibnamefont {Chang}},
  \bibinfo {author} {\bibfnamefont {S.~M.}\ \bibnamefont {Albrecht}}, \bibinfo
  {author} {\bibfnamefont {M.~H.}\ \bibnamefont {Madsen}}, \bibinfo {author}
  {\bibfnamefont {E.}~\bibnamefont {Johnson}}, \bibinfo {author} {\bibfnamefont
  {J.}~\bibnamefont {Nyg{\aa}rd}}, \bibinfo {author} {\bibfnamefont {C.~M.}\
  \bibnamefont {Marcus}},\ and\ \bibinfo {author} {\bibfnamefont {T.~S.}\
  \bibnamefont {Jespersen}},\ }\bibfield  {title} {\bibinfo {title} {Epitaxy of
  semiconductor--superconductor nanowires},\ }\href
  {https://doi.org/10.1038/nmat4176} {\bibfield  {journal} {\bibinfo  {journal}
  {Nature Materials}\ }\textbf {\bibinfo {volume} {14}},\ \bibinfo {pages}
  {400} (\bibinfo {year} {2015})}\BibitemShut {NoStop}%
\bibitem [{\citenamefont {Larsen}\ \emph {et~al.}(2015)\citenamefont {Larsen},
  \citenamefont {Petersson}, \citenamefont {Kuemmeth}, \citenamefont
  {Jespersen}, \citenamefont {Krogstrup}, \citenamefont {Nyg\aa{}rd},\ and\
  \citenamefont {Marcus}}]{larsen2015}%
  \BibitemOpen
  \bibfield  {author} {\bibinfo {author} {\bibfnamefont {T.~W.}\ \bibnamefont
  {Larsen}}, \bibinfo {author} {\bibfnamefont {K.~D.}\ \bibnamefont
  {Petersson}}, \bibinfo {author} {\bibfnamefont {F.}~\bibnamefont {Kuemmeth}},
  \bibinfo {author} {\bibfnamefont {T.~S.}\ \bibnamefont {Jespersen}}, \bibinfo
  {author} {\bibfnamefont {P.}~\bibnamefont {Krogstrup}}, \bibinfo {author}
  {\bibfnamefont {J.}~\bibnamefont {Nyg\aa{}rd}},\ and\ \bibinfo {author}
  {\bibfnamefont {C.~M.}\ \bibnamefont {Marcus}},\ }\bibfield  {title}
  {\bibinfo {title} {Semiconductor-nanowire-based superconducting qubit},\
  }\href {https://doi.org/10.1103/PhysRevLett.115.127001} {\bibfield  {journal}
  {\bibinfo  {journal} {Phys. Rev. Lett.}\ }\textbf {\bibinfo {volume} {115}},\
  \bibinfo {pages} {127001} (\bibinfo {year} {2015})}\BibitemShut {NoStop}%
\bibitem [{\citenamefont {de~Lange}\ \emph {et~al.}(2015)\citenamefont
  {de~Lange}, \citenamefont {van Heck}, \citenamefont {Bruno}, \citenamefont
  {van Woerkom}, \citenamefont {Geresdi}, \citenamefont {Plissard},
  \citenamefont {Bakkers}, \citenamefont {Akhmerov},\ and\ \citenamefont
  {DiCarlo}}]{lange2015}%
  \BibitemOpen
  \bibfield  {author} {\bibinfo {author} {\bibfnamefont {G.}~\bibnamefont
  {de~Lange}}, \bibinfo {author} {\bibfnamefont {B.}~\bibnamefont {van Heck}},
  \bibinfo {author} {\bibfnamefont {A.}~\bibnamefont {Bruno}}, \bibinfo
  {author} {\bibfnamefont {D.~J.}\ \bibnamefont {van Woerkom}}, \bibinfo
  {author} {\bibfnamefont {A.}~\bibnamefont {Geresdi}}, \bibinfo {author}
  {\bibfnamefont {S.~R.}\ \bibnamefont {Plissard}}, \bibinfo {author}
  {\bibfnamefont {E.~P. A.~M.}\ \bibnamefont {Bakkers}}, \bibinfo {author}
  {\bibfnamefont {A.~R.}\ \bibnamefont {Akhmerov}},\ and\ \bibinfo {author}
  {\bibfnamefont {L.}~\bibnamefont {DiCarlo}},\ }\bibfield  {title} {\bibinfo
  {title} {Realization of microwave quantum circuits using hybrid
  superconducting-semiconducting nanowire {Josephson} elements},\ }\href
  {https://doi.org/10.1103/PhysRevLett.115.127002} {\bibfield  {journal}
  {\bibinfo  {journal} {Phys. Rev. Lett.}\ }\textbf {\bibinfo {volume} {115}},\
  \bibinfo {pages} {127002} (\bibinfo {year} {2015})}\BibitemShut {NoStop}%
\bibitem [{\citenamefont {Golubov}\ \emph {et~al.}(2004)\citenamefont
  {Golubov}, \citenamefont {Kupriyanov},\ and\ \citenamefont
  {Il'ichev}}]{golubov2004}%
  \BibitemOpen
  \bibfield  {author} {\bibinfo {author} {\bibfnamefont {A.~A.}\ \bibnamefont
  {Golubov}}, \bibinfo {author} {\bibfnamefont {M.~Y.}\ \bibnamefont
  {Kupriyanov}},\ and\ \bibinfo {author} {\bibfnamefont {E.}~\bibnamefont
  {Il'ichev}},\ }\bibfield  {title} {\bibinfo {title} {The current-phase
  relation in {Josephson} junctions},\ }\href
  {https://doi.org/10.1103/RevModPhys.76.411} {\bibfield  {journal} {\bibinfo
  {journal} {Rev. Mod. Phys.}\ }\textbf {\bibinfo {volume} {76}},\ \bibinfo
  {pages} {411} (\bibinfo {year} {2004})}\BibitemShut {NoStop}%
\end{thebibliography}

%

\end{document}